# Modeling of high-pressure transient gas-liquid flow in M-shaped jumpers of subsea gas production systems

A. Yurishchev[✉], N. Brauner, A. Ullmann

*School of Mechanical Engineering, Tel-Aviv University, Tel-Aviv 6997801, Israel*

## Abstract

Two-phase flow with low liquid loads is common in high-pressure natural gas offshore gathering and transmission pipelines. During gas production slowdowns or shutdowns, an accumulation of liquid in the lower sections of subsea pipelines may occur. This phenomenon is observed in jumpers that connect different units in deep-water subsea gas production facilities. The displacement of the accumulated liquid during production ramp-up induces temporal variations in pressure drop across the jumper and forces on its elbows, resulting in flow-induced vibrations (FIV) that pose potential risks to the structural integrity of the jumper. To bridge the gap between laboratory experiments and field conditions, transient 3D numerical simulations were conducted using the OpenFOAM software. These simulations facilitated the development of a mechanistic model to elucidate the factors contributing to increased pressure and forces during the liquid purging process. The study examined the influence of gas pressure level, pipe diameter, initially accumulated liquid amount, liquid properties, and gas mass flow rate on the transient pressure drop and the forces acting on the jumper's elbows. The critical gas production rate required for complete liquid removal of the accumulated liquid was determined, and scaling rules were proposed to predict the effects of gas pressure and pipe diameter on this critical value. The dominant frequencies of pressure and force fluctuations were identified, with low-pressure systems exhibiting frequencies associated with two-phase flow phenomena and high-pressure systems showing frequencies attributed to acoustic waves.

[✉] Corresponding author
*E-mail address*: yurishchev@mail.tau.ac.il (Alexander Yurishchev)

# 1 Introduction

Subsea jumpers are components in deep-water offshore gas production systems. These structures serve as connection systems, linking various pipeline points. For example, they can connect the wellhead to the Infield Gathering Manifold to the export flow lines through Pipeline End Terminations. The jumpers are built in various shapes, such as U-shape, M-shape, S-shape, L-shape, and Z-shape. These configurations are achieved by utilizing steel bends, tees, and elbows as required. The jumpers are designed to accommodate high static and dynamic loads that might be caused, for instance, by high internal pressure, thermal effects, internal flow patterns and external subsea currents.

The presence of a liquid phase, formed by gas condensates, produced reservoir water, and injected hydrate inhibitors, such as mono-ethylene glycol (MEG) solutions (e.g.,Cha et al., 2013; Liu et al., 2021; Badi et al., 2023; Hamidpour et al., 2023) results in two-phase flow phenomena that may frequently challenge the natural gas transportation. The hilly terrain offshore surface forces the transportation pipelines to include low sections, where the liquid phase may accumulate during low gas production rate stages or complete shutdown (e.g., Hamami Bissor et al., 2020; Duan et al., 2022; Liao et al., 2023). Naturally, subsea jumpers include low sections as well, thus they may encounter the same issue of liquid accumulation. In contrast to hilly terrain pipelines that consist of slightly inclined sections, typical jumpers often include vertical sections, so a high liquid amount displacement may be associated with complex flow patterns (e.g., churn, slug flow). Proper gas transportation relies on safe liquid phase removal from the jumpers during the production restart and preventing its further accumulation during the continuous operation. Consequently, the determination of the critical (minimal) gas flow rate needed to ensure these conditions is crucial.

The critical gas flow rate needed to drive the liquid phase upwards in vertical or off-vertical pipes was investigated in the two-phase flow literature. The critical conditions in such systems are usually associated with force balance on the entrained droplets into the gas core (e.g., Turner et al., 1969; Coleman et al., 1991; Li et al., 2002; Guo et al., 2006; Sawant et al., 2008), or with the prevention of liquid flow reversal in annular flow (e.g., Hewitt et al., 1985; Zabaras et al., 1986; Belfroid et al., 2008; Ma et al., 2023). Avoiding liquid flow reversal is often correlated with the transition to annular flow pattern and liquid annulus stabilization. Ohnuki and Akimoto, 2000; Prasser et al., 2005; Skopich et al., 2015; Sharaf et al., 2016; Vieira et al., 2021; Wang et al., 2023 investigated the flow characteristics in atmospheric air-water systems considering various pipe diameters, $D = 51 \div 200$ mm. The results suggested that flow pattern transitions depend on the pipe diameter. For instance, Ohnuki and Akimoto, 2000 revealed that the churn flow is dominant in large-diameter pipes, whereas slug flow was usually reported for small-diameter pipes. However, the obtained knowledge based on those, and similar studies might be inapplicable to the phenomena anticipated at typical conditions in subsea jumpers due to significantly higher internal pressures, different gas properties and larger pipe internal diameters. Indeed, the flow pattern transitions observed in an experimental study by Omebere-Iyari et al., 2007 were significantly different from those reported for lower pressures and pipe diameters. This work described the conditions of naphtha-nitrogen two-phase flow in a relatively large, 189 mm diameter pipe at 20 and 90 bars internal pressures. In practice, jumpers may experience even higher pressure and consist of larger diameter sections. In more recent works, Li et al., 2018; Shibata et al., 2022, flow regime transitions were investigated for internal pressures of 1-6 MPa and 3 MPa, respectively. However, they focused on the bubbly to slug regime transition, which is relevant to high liquid loads and relatively low gas flow rates.

Operation conditions, such as gas pressure, gas flow rate along with accumulated liquid amount and jumper's geometry may affect the two-phase flow patterns developed in the jumpers. Periodic flow regimes (e.g., churn and slug flows) could appear during the transient stage of the accumulated liquid displacement upon the gas production start-up. The fluid-structure interactions associated with those cyclic flow patterns may promote potentially harmful Flow-Induced-Vibrations (FIV). These should be considered along with Vortex-Induced-Vibrations (VIV) due to the external deep-sea water currents. The VIV were investigated and reported in the literature, not only for a more general context of flow over cylinders (e.g., Xie et al., 2012; Mannini et al., 2014; Wang et al., 2019), but also exclusively for subsea jumpers (e.g., Wang et al., 2013; Bruschi et al., 2015; Zheng et al., 2015; Gross et al., 2018; Qu et al., 2022). Prolonged transient loads applied on the jumper may lead to a resonance reaction if the load fluctuations' dominant frequency is about the structure's natural frequency. Therefore, the internal two-phase FIV and their effects on the subsea jumper structure should also be investigated to ensure safe gas production and transportation. The FIV and their effects on subsea jumpers are usually explored in association with the slug flow regime. The periodic nature of the slug flow may cause the jumper to vibrate. Van der Heijden et al., 2014 introduced a three-beam-model based on Euler-Bernoulli beams with bending stiffness and mass to mimic U-shaped jumper geometry. The model enables calculating the displacements of the jumper due to the slug flow loads. More realistic two-phase flow characteristics and their effects on the structure may be achieved in 3D numerical simulations, FSI (Fluid Structure Interaction). Kim and Srinil, 2018 investigated numerically the loads applied on the M-shaped jumper's bends, showing potential multiple resonance conditions and structure deformations. In a similar study, Li et al., 2023 combined experiments and numerical simulations to examine flow patterns and recognize risky scenarios under different mixture velocities and volume fractions of the liquid phase. Yuan et al., 2023 tested different gas contents and showed that the gas content of 50% will probably result in slug flow, but for 70%, the annular flow is dominant. Also, the effects of the horizontal and vertical section lengths of the jumper were revealed. Other jumper configurations, such as Z- and U-shapes, were also addressed in the literature (e.g., Li et al., 2022; Zhu et al., 2022, 2023). Those studies reported the structure vibration characteristics that were associated with two-phase flow patterns.

The jumper structures typically include several elbows that are subjected to periodic forces due to the two-phase flow. Some studies (e.g., Abdulkadir et al., 2011; Miwa et al., 2014; Saidj et al., 2014; Garcia et al., 2023) reported on the two-phase flow through horizontal and vertical bends. Cargnelutti et al., 2010 introduced a model based on the momentum change of a slug to predict the forces acting on a 90° bend of a small-scaled tube ($D$ = 6 mm). The model predictions were in good agreement with the reported experimental results for the slug flow regime. Belfroid et al., 2016 extended the research to a larger pipe diameter ($D$ = 152.4 mm) and reported force fluctuation dominant frequencies.

Apparently, most of the previous publications that studied gas-liquid flows in jumpers concentrated on high liquid loads that usually led to periodic slug flow patterns. Those conditions are typical for continuous gas production with large amounts of liquids in the feed. Yet, the two-phase flow dynamics during the production start-up, which is characterized by displacing a finite amount of liquid that accumulated in the lower section, lacks attention in the literature. Such conditions were investigated experimentally and numerically in our previous study, Yurishchev et al., 2024. However, in that work, we introduced a scaled-down M-shaped jumper that operates at atmospheric pressure and, therefore, does not represent the dynamics in real gas production, which is characterized by high-pressure conditions and large-diameter pipes.

In the present study, we reveal the effects of gas mass flow ramp-up rate, initial amount of the accumulated liquid, gas pressure level, liquid properties, and pipe diameter on the transient response of pressure drop on the jumper and the forces acting on the riser's upper elbow. In addition, the critical (minimal) gas mass flow rate required for purging the liquid was determined by numerical simulations for various operational conditions, and its dependency on the gas pressure level and pipe diameter was demonstrated and compared to suggested scaling rules. A modified mechanistic model based on the model introduced by Yurishchev et al., 2024 used to predict the transient pressure variation and force acting on the elbow. The CFD simulation and the model are used to explore the dominant pressure fluctuation frequencies during the transient process and to identify their source.

## 2 Numerical model

A three-dimensional numerical model was employed to simulate the transient two-phase flow phenomena of liquid and compressible gas during the displacement of the liquid and its removal from the jumper. Numerical solutions were obtained using OpenFOAM, an open-source CFD software. A geometry of an M-shaped jumper is considered, which is scaled by its internal pipe diameter, $D$ (see **Figure 2.1**). In our previous study, Yurishchev et al., 2024, the validity of the numerical model (adequate solver, turbulence model, boundary conditions, discretization schemes, etc.) has been established based on the experimental results acquired using an atmospheric experimental facility of M-shaped jumper (internal pipe diameter $D$ = 50 mm). The relevant numerical parameters are described in the following sections. It was shown that the numerical simulation results reasonably agree with the experimental data on (1) the critical gas flow rate needed for the removal of the accumulated liquid in the lower horizontal section of the jumper, (2) pressure drop on the jumper, and (3) loads applied on the jumper's elbows. Therefore, the current study refers to that previous work and expands the investigation to reveal the effects of different pipe internal diameters, gas mass flow ramp-up rate, accumulated liquid amount, gas pressure level, pipe diameter, and liquid properties (see **Table 2.1**) on the flow phenomena for a set of higher than atmospheric gas pressures for air and methane (**Table 2.2** and **Table 2.3**, respectively).

**Table 2.1**: Liquids' physical properties

| ***Liquid Property*** | ***Liquid*** | | |
|---|---|---|---|
| | Water | MEG50% | MEG100% |
| Density - $\rho_L$ [kg/m$^3$] | 998 | 1079 | 1130 |
| Dynamic Viscosity - $\mu_L$ [mPa·s] | 1 | 4.6 | 16.5 |
| Surface Tension - $\sigma$ [mN/m] | 72 | 55 | 48 |

**Table 2.2:** Air physical properties for different pressures (1 ÷ 5 atm).

| ***Air Property*** | ***Outlet Pressure – $P_O$* [atm]** | | |
|---|---|---|---|
| | 1 | 2 | 5 |
| Critical Temperature - $T_c$ [K] | | 130.14 | |
| Critical Pressure - $P_c$ [MPa] | | 3.43 | |
| Acentric factor - $\omega$ | | 0.0337 | |
| Molar Mass - $M$ [kg/kmol] | | 28.96 | |
| Dynamic Viscosity - $\mu_G$ [$\mu$Pa·s] | | 18.1 | |
| Density - $\rho_G$ [kg/m$^3$] | 1.18 | 2.36 | 5.91 |

**Table 2.3**: Methane physical properties for different pressures (20 ÷ 300 atm).

| ***Methane Property*** | ***Outlet Pressure – $P_O$* [atm]** | | | | |
|---|---|---|---|---|---|
| | 20 | 50 | 100 | 200 | 300 |
| Critical Temperature - $T_c$ [K] | 190.56 | | | | |
| Critical Pressure - $P_c$ [MPa] | 4.59 | | | | |
| Acentric factor - $\omega$ | 0.0115 | | | | |
| Molar Mass - $M$ [kg/kmol] | 16.04 | | | | |
| Dynamic Viscosity - $\mu_G$ [$\mu$Pa·s] | 10.91 | 11.72 | 13.94 | 20.83 | 26.13 |
| Density - $\rho_G$ [kg/m$^3$] | 14.59 | 39.11 | 86.64 | 178.37 | 216.06 |
| Compressibility factor - $Z$ | 0.956 | 0.893 | 0.806 | 0.781 | 0.921 |
| Effective Molar Mass ($M/Z$)- $M_{eff}$ [kg/kmol] | 16.77 | 17.96 | 19.91 | 20.56 | 17.43 |
| Speed of sound – $C_G$ [m/s] | 428.25 | 420.21 | 426.04 | 535.11 | 667.71 |

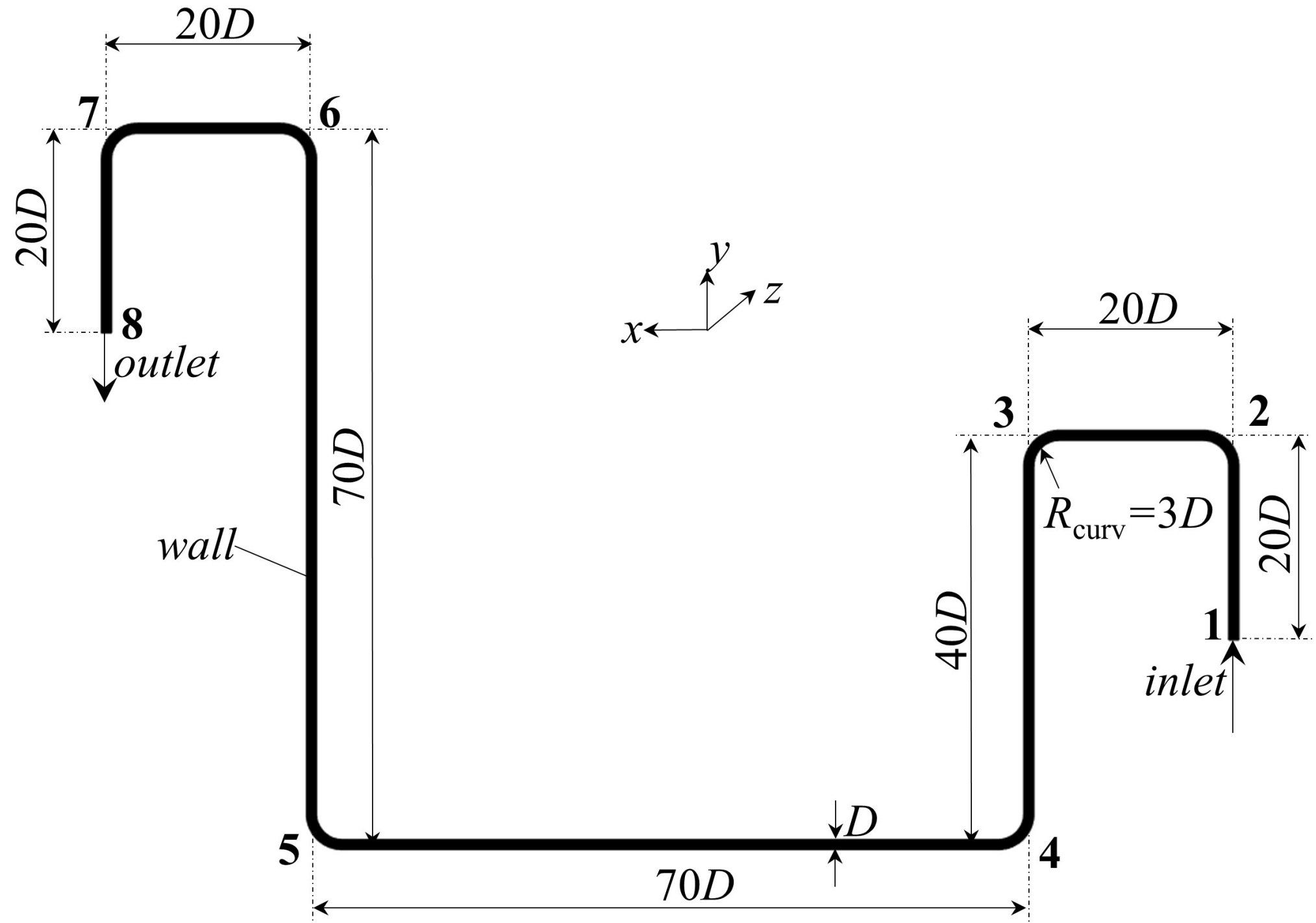

**Figure 2.1:** M-shaped jumper geometry scaled by the internal pipe diameter ($D$).

## 2.1 Numerical Solver

To simulate the flow phenomena, the "compressibleInterFoam" solver was utilized. The solver is well-suited for modeling two-phase compressible, non-isothermal, immiscible fluids. It implements a Volume of Fluid (VOF) approach based on interface capturing methods developed by Hirt and Nichols, 1981 and implemented in OpenFOAM by Roenby et al., 2017. In the VOF method, the phase fraction is assigned a value of 1 in elements exclusively occupied by one phase (e.g., liquid) and 0 in elements containing the second phase (e.g., gas). The interface between the phases is defined by phase fractions ranging from 0 to 1.

The governing equations are the continuity, momentum, and energy equations. Additionally, the VOF model provides an extra equation that governs the volume fraction. Adjusting the "compression coefficient" in the volume fraction equation, $C_{\alpha}$, to a value of 2 (the default value is 1) enhanced the sharpness of the interface, as described by Magnini et al., 2018. To simulate the turbulent flow, the Realizable $\kappa$-$\varepsilon$ turbulence model is used, which requires additional transport equations for turbulence kinetic energy ($\kappa$) and turbulence dissipation rate ($\varepsilon$). Yurishchev et al., 2024 found this model to be much more reliable than the standard $\kappa$-$\varepsilon$ model in the prediction of experimental results, with a similar quality to that of the $\kappa$-$\omega$ SST simulations, while requiring considerably lower computational resources. Therefore, employing this turbulence model should provide reliable predictions and insights for the intended parametric investigations (e.g., various gas pressures and pipe diameters). The PISO algorithm was employed for pressure-velocity coupling. The algorithm predicts the velocity field implicitly and utilizes an explicit first-order pressure correction. The first-order Euler scheme was applied to time derivatives in order to enhance solution stability. Gradients were computed using the second-order Gauss linear scheme. The interface compression was applied to reduce interface smearing. The second-order Gauss vanLeer interfaceCompression scheme was used for the convective terms in the volume fraction equation, and Gauss linear scheme was used for the convection terms in other transport equations. The interpolation scheme, responsible for determining the variable value at a face based on the values at the cells' centroids on either side of that face, was set to a linear interpolation method.

To represent the variation of the real gas density with the pressure and temperature, the compressibility factor, $Z$, was introduced in the equation of state, $\rho_G=P_G M/(Z\bar{R}T)$, where $\bar{R}$ is the universal gas constant, $\rho$ is the gas density, $P_G$ is the gas pressure, $T$ is the gas temperature and $M$ is the gas molar mass. The compressibility factor for air for tested outlet pressure range $1 \div 5$ atm is ~ 1 (e.g., Jones, 1983). However, the compressibility factor for methane varies with pressure (see **Table 2.3**). To simplify the calculations, we assumed that the compressibility factor $Z$ is constant for each tested case and determined it based on the outlet pressure level and temperature. We show later that the amplitude of the pressure variations for tested cases with methane is rather small compared to the outlet pressure level and that the temperature variations are negligible. Thus, the assumption of constant $Z$ is practically justified. Accordingly, the gas effective molar mass ($M/Z$, see **Table 2.3**) is applied to the ideal gas equation of state in "compressibleInterFoam" solver settings. To further validate the constant $Z$ assumption, the effects of applying other equations of state for gas available in OpenFOAM (e.g., the Peng and Robinson, 1976 real gas model) were investigated for several cases. The impact on the simulation results was found to be insignificant, while the calculation time increased.

## 2.2 Solution Stability and Convergence Monitors

The convergence of the numerical solution was monitored by examining the residual values. A residual threshold of $10^{-6}$ was set for the velocity, $10^{-7}$ for the pressure, and $10^{-5}$ for the temperature, turbulent kinetic energy ($\kappa$), turbulence dissipation rate ($\varepsilon$), specific turbulence dissipation rate ($\omega$), and phase fraction. Parallel computing was utilized to facilitate the solution process, employing 24 - 96 processors for the solution acceleration. To ensure solution stability, an adjustable time step approach was employed, limiting the Courant number to 0.5. Temporal discretization independence tests were conducted to evaluate the effect of decreasing the maximum Courant number to 0.25 on solution accuracy. However, it was found that reducing the Courant number did not significantly improve solution accuracy. Conversely, increasing the Courant number beyond 1 often led to solution divergence.

## 2.3 Initial and Boundary Conditions

The boundary conditions applied in the study are as follows. (1) The calculation domain *inlet* is subjected to a time-dependent mass flow rate and a constant temperature $T = 280$ K; (2) The domain *wall* is assumed to have a no-slip condition, i.e., zero fluid velocity, and adiabatic wall as the thermal condition; (3) The *outlet* maintains a constant pressure, $P_O$. At time $t = 0$, a certain amount of liquid is placed in the lower horizontal part of the jumper. No gas flow enters the jumper, and the pressure distribution is hydrostatic. The initial configuration of the liquid layer is determined through simulation runs conducted under zero inlet flow conditions. During this phase, the layer stabilizes under the influence of gravitational and surface tension forces. The contact angle was set to $\theta$=30° ensuring an accurate representation of its initial shape and the corresponding pressure field. The contact angle value corresponds to the receding angle of water droplet on a wetted Perspex surface obtained experimentally by Yurishchev et al., 2023. The same contact angle was set for atmospheric and high-pressure system simulations. It is worth noting that industrial jumpers are typically made of steel, and the liquid is not pure water, where the value of the contact angle might be different. Then, for $t > 0$, the gas mass flow, $\dot{m}_{G|in}(t)$, at the inlet changes gradually from zero to a prescribed final constant mass flow rate, $\dot{m}_{G|in,F}$. A realistic smoothed flow-rate variation, with an adjustable time constant, $\tau$: $\dot{m}_{G|in}(t) = \dot{m}_{G|in,F}(1\text{-}e^{-t/\tau}\text{-}t/\tau e^{-t/\tau})$, was used. This mass flow rate corresponds to gas superficial velocity, denoted as $U_{GS}$, and gas volumetric flow rate at Standard Temperature and Pressure (STP), denoted as $Q^0_G$. It is worth noting that for relatively low gas pressures, the pressure at the inlet rises significantly during the transient liquid removal process. Therefore, it is essential to consider the mass flow rate of the gas as a boundary condition, while accounting for its compressibility rather than relying solely on the volumetric flow rate or gas velocity as a boundary condition imposed on the *inlet*. The initial pressure $P_O$ represents the state of the production shutdown when the pressure is set by no-flow conditions downstream to the jumper. The pressure then consists of the separator pressure and hydrostatic pressure drop in the downstream pipeline. Evidently, the pressure at the jumper's outlet should increase during the start-up operation. Nevertheless, the pressure rise at the outlet is relatively slow compared to the simulation time and is assumed to be constant.

## 2.4 Grids generation procedure

The grid generation for the numerical simulation was guided by (1) the near wall restrictions of the Realizable $\kappa$-$\varepsilon$ turbulence model and (2) the need to adequately resolve the phases' interface capture, as elaborated in detail in previous studies (Yurishchev et al., 2022, 2024). Briefly, the Realizable $\kappa$-$\varepsilon$ turbulence model requires the centroid of the cells adjacent to the walls to be located at $y^+ > 30$ to avoid false predictions, where $y^+$ is the dimensionless normal distance from the wall that in this range is associated with the log velocity profile region. The correctness of the theoretical calculations and the $y_p$ value used (see its calculation below) were tested by obtaining the actual $y^+$ values in the simulations and confirming that they were in the recommended range of $30 < y^+ < 300$. Concurrently, the most suitable grid density in the axial and lateral directions was selected based on the spatial convergence tests. The verified grid was then used to perform wide parametric studies. Finally, temporal convergence tests were performed with the selected grid, concluding that a maximal Courant number of 0.5 ensures simulation stability and produces reliable results, as stated in **Section 2.2.**

We used the friction coefficient, $C_f = 0.046\mathrm{Re}^{-0.2}$, and the shear velocity $u_\tau = (\tau_w/\rho_G)^{0.5}$, where $\tau_w = 0.5C_f\rho_G U_{GS}^2$, to calculate the theoretical *dimensional* normal distance from the wall, $y_p = y^+\mu_G/\rho_G u_\tau \propto D^{1-n}\mu_G^n(\rho_G U_{GS})^{-n}$, where $n \approx 0.9$. The calculations suggest that for a given pipe internal diameter $D$, the gas and its superficial velocity range of $U_{GS}/U_{GS|Crit} > 0.25$, the wall-adjacent element thickness can be the same for the whole tested gas pressure range of $1 \div 5$ atm for air and $20 \div 300$ atm for methane. In practice, this allows us to use the same near-wall meshing for a relatively wide gas-pressure range. Yurishchev et al., 2024 presented spatial and temporal convergence tests, and validation against experimental data for an atmospheric air-water system with pipe diameter $D = 50$ mm. So, the same grid is used in this work to study the flow phenomena with air as the gaseous phase in the range of $P_O = 2 \div 5$ atm. For high pressure methane water system, we performed grid independence study based on four grids (see **Table A.1**) and time independence based on Courant number limitations to 0.25, 0.5 and 1. We showed the convergence of the critical gas velocity needed to remove the accumulated liquid from the system, and the time variation of the pressure drop on the jumper and of the force acting on the riser's upper elbow. In addition, we estimated the discretization error. However, the $y_p$ calculations also suggest that the near-wall meshing should differ for air and methane.

An increase in the pipe diameter also requires adjustment of the near-wall treatment. Thus, the spatial and temporal convergence of the simulations of the methane-liquid system and the discretization error estimation are demonstrated in **Appendix 0**.

### 2.5 Forces on elbows

The 3D simulation results can be used to calculate the forces acting on the structure elbows. The force component acting on an elbow can be calculated through the integration of the gauge pressure above the external pressure, $P_{ex}$ and the viscous shear stresses acting on the elbow surface:

$$\begin{cases} F_x = \iint (p_s - P_{ex})\vec{\imath}\, dA_s + \iint (\vec{\imath}\cdot\bar{\bar{\tau}}\cdot\vec{n})\, dA_s \\ F_y = \iint (p_s - P_{ex})\vec{\jmath}\, dA_s + \iint (\vec{\jmath}\cdot\bar{\bar{\tau}}\cdot\vec{n})\, dA_s \end{cases} \tag{2.1}$$

Alternatively, these force components can be deduced from a momentum balance on the fluid flowing through the elbow:

$$\begin{cases} \frac{\partial}{\partial t}\iiint \rho u_x dV_{el} + \iint \rho u_x^2 dA_c = -F_x - \iint (p_{out} - P_{ex}) dA_c \\ \frac{\partial}{\partial t}\iiint \rho u_y dV_{el} - \iint \rho u_y^2 dA_c = -F_y + \iint (p_{in} - P_{ex}) dA_c - \iiint \rho g dV_{el} \end{cases} \tag{2.2}$$

where $F_x$, $F_y$, are the force components applied on the elbow (the force applied on the liquid are $(-F_x + P_{ex}A_c)$, $(-F_y + P_{ex}A_c)$; $\rho$, $u_x$, $u_y$ are the local and instantaneous fluid density and velocity components, respectively; $p_{in}$, $p_{out}$ are the local instantaneous pressures at the inlet and the outlet of the elbow, respectively; $\bar{\bar{\tau}}$ is the viscous stress tensor; $V_{el}$ is the elbow volume; $A_c$, $A_s$ are the flow cross-sectional areas at its inlet/outlet and the elbow wetted area, respectively.

Obviously, both the above ways of calculating the force should yield the same result. However, the calculation via **Eq. 2.1** is simpler, in particular when omitting the wall shear contribution, which was found to be insignificant relatively to the other components (Yurishchev et al., 2024). Note that to focus solely on evaluating the force acting on the elbow due to the fluid flow through it, the static force resulting from the difference between the internal and external pressure under no-flow conditions is eliminated. This was done by equating the external pressure to the outlet pressure, $P_{ex}=P_O$, which is also taken as the initial (no-flow) pressure level in the system.

# 3 Mechanistic modeling

## 3.1 Formulation

Yurishchev et al., 2024 discussed the importance of the pressure variation in determining the loads applied to the pipeline structure during the transient process of the liquid displacement.. The experimental observations and numerical simulations allowed us to formulate a mechanistic model. The proposed model assumes that the gas does not penetrate the liquid phase and that the (incompressible) liquid moves as a continuous plug. With these assumptions, the model aims to provide insights into the pressure variations under different operational conditions, including gas mass flow rate, accumulated liquid amount, physical properties of the fluids involved, and the geometry of the jumper. It was demonstrated that the model predicts the time variations of the measured pressure drop and forces for relatively high initial liquid amounts and air velocities, where the assumption of unaerated liquid plug can be justified. This assumption is apparently less applicable to cases of low liquid amounts and gas velocities. Thus, a similar model is considered in this study for investigating the effects of operational conditions relevant to industrial jumpers. However, the numerical simulation results (see **Section 4** for a detailed discussion) showed that for high-pressure systems, methane at $P_O$ > 20 atm, the gas penetration into the liquid cannot be neglected. Accordingly, the gas-liquid flow in the riser during the liquid displacement can be characterized by the propagation of an aerated slug. Therefore, some modification of the original model formulation is needed to describe the liquid removal process for these cases.

**Figure 3.1** illustrates a typical liquid displacement process observed in the numerical simulations (LHS); and is assumed in the mechanistic model (RHS) for methane-liquid high-pressure systems. In those cases, we considered a typical scenario where the accumulated liquid volume is smaller than the volume of the jumper's horizontal section. The liquid volume is given by $\varepsilon_{i\text{-}hor}V_h$, where $\varepsilon_{i\text{-}hor}$ is the initial holdup in the horizontal section, and $V_h$ is the volume of the horizontal section. At $t = 0$, in the numerical simulations, the prescribed liquid amount rests in the lower horizontal section under the gravitational force (**Figure 3.1.a**). With the onset of gas flow, the liquid is pushed towards the riser's lower elbow. In contrast, in the mechanistic model, the liquid is assumed to be gathered immediately as a liquid plug in the horizontal section near the riser's lower elbow. The plug volume is then equal to the initial accumulated liquid volume $\varepsilon_{i\text{-}hor}V_h$. Then, at $t = t_1 > 0$ (**Figure 3.1.b**), the gas flow drives the liquid towards the riser, penetrates the liquid to create an aerated slug at the riser's lower section, and accelerates it downstream.

As long as liquid remains in the horizontal section, the length of the aerated slug increases over time. Finally, at $t = t_2 > t_1$ (**Figure 3.1.c**), all the liquid has been removed from the horizontal section and is continuously displaced as an aerated slug of a constant length towards the jumper's outlet. The numerical simulations enable estimating the aeration level of the slug, $\alpha = V_{G,S}/V_S$, where $V_{G,S}$ and $V_S$ are the gas and the total volumes of the aerated slug, respectively.

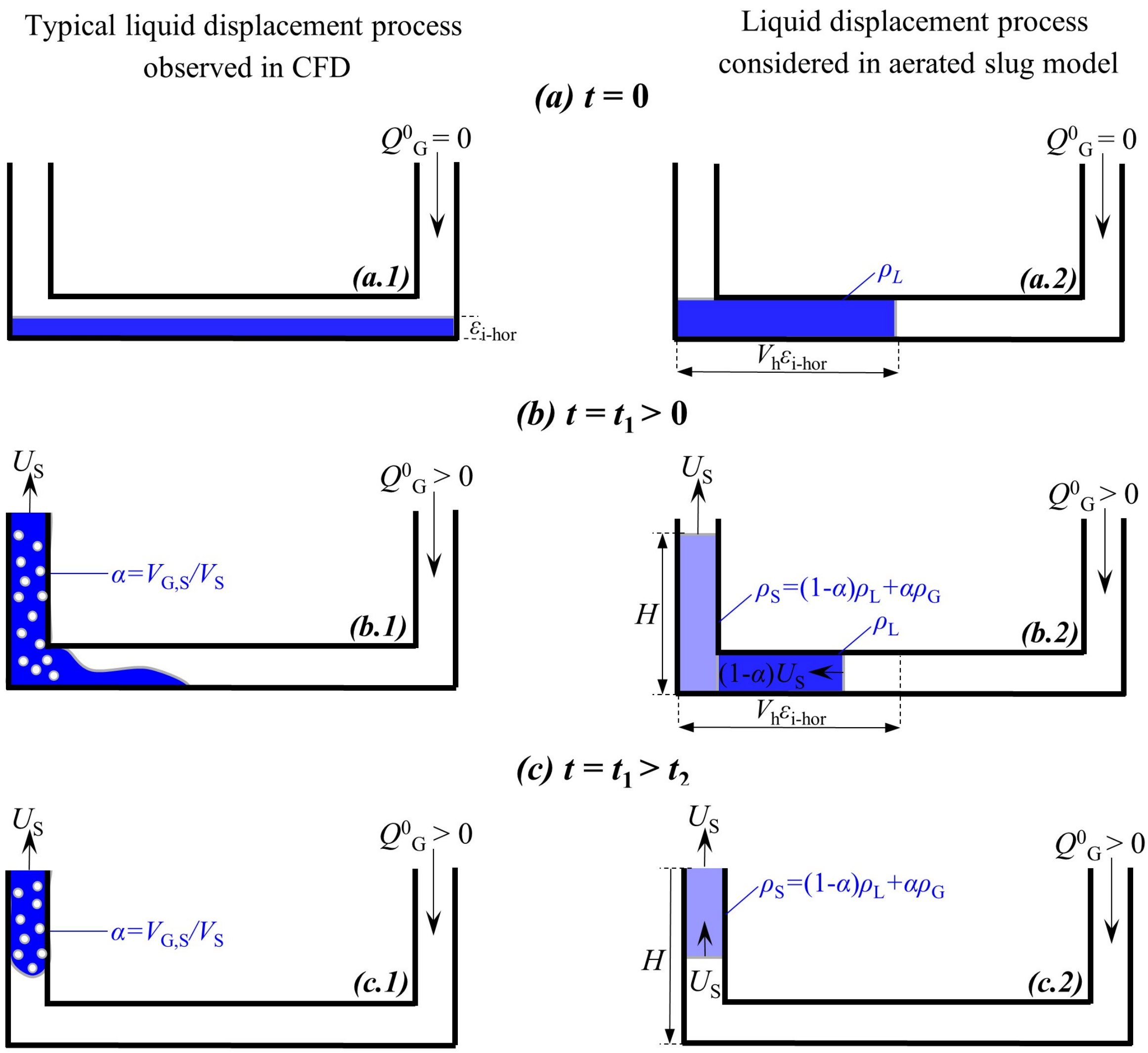

**Figure 3.1:** An illustration of a typical liquid displacement process as observed in numerical simulations (LHS) and liquid displacement process assumed in the mechanistic model (RHS).

**Figure 3.2** shows the scheme of the jumper geometry and the gas-liquid dynamics considered in the current model. The geometry of the jumper was simplified by assuming the outlet is at point 7 (i.e., ignoring the downcomer section between points 7 and 8, **Figure 2.1**). The model allows the determination of three key parameters: (1) the displacement of the liquid in the riser, $H(t)$; (2) the velocity of the aerated liquid slug, $U_S(t)$; and (3) the gas pressure upstream the liquid plug/slug, $P_G(t)$. These are obtained by solving a system of three equations derived from fundamental principles.

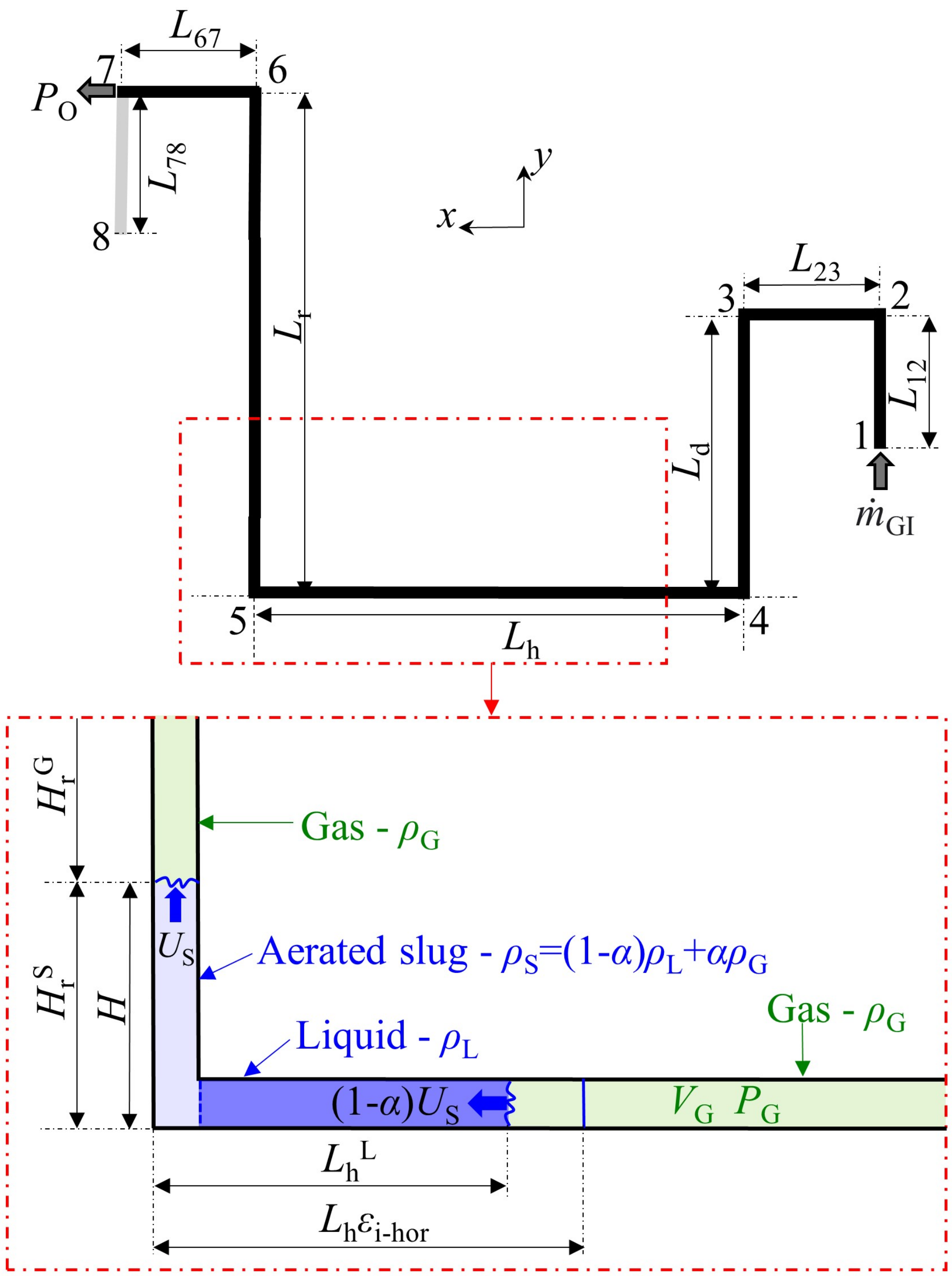

**Figure 3.2**: Scheme of the jumper geometry and gas-liquid dynamics considered in the mechanistic model.

The velocity of the liquid slug front reads:

$$\frac{dH}{dt} = U_S = U_S^L = U_S^G; \qquad H(t) = 0 \tag{3.1}$$

where $H$ is the aerated slug front displacement relative to the initial position, $U_S$ is its velocity, and $U^L{}_S$ and $U^G{}_S$ are the liquid and gas velocities in the aerated slug, respectively, which are equal when no-slip between the phases is assumed.

Momentum balance on the slug (aerated and non-aerated parts) reads:

$$A(P_G - P_O - \rho_G g H_r^G) - A\rho_S g H_r^S - A\rho_G g H_r^G - A\Delta P_f = \rho_S A L_S \frac{dU_S}{dt} + \rho_L A L_L^h (1-\alpha)\frac{dU_S}{dt} \qquad \textbf{3.2}$$

which after rearrangement reads:

$$\frac{dU_S}{dt} = \frac{P_G - P_O - \rho_G g H_r^G - \rho_S g H_r^S - \Delta P_f}{\rho_S L_S + (1-\alpha)\rho_L L_h^L}; \qquad U_S(t=0) = 0$$

where $P_G$ is gas pressure (experienced by the liquid (plug/slug) tail), $P_O$ is the pressure at the outlet (experienced by aerated slug front), $\rho_G$ and $\rho_L$ are the gas and the liquid densities, respectively, $\rho_S=(1-\alpha)\rho_L+\alpha\rho_G$ is the aerated slug effective density, $g$ is the gravitational acceleration, $L_S$ is the aerated slug length, $L_h^L$ is the liquid plug length in the horizontal section, and $A$ is the pipe cross-sectional area. $H^S_r$ and $H^G_r$ are hydrostatic heads due to the sections of the riser occupied by the aerated slug and gas, respectively. The frictional pressure drop, $\Delta P_f$ is calculated based on the wall shear obtained by applying the Blasius correlation for the friction factor:

$$\Delta P_f = \frac{2L_S}{D} f_S \rho_S U_S^2 + \frac{2L_h^L}{D} f_L \rho_L \left((1-\alpha)U_S\right)^2; \qquad f_{S|L} = \begin{cases} \text{Turbulent: } 0.046/\text{Re}_{S|L}^{0.2} \\ \text{Laminar: } 16/\text{Re}_{S|L} \end{cases} \qquad \textbf{3.3}$$

where $f_S$, $f_L$ are the friction factors for the aerated slug and liquid plug, respectively.

The liquid plug length in the horizontal section is $L_h^L = max\{L_h^{L0}-(1-\alpha)H,\ 0\}$, where $L_h^{L0}=L_h\varepsilon_{i\text{-}hor}$ is the initial liquid plug length in the horizontal section. $L_h^L=0$ corresponds to the stage when the liquid plug tail reaches the riser's lower elbow (Point 5 in **Figure 3.2**), and the entire liquid mass already propagates in the riser (and further downstream) as an aerated slug. Accordingly, the aerated slug length is $L_S=min\{H,\ L_h\varepsilon_{i\text{-}hor}/(1-\alpha)\}$.

A mass balance on the gas occupying the domain upstream of the liquid tail reads:

$$\frac{dm_G}{dt} = \frac{1}{ZRT}\frac{d(P_G V_G)}{dt} = \frac{1}{ZRT}\left[P_G\frac{dV_G}{dt} + V_G\frac{dP_G}{dt}\right] = \dot{m}_{GI} - \dot{m}_{GO}$$

which after rearrangement reads: **3.4**

$$\frac{dP_G}{dt} = \frac{ZRT\dot{m}_{GI} - P_G A U_S}{A L_G}; \qquad P_G(t=0) = P_O + L_r\rho_G g$$

where $m_G$ is gas mass upstream the liquid tail, $\dot{m}_{GI}$ is the inlet gas mass flow rate, $\dot{m}_{GO}$ is the gas mass flow rate invested in the slug aeration process,($R$ is the specific gas constant). $V_G=AL_G$ is the gas volume in the domain upstream of the liquid tail, where $L_G=max\{L_G^0+(1-\alpha)H,\ L_G^0+H-L_h\varepsilon_{i\text{-}hor}\alpha/(1-\alpha)\}$, and $L_G^0 = L_{12}+L_{23}+L_d+L_h(1-\varepsilon_{i\text{-}hor})$ is the initial gas domain length. First, the liquid plug tail (and the gas domain front) velocity in the horizontal section is $(1\text{-}\alpha)U_S$, and its displacement relative to the initial state is $(1\text{-}\alpha)H$. The gas mass flow rate invested in the slug aeration process is then $\dot{m}_{GO}=A\rho_G\alpha U_S$. However, once the entire liquid has been displaced from the horizontal section and the gas domain front reaches the riser's lower elbow (Point 5 in **Figure 3.2**), the slug length reaches its final value ($\dot{m}_{GO}$=0), and it is further displaced downstream at velocity $U_S$. From this point on, the slug tail further displacement is $H\text{-}L_h\varepsilon_{i\text{-}hor}/(1\text{-}\alpha)$. The aerated slug tail leaves the riser when $H\text{-}L_h\varepsilon_{i\text{-}hor}/(1\text{-}\alpha)=L_r$, where $L_r$ is the riser's length. Note that the change in $\dot{m}_{GO}$ does not affect the final expression for $dP_G/dt$, since $dV_G/dt=AdL_G/dt=A(1\text{-}\alpha)U_S$ during the aeration stage (when $\dot{m}_{GO}=A\rho_G\alpha U_S$), and $dV_G/dt=AU_S$ when $\dot{m}_{GO}$=0. The contribution of the two terms is $P_GAU_S$ in **Eq. 3.4** in both scenarios. In fact, once the slug tail leaves the riser lower elbow and the aerated slug length has reached its maximal length, the further tracking of its front and tail locations is the same as that presented in Yurishchev et al., 2024 for an unaerated liquid plug. Note that for low aeration levels, where $\alpha = 0$ can be considered (typical for low gas pressure systems and high gas flowrates and/or high initial liquid amounts, which initially also occupy part of the riser and downcomer sections), the model reduces to its original formulation introduced by Yurishchev et al., 2024.

## 3.2 Forces on the riser elbows

The instantaneous gas pressure and aerated slug velocity predictions are essential for estimating the forces exerted on the jumper elbows due to the liquid passage. Of particular interest are the forces acting on the upper elbow of the riser, indicated as Point 6 in **Figure 3.2**. The time variation of the force components (i.e., $F_x$, $F_y$) applied on the riser upper elbow are obtained by momentum balance on the fluid passing the elbow:

$$F_{x|y} = (P_6 - P_O)A + \rho^* U_S^2 A \qquad \textbf{3.5}$$

where $\rho^*$ is the aerated slug or the gas density. The aerated slug density $\rho_S$ is applied while the slug passes through the elbow; and the gas density $\rho_G$ is applied while the gas flows through the elbow, namely before the arrival and after the departure of the aerated slug. The numerical simulation results (see **Section 4.2.2** for a detailed discussion) showed that in cases of high-pressure gas (methane at $P_O > 20$ atm), the contribution of the gas flow in the inertia term cannot be ignored. In fact, **Eq. 3.5** represents a simplification of the actual momentum balance, **Eq. 2.2**. The mechanistic model assumes that the elbow's volume is small, so the gravity force and the time variation of the liquid momentum within the elbow volume are neglected. Also, the force resulting from the wall shear, which is, in principle, included in the force exerted by the walls on the fluid, is practically negligible. These assumptions were justified by Yurishchev et al., 2024.

### 3.3 Analysis of the system dynamics

The two-phase flow dynamics during the liquid displacement are modeled by a set of non-linear differential equations (**Eq. 3.1**, **3.2** and **3.4**). To characterize the system dynamics, the equations are linearized around the steady-state solution at $t$=0 (see **Appendix B**). The linear model shows that the system dynamics corresponds to a second-order underdamped system with the following time constant $\tau$ and a damping parameter, $\xi$:

$$\tau = \left( \frac{\frac{\rho_L}{\Delta\rho} \frac{{L_h^L}^0}{g}}{1 + \frac{P_G^0}{(1-\alpha)\Delta\rho g L_G^0}} \right)^{0.5} \qquad \textbf{3.6}$$

$$\xi_{\text{Laminar}} = \frac{16\mu_L}{D^2} \left( \frac{\frac{{L_h^L}^0}{g\rho_L\Delta\rho}}{1 + \frac{P_G^0}{(1-\alpha)\Delta\rho g L_G^0}} \right)^{0.5} ; \xi_{\text{Turbulent}} = 0 \qquad \textbf{3.7}$$

If turbulent flow is considered during the initial stages of the liquid movement, then the damping parameter $\xi$=$\xi_{\text{Turbulent}}$=0 and $\omega_n$=$\omega_r$. However, for laminar flow, $\xi = \xi_{\text{Laminar}} \neq 0$ (see **Eq. 3.7**). Note that for cases of high-pressure methane, $P_O$ > 20 atm, $P^0_G/(1-\alpha)\Delta\rho g L^0_G >> 1$, thus the damping parameter, time constant and natural frequency are simplified to:

$$\xi_{\text{Laminar}} = \left( \frac{256(1-\alpha){L_h^L}^0 L_G^0 \mu_L^2}{\rho_L P_G^0 D^4} \right)^{0.5} ; \tau = \frac{1}{\omega_n} = \frac{1}{2\pi f_n} = \left( \frac{(1-\alpha)\rho_L {L_h^L}^0 L_G^0}{P_G^0} \right)^{0.5} \qquad \textbf{3.8}$$

For the jumpers considered in this study ($D$ = 50 ÷ 150 mm, geometry scaled according to **Figure 2.1** and the various gas pressures), the dumping parameter is $\xi$<<1. Consequently, the system is always underdamped for the operational conditions tested, and the resonance frequency, $\omega_r = 2\pi f_r$, is practically equal to the natural frequency, $\omega_n = 2\pi f_n$. This frequency can be considered when referring to the gas pressure fluctuations. The fluctuations might be amplified in case the input gas flow rate, $\dot{m}_{GI}(t)$, is subject to even small changes at this or close to this frequency. In a dimensionless representation of the natural frequency, the following relation between the Strouhal number and the Euler number is obtained:

$$\text{St} = \frac{1}{2\pi} \text{Eu}^{0.5} \qquad \textbf{3.9}$$

where:

$$\text{St} = \frac{f_n \left( {L_h^L}^0 L_G^0 \right)^{0.5}}{U_{GS}} ; \ \text{Eu} = \frac{P_G^0}{(1-\alpha)\rho_L U_{GS}^2} \qquad \textbf{3.10}$$

## 3.4 Prediction of the critical gas flowrate

The critical (minimal) gas velocity or flow rate needed to carry the liquid upwards in a vertical pipe can be estimated based on criteria available in the literature. The predictions of some of them demonstrated reasonable agreement with the experimental data in an atmospheric system (Yurishchev et al., 2024), where the gas critical velocity was defined as the minimal needed to remove the accumulated liquid from the system. In the present study, we tested whether the same criteria might be valid for upscaling.

The first criterion tested refers to the critical gas velocity for *avoiding the flow reversal in a continuous film*, (Wallis, 2020), the gas superficial velocity at in-situ conditions, $U_{GS}$, should exceed the following threshold:

$$U_{\mathrm{GS}}\sqrt{\frac{\rho_{\mathrm{G}}}{\Delta\rho g \sin\beta\, D}} \geq C \qquad \mathbf{3.11}$$

where $\Delta\rho = \rho_L - \rho_G$, $\beta$ is the pipe inclination angle, and $C$ is a constant = 0.8 ÷ 1.

Applying this criterion for vertical flow and real gas equation of state for gas, $\rho_G = P_G M/Z\bar{R}T$, the critical gas volumetric flowrate at STP (standard temperature and pressure conditions), $Q^0_{G|Crit}$ reads:

$$Q^0_{\mathrm{G|Crit|FRC}} = \frac{\pi D^2}{4}\frac{\bar{R}T^0}{P^0_{\mathrm{G}}}\, C\sqrt{\Delta\rho g D\frac{P_{\mathrm{G}}}{ZM\bar{R}T}} \qquad \mathbf{3.12}$$

where *FRC* stands for "*F*low *R*eversal *C*riterion".

Assuming that the process is isothermal and $\Delta\rho \approx \rho_L$, the following simple scaling for $Q^0_{G|Crit}$:

$$Q^0_{\mathrm{G|Crit|FRC}} \propto (ZM)^{-0.5} P_{\mathrm{G}}^{0.5} D^{2.5} \qquad \mathbf{3.13}$$

Apparently, **Eq. 3.13** suggests that the critical gas volumetric flowrate at *STP* conditions, increases with both the gas pressure and internal pipe diameter. However, based on **Eq 3.11**, the critical gas volumetric flowrate at *in-situ* conditions, $Q_{G|Crit}$, decreases with the pressure:

$$Q_{\mathrm{G|Crit|FRC}} = \frac{\pi D^2}{4} C\sqrt{\frac{\Delta\rho g D Z\bar{R}T}{P_{\mathrm{G}}M}} \rightarrow Q_{\mathrm{G|Crit|FRC}} \propto (Z/M)^{0.5} P_{\mathrm{G}}^{-0.5} D^{2.5} \qquad \mathbf{3.14}$$

Another criterion for the establishment of stable concurrent upward annular flow refers to the gas velocity required to break the liquid into non-deformable drops, Brauner, 2003. The criterion is based on Brodkey, 1969 correlation for the critical Weber number resulting in drop breakage due to the gas inertia, and the corresponding maximal drop size. The upward flow stability can be maintained when the maximal drop size is smaller than that of deformable drops. Based on this criterion, it was shown that the gas volumetric flowrate at in-situ conditions should exceed the following threshold:

$$Q_{\mathrm{G}} \geq \frac{\pi D^2}{4}\left(\frac{360\sigma\rho_{\mathrm{L}}g}{\rho_{\mathrm{G}}^2}\right)^{0.25} \quad \mathbf{3.15}$$

where $\sigma$ is the surface tension.

The corresponding critical gas volumetric flow rate at STP conditions is:

$$Q^0_{\mathrm{G|Crit|GIC}} = \frac{\pi D^2}{4}\frac{\bar{R}T^0}{P^0_{\mathrm{G}}}(360\sigma\rho_{\mathrm{L}}g)^{0.25}\left(\frac{P_{\mathrm{G}}}{ZM\bar{R}T}\right)^{0.5} \propto (ZM)^{-0.5}P_{\mathrm{G}}^{0.5}D^2 \quad \mathbf{3.16}$$

where *GIC* stands for "*G*as *I*nertia *C*riterion".

Yurishchev et al., 2024 determined that once the criterion for breaking the liquid into drops is met, the gas volumetric flow rate is already sufficiently high to lift the drops upwards. Also, for the atmospheric air-water system, both criteria (*FRC* and *GIC*) predict practically the same critical gas flow rates, which agree with the experimental data obtained in an atmospheric jumper. In addition, the criteria suggest the same simple scaling rule for the gas pressure, $\propto Z^{-0.5}P_{\mathrm{G}}^{0.5}$, whereas the dependency on the internal pipe diameter is weaker in the *GIC* criterion, $\propto D^2$, compared to the *FRC* criterio*n*, $\propto D^{2.5}$.

The compressibility factors, *Z*, for air and methane were calculated using Peng-Robinson equation of state for real gas, (Peng and Robinson, 1976). The critical values (and the acentric factors) for the gases are presented in **Table 2.2** and **Table 2.3.**

The scaling rules presented in **Eqs. 3.13**, **3.14** and **3.16** are applied by considering the $Q^0_{\mathrm{G|Crit|ref}}$ values obtained in the experiments conducted with the air-water system at atmospheric pressure ($P_{\mathrm{O}}$ = 1 atm) and $D$ = 50 mm pipe diameter as a reference in the following way (based on **Eq. 3.13** for example):

$$Q^0_{\mathrm{G|Crit|M}} = Q^0_{\mathrm{G|ref}}\left(\frac{M_{\mathrm{ref}}}{Z_{\mathrm{M}}M_{\mathrm{M}}}P_{\mathrm{G|M}}\right)^{0.5}\left(\frac{D_{\mathrm{M}}}{D_{\mathrm{ref}}}\right)^{2.5} \quad \mathbf{3.17}$$

where subscript "ref" indicates the atmospheric air-liquid system experimental data and subscript "M" refers to the methane-liquid system.

Similar scaling is conducted based on **Eq. 3.14** for $Q_{\mathrm{G|Crit|FRC}}$ and on **Eq. 3.16** for $Q^0_{\mathrm{G|Crit|GIC}}$.

# 4 Results and discussion

## 4.1 Critical gas flow rate

A comprehensive set of numerical simulations was executed to investigate the effects of the gas pressure and pipe diameter on the critical gas volumetric flow rates and the validity of the scaling rules presented in **Section 3.4**. The simulations were carried out for air and methane at various gas pressures. $P_O = 1 \div 5$ atm and $20 \div 300$ atm, for air and methane, respectively, and for internal pipe diameters of $D = 50 \div 150$ mm.

**Figure 4.1** presents the comparison between the numerical simulation results and the suggested scaling rules regarding the gas pressure variations (**Eq. 3.13** and **3.14**) for a given pipe diameter, $D = 50$ mm. The numerical simulations and scaling rules show the same trends for $Q^0_{G|Crit}$ and $Q_{G|Crit}$. While $Q^0_{G|Crit}$ increases with the gas pressure, $Q_{G|Crit}$ decreases with gas pressure. Using the air data with atmospheric pressure, which was validated by experiments, and considering the molar weight and compressibility factor differences between air and methane (see **Eq. 3.17**), the scaling rules for gas pressure, $P_G$, yield similar values for the critical flow rates to those obtained in the numerical simulations. Moreover, as shown in **Figure 4.2.a**, the same scaling rule is also valid for predicting the gas pressure effects in larger pipe diameters, $D = 100$ and 150 mm.

The effect of the pipe diameter on the critical gas volumetric flowrates is presented in **Figure 4.2.b**, which shows the numerical simulation results and the predictions obtained by the scaling rules (**Eq. 3.13** and **3.16**). The effect of increasing the pipe diameter is demonstrated for methane at three given pressures, $P_O = 50$ atm, 100 atm, and 300 atm. Apparently, the scaling rule associated with flow reversal, **Eq. 3.13** ($\propto D^{2.5}$, *FRC*), is more appropriate than the rule referred to drops breakage due to the gas inertia, **Eq. 3.16** ($\propto D^2$, *GIC*), for the tested diameters range, $D = 50$, 100 and 150 mm. The latter criterion underestimates the numerical simulation results (see the example for $P_O = 50$ atm, gray solid line, **Figure 4.2.b**). These results suggest that the minimum gas flow rate required to ensure the removal of the liquid from the jumper must meet the stricter criterion of preventing the flow reversal of the liquid, *FRC* criterion.

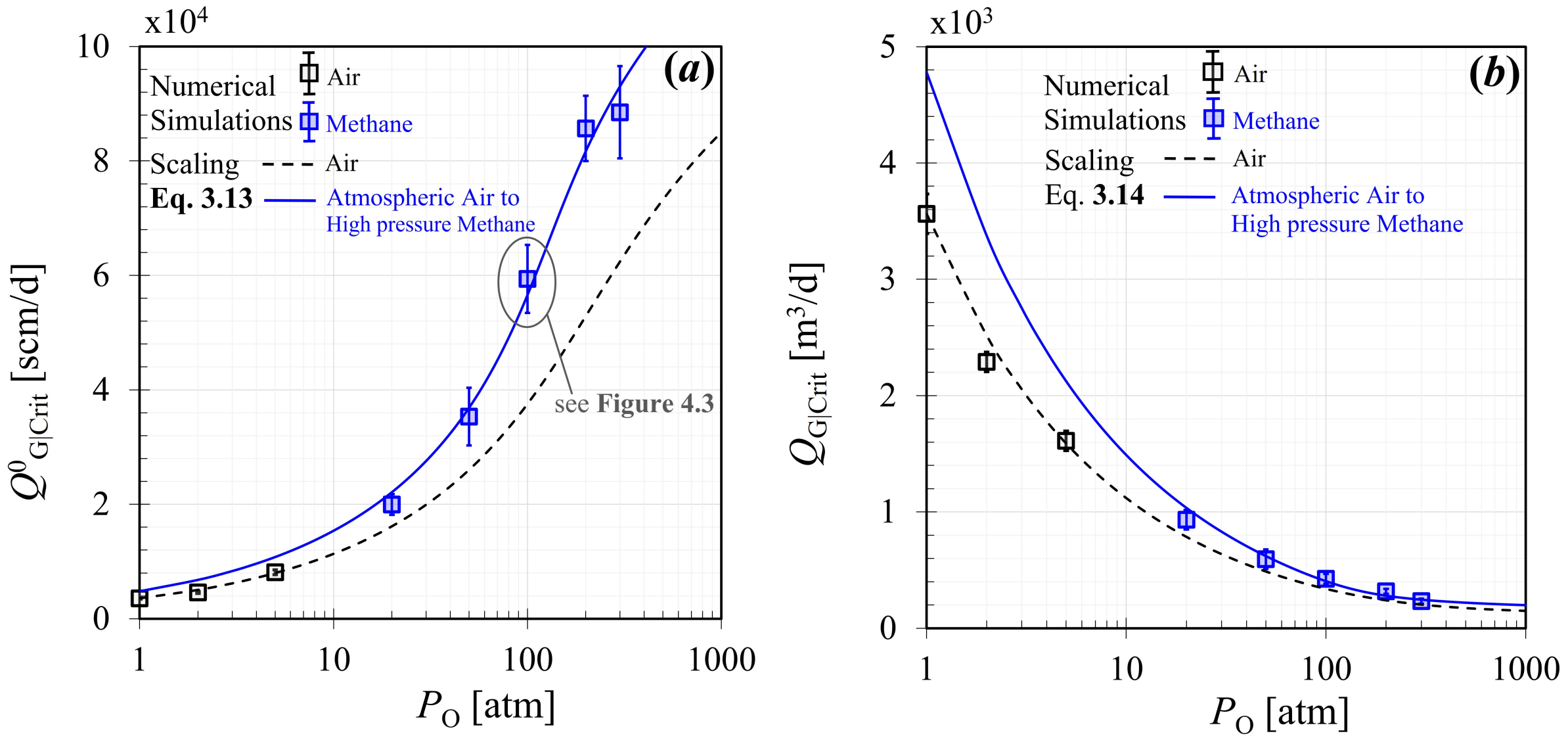

**Figure 4.1**: Results of numerical simulations (□) and scaling rules (---) regarding the critical gas volumetric flowrate as a function of gas pressure, $P_O$ for air (dashed black) and methane (dashed blue) with $D = 50$ mm at (*a*) STP conditions, $Q^0_{G|Crit}$ and (*b*) in-situ conditions, $Q_{G|Crit}$. Reference is given in *(a)* to **Figure 4.3**, which explains the flow phenomena at sub- and super-critical conditions for a specific case (methane, $P_O = 100$ atm).

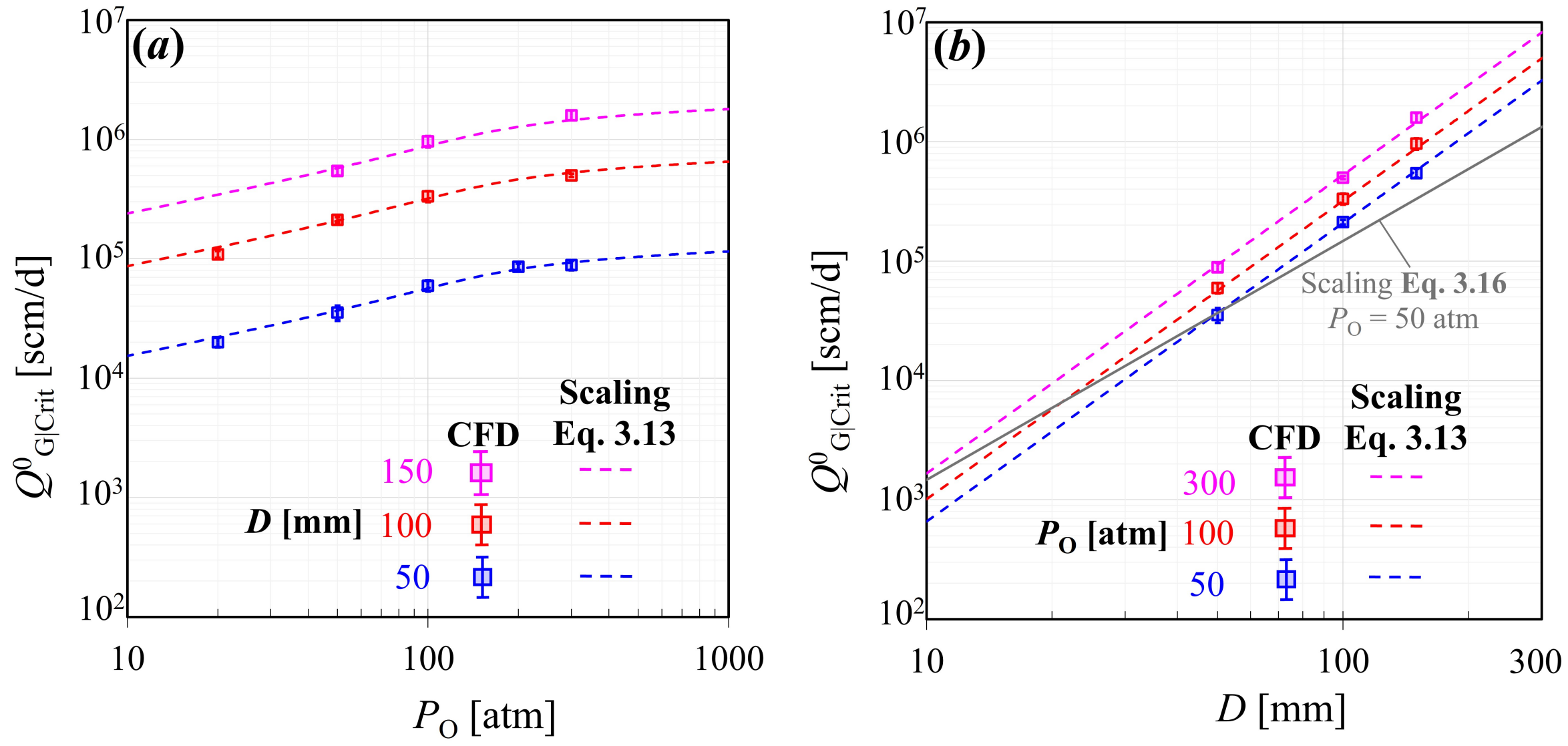

**Figure 4.2:** Results of numerical simulations (□) and scaling rules associated with the flow reversal criteria (---) regarding the critical gas volumetric flowrate of methane as a function of (*a*) gas pressure for various given pipe diameters and (*b*) pipe diameter for various outlet gas pressures. A solid gray line (—) in (*b*) represents the scaling rule, which refers to drop breakage due to gas flow inertia (*GIC*) for $P_O = 50$ atm.

The numerical simulations allow examining the flow characteristics for sub- and super-critical conditions to better understand the liquid displacement process (see **Figure 4.3**). In sub-critical conditions (see **Figure 4.3.a**), upon initiating the gas flow, a liquid plug is formed, which is displaced, and accelerated toward the outlet through the riser. Some liquid volume may be removed from the system during this initial stage. However, at longer flowtimes, the gas penetrates the liquid plugs that occasionally form. The liquid that remains, mostly on the riser's walls, loses its momentum and falls downward under the force of gravity. Eventually, the liquid left in the system circulates in the riser, with a clear backflow in the liquid film near the riser's walls (e.g., **Figure 4.3.c**). The liquid cannot be purged from the jumper under these flow conditions. On the other hand, in super-critical conditions, the initially accelerated liquid plug leaves the system, similarly to the initial stages taking place under sub-critical conditions. However, at longer flowtimes, the gas flow dries the lower riser's elbow, and the tail of the liquid film propagates upwards in the riser (see **Figure 4.3.b**), indicating that the liquid is being removed from the system. Moreover, inspecting the flow in the riser's pipe cross-section reveals no liquid backflow near the walls, even where the liquid film is relatively thick (e.g., **Figure 4.3.d**). Occasionally, some drops disintegrate from the liquid annulus into the gas core flow and are elevated upwards. However, this mechanism seems to have a rather weak impact on the liquid removal process, as a relatively small amount of liquid is being removed from the system as drops. This observation of the simulation results support using the backflow scaling rule (**Eq. 3.13**) to estimate the critical gas flow rate in the tested range of pipe diameters.

Consistent with the predictions of the above scaling rules, the simulations of the liquid purge out carried with water-MEG solutions, where the liquid viscosity varies by an order of magnitude, indicate that the critical gas is practically not affected by the liquid viscosity in the tested range. Also, the critical velocity is unaffected by the initial amount of accumulated liquid and the gas flow ramp-up rate. The latter affect the flow phenomena during the initial stages of the liquid displacement, however not the critical gas velocity for complete removal of the liquid.

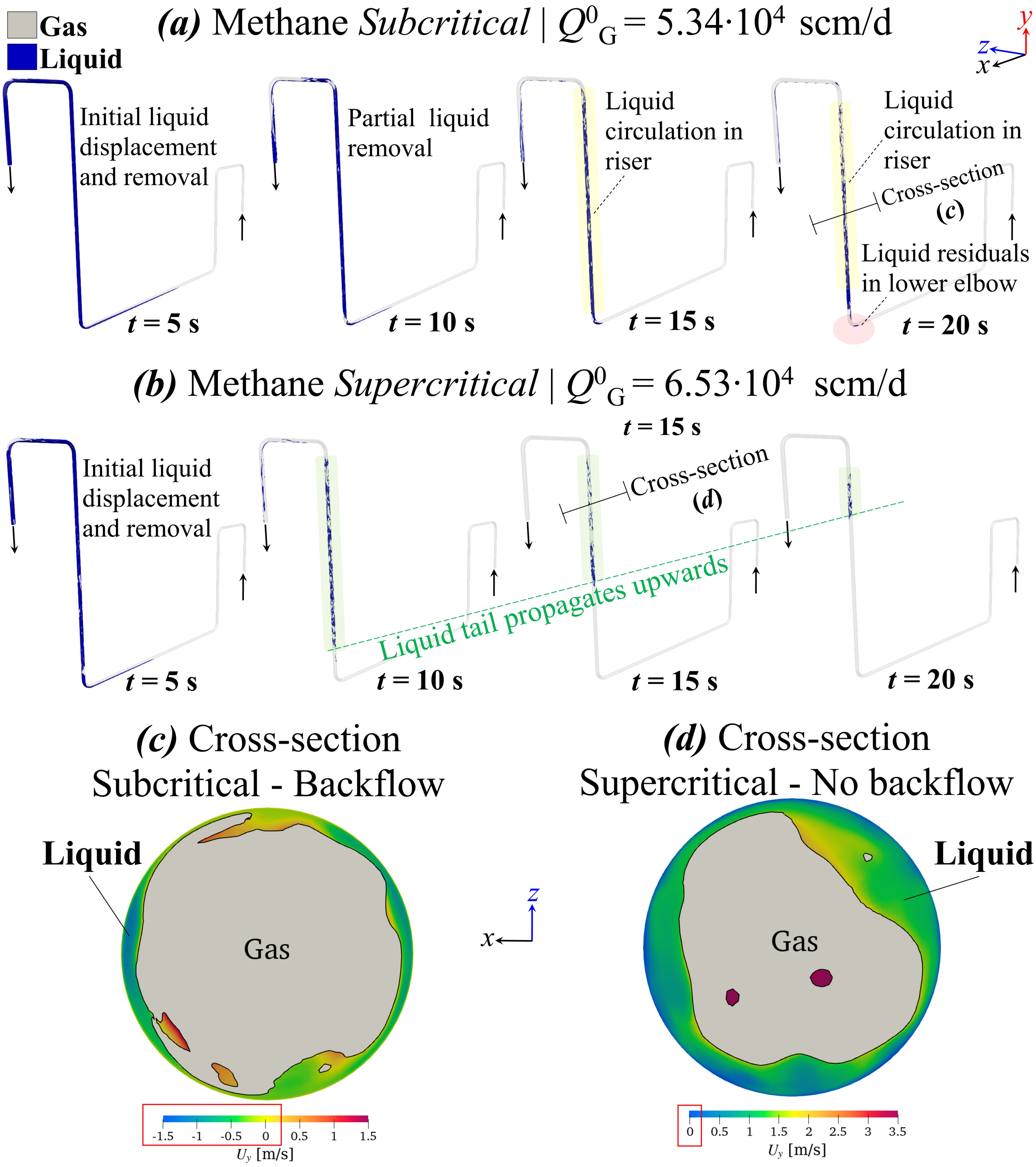

**Figure 4.3**: Flow characteristics for (*a*) sub- and (*b*) super- critical conditions, methane, $P_O$ = 100 atm, $D$ = 50 mm. (*c*) and (*d*) show the axial velocity in the cross-sections indicated in (*a*) - 35$D$ downstream the riser lower elbow and (*b*) - 50$D$ downstream the riser lower elbow, respectively. Grey: $\alpha$>0.5.

### 4.2 Pressure and force responses

Upon increasing the gas flow rate according to start-up protocol, the accumulated liquid resting in the lower horizontal section of the jumper is displaced and enters the jumper riser. The pressure drop on the jumper rises as a result of the liquid acceleration, hydrostatic pressure, and frictional pressure losses. The complex two-phase flow phenomena and the gas compressibility might produce pressure fluctuations during the liquid residence in the riser, and/or the riser's elbows. Also, the formed liquid plug/slug is accelerated upon the increase of the gas flow rate and hits the riser's upper elbow. This elbow experiences the highest loads during liquid displacement since the liquid arrives already at a relatively high velocity. Moreover, the transient forces exerted on the upper elbow are translated to the moments acting on the jumper's lower elbows.

In this section, we analyze the pressure response to determine the maximal pressure drop rise and the time variation of the forces acting on the upper elbow of the riser due to the gas-liquid flow during the liquid displacement. The periodic and potentially dangerous fluctuations, which might lead to FIV during the production start-up process, are explored. The following discussion deals with the typical phenomena in relatively low-pressure air-water systems, **Section 4.2.1** and then in high-pressure methane-liquid systems, **Section 4.2.2**. The pressure drop and force variations during the liquid displacement in those two systems are different and, therefore, discussed separately. The discussion focuses on two scenarios where the gas flow rate is *sub*- and *super*-critical, $Q^0_G/Q^0_{G|Crit} = 0.5$ and 1.2, respectively.

### *4.2.1 Air Liquid*

The numerical simulations reveal that the time variations of the pressure drop can be divided into two stages (see **Figure 4.4.upper row**). The first is the initial liquid removal stage, when the pressure usually rises to its peak due to the hydrostatic pressure and the liquid acceleration from rest. The peak values are naturally higher for the higher flow rates. The second stage is associated with the late displacement of residue liquid chunks. This stage is characterized by a cyclic pressure variation due to the periodic elevation of the liquid chunks by the gas flow in the riser section. The pressure fluctuations amplitude is more significant for a low (*sub*critical) gas flow rate, $Q^0_G/Q^0_{G|Crit} = 0.5$, than for a high (*super*critical), $Q^0_G/Q^0_{G|Crit} = 1.2$. At the supercritical gas flow rate, the fluctuations decay rapidly as the liquid is removed from the system. FFT analysis of the pressure variations (see **Figure 4.4.lower row**) shows a dominant frequency of ~ 1 Hz. The pressure fluctuation frequencies appear to be insensitive to the examined air pressure and flow rates. To support the link between the pressure and liquid holdup fluctuations, **Figure 4.6** demonstrates the time variation of the holdup at two locations along the riser and the corresponding frequency spectrum. The time shift between the holdup variations in the two locations suggests that the oscillations result from the irregular displacement of liquid chunks in the riser. Notably, the dominant frequency remains consistent at both locations and is practically identical to the dominant pressure-fluctuation frequency (refer to **Figure 4.4.c.1**).

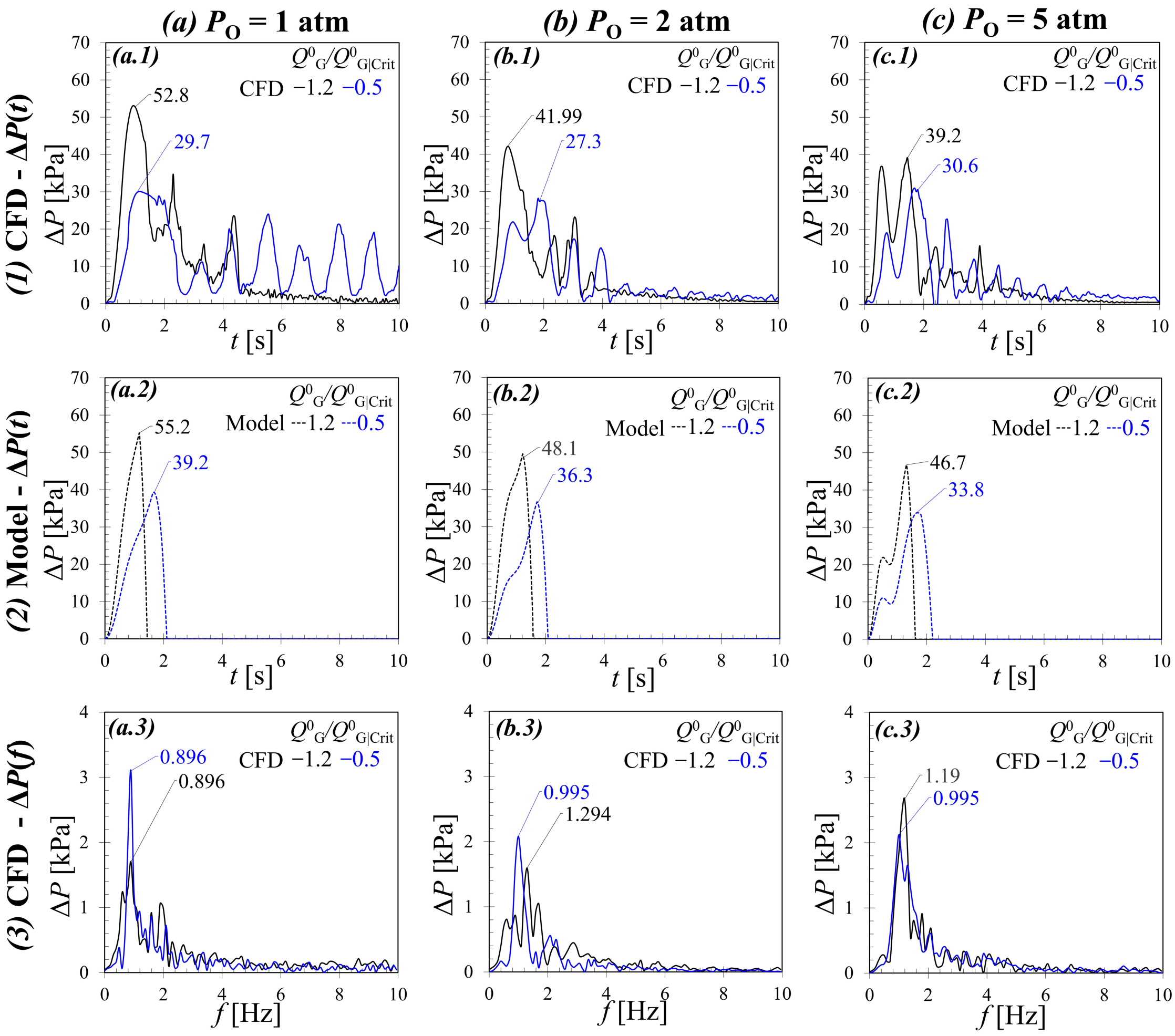

**Figure 4.4**: Pressure drop time variation during the accumulated liquid displacement, obtained by *(1)* the numerical simulations and *(2)* the mechanistic model. *(3)* Representation of the pressure drop fluctuations obtained from the numerical simulations in the frequency domain. The results presented for $Q^0_G/Q^0_{G|Crit}$ = 0.5 and 1.2 ($D$ = 50 mm, $\varepsilon_{i\text{-}hor}$ = 75%, $t_{ramp}$ = 2 s) and for various outlet air pressures: $P_O$ = *(a)* 1, *(b)* 2 and *(c)* 5 atm.

The mechanistic model (**Eq. 3.1**, **3.2**, and **3.4**) is found useful for predicting the initial pressure peak for atmospheric and higher outlet pressure when an unaerated slug, as implied by the simulations under these conditions; see also **Figure 4.5** below, is considered. In fact, upon introducing $\alpha$ = 0, the model reduces to the original plug model introduced by Yurishchev et al., 2024, which agreed with experimental results obtained in an atmospheric system. As shown, the model predictions are in good agreement with the numerical simulation results (compare **Figure 4.4**.**middle row** to the **upper row**). Moreover, **Figure 4.5** suggests that the flow pattern in the numerical simulations during the liquid removal resembles well enough the basic assumptions of the mechanistic model. Namely, the liquid first assembles to form a plug, and then, most of it leaves the system in that shape. The figure shows that, over time, the rear part of the plug may experience gas penetration. However, it seems that this has a mild effect on the pressure drop peak, which is in reasonable agreement with the model predictions. Naturally, the model that assumes fluid movement as a single plug cannot predict the final stage of liquid purging predicted by the numerical model, in which the breakdown of the liquid plug tail to water chunks results in pressure fluctuations. The frequency of the pressure fluctuation predicted by the linearized model (**Eq. 3.6**) is 0.87, 1.19, and 1.85 Hz for $P_O$=1, 2, and 5 atm, respectively, barely detectable in the model response because the liquid is purged in about 2 s.

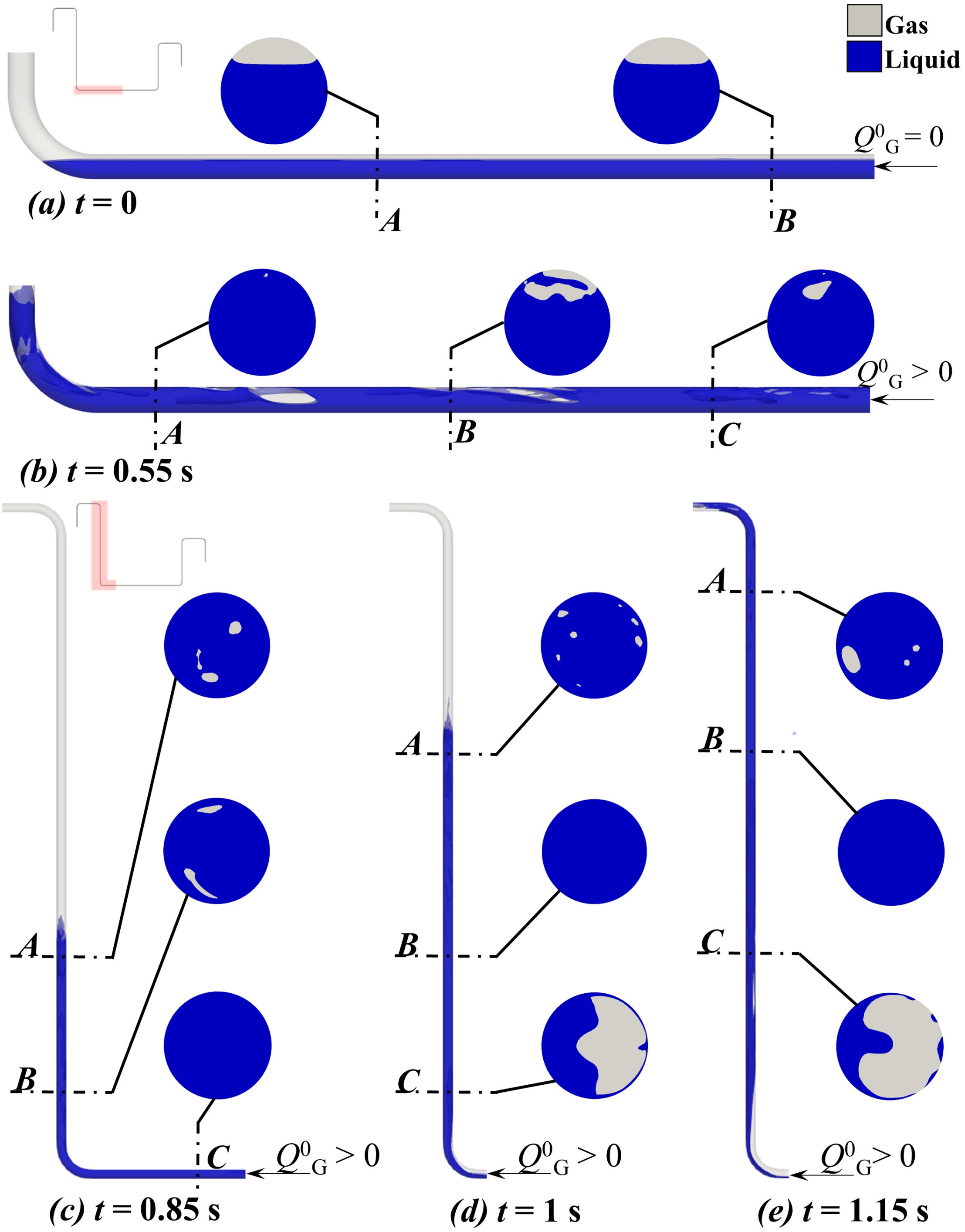

**Figure 4.5:** Demonstration of accumulated liquid displacement by the airflow acquired from the numerical simulation for the case of $P_O$ = 2 atm, $D$ = 50 mm, $\varepsilon_{i\text{-hor}}$ = 75%, $Q^0_G/Q^0_{G|Crit}$ = 1.2 and $t_{ramp}$ = 2 s. The air and water distributions at various cross-sections substantiate the mechanistic model's assumption of liquid plug displacement (blue $\alpha$<0.5, gray $\alpha$>0.5).

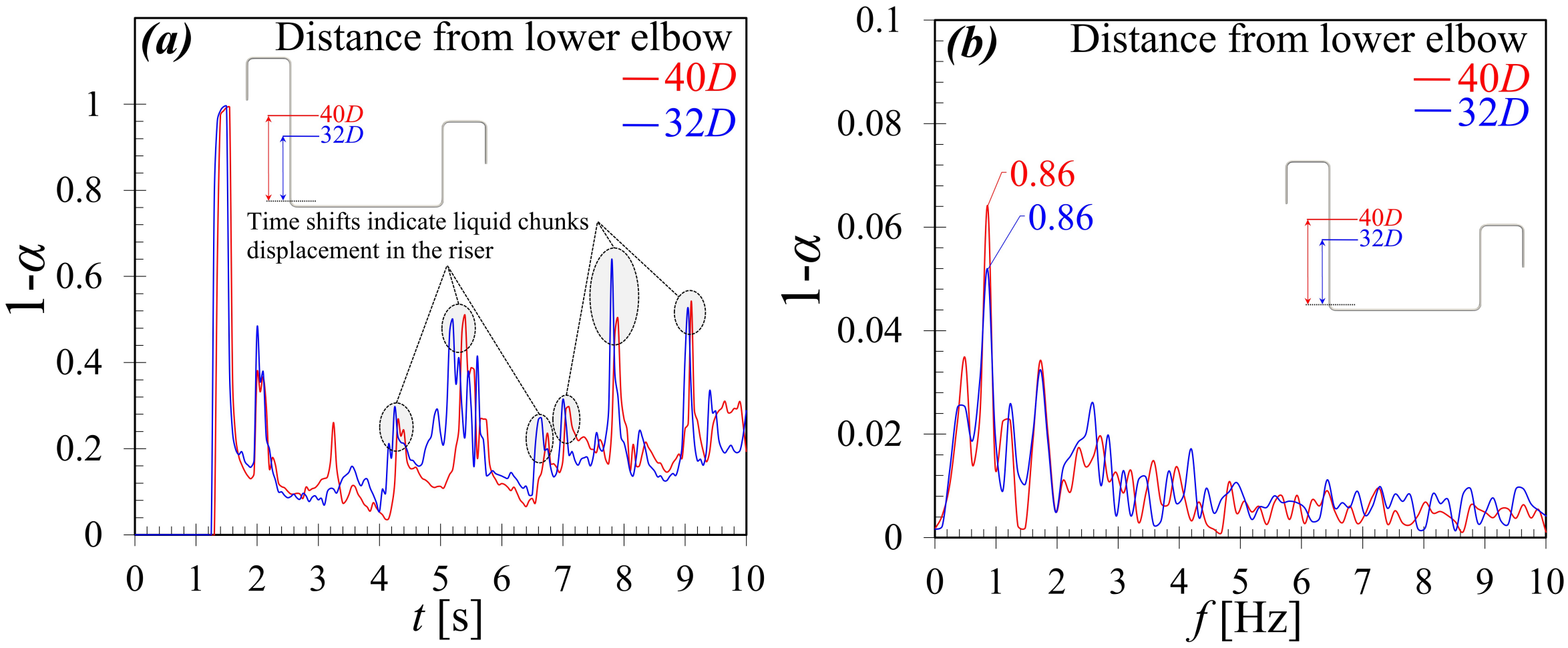

**Figure 4.6:** *(a)* The time variation of the holdup at two locations along the riser and *(b)* the corresponding frequency spectrum. The results presented for $P_O$ = 1 atm, $D$ = 50 mm, $\varepsilon_{i\text{-hor}}$ = 75%, $Q^0_G/Q^0_{G|Crit}$ = 0.5 and $t_{ramp}$ = 2 s.

The pressure drop variation obtained from the mechanistic model is associated with three components – hydrostatics, liquid friction, and liquid acceleration (see **Figure 4.7**). Apparently, each component has its unique contribution to the total pressure drop, and none of them can be ignored. The main contribution in the early stage of the pressure rise is associated with liquid acceleration from rest. Later, when the liquid plug rises in the riser and gains velocity, the hydrostatic and friction components increase. Finally, the three components vanish as the plug leaves the system. Note that, for the supercritical gas flow rate and the operating conditions to which **Figure 4.4**, **Figure 4.5**, and **Figure 4.7** refer, the liquid purging can be achieved even before the 2 s ramp-up time is completed.

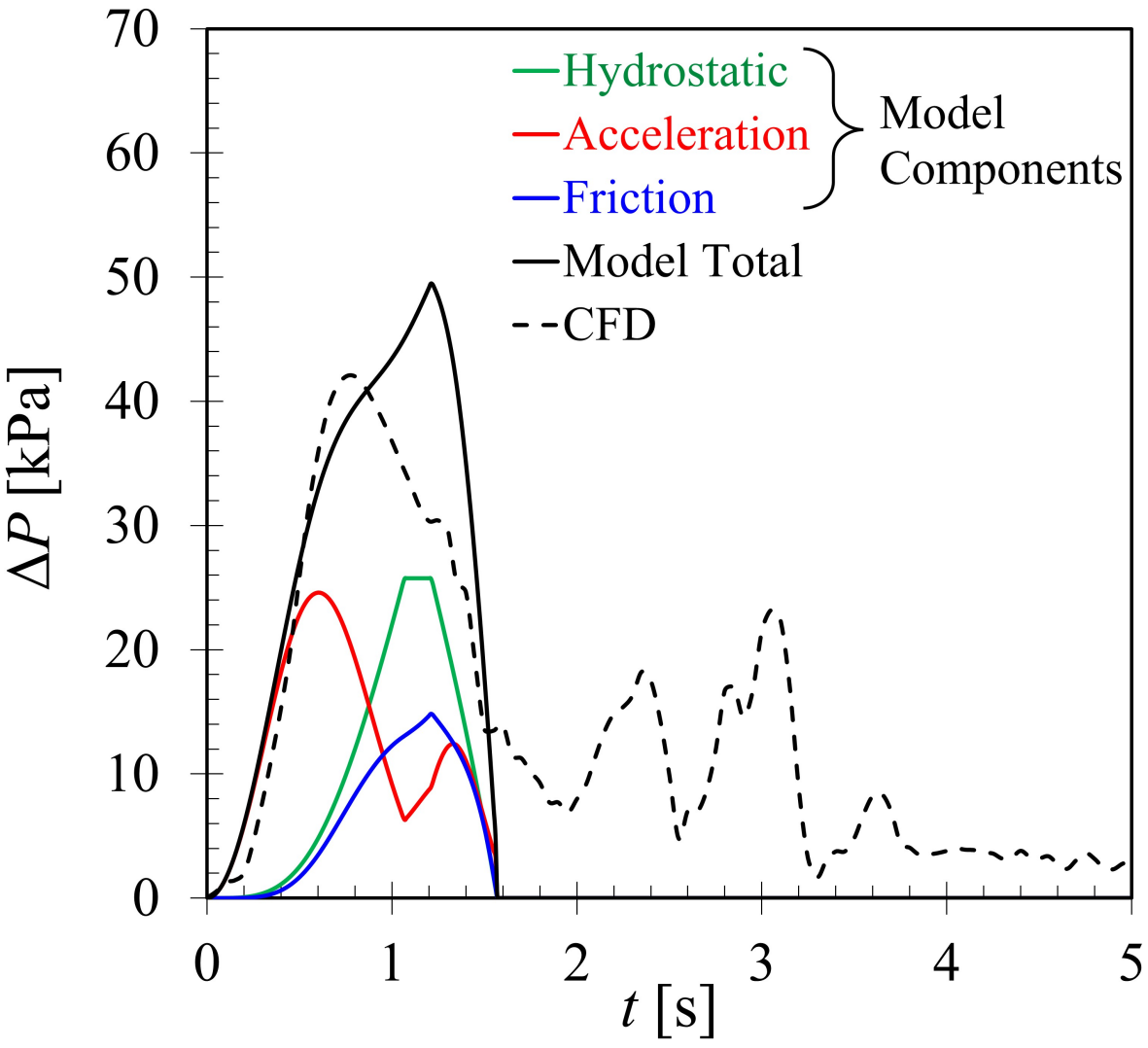

**Figure 4.7:** Comparison between the numerical simulation data and mechanistic model predictions for the pressure drop variation during the accumulated liquid displacement in case of initial air pressure $P_O$ = 2 atm, $Q^0_G/Q^0_{G|Crit}$ = 1.2, $D$ = 50 mm, $\varepsilon_{i\text{-hor}}$ = 75%, and $t_{ramp}$ = 2 s. The pressure drop components are shown to demonstrate each contribution.

**Figure 4.8** shows the time variation of the $F_x$ force component acting on the riser's upper elbow during the accumulated liquid displacement. A rather similar variation of the force is obtained by the simulations and by the model (see **Figure 4.8.1**). In the initial short term, a high-force peak was documented in both cases, $Q^0_G/Q^0_{G|Crit}$ = 0.5 and 1.2, with higher values for the latter. This peak is associated with the accelerated liquid passage through the elbow. Following the initial peak, a series of lower peaks were observed in the simulations. These are due to the passage of liquid residue chunks that are created as a result of the gas penetration into the rear part of the initially assembled liquid plug. Those chunks impact the elbow, producing the low-force peaks with dominant frequency of ~ 1 Hz. The mechanistic model, which assumes that a single liquid plug is formed, does not produce any later low-force peaks that follow the initial high-force peak.

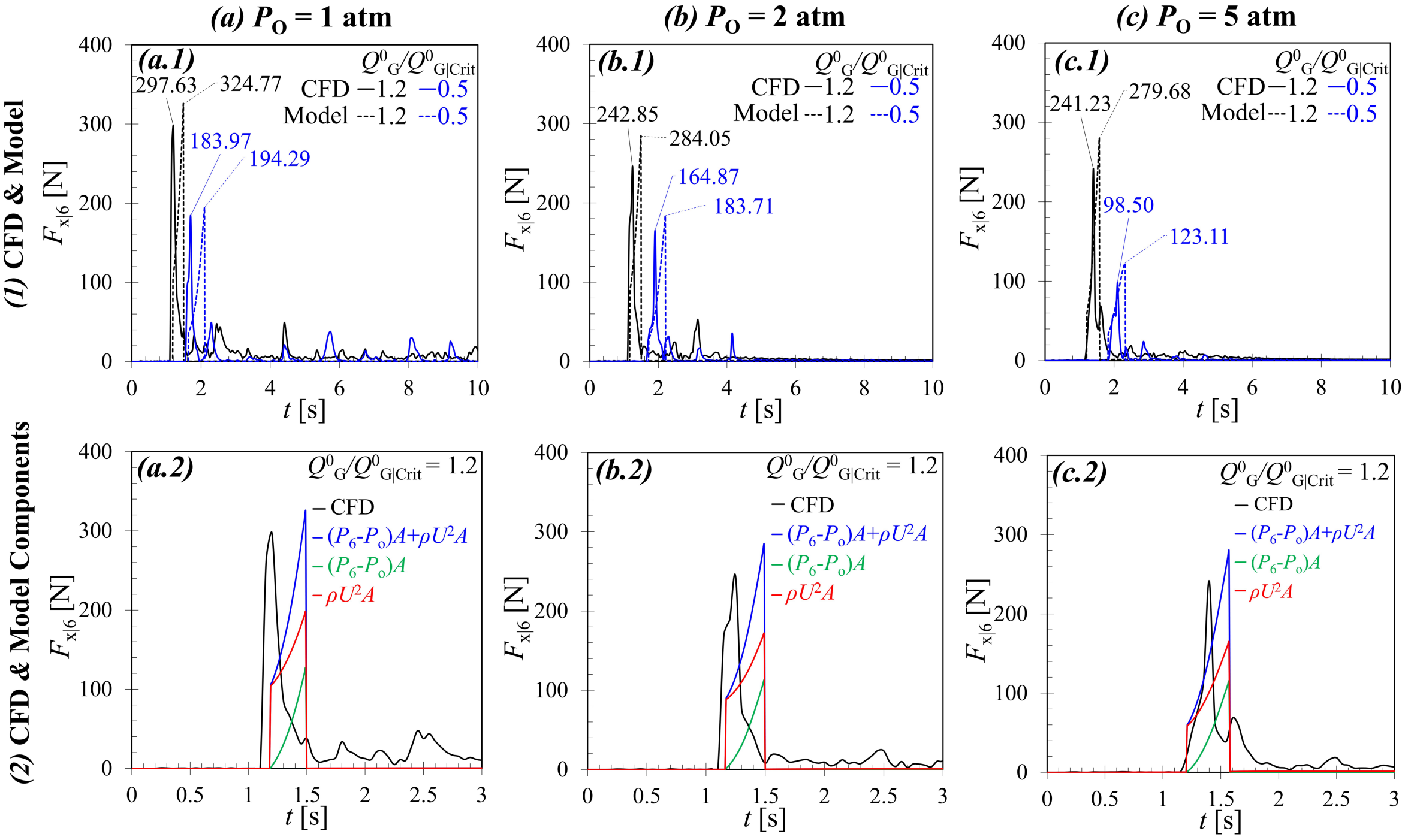

**Figure 4.8:** *(1)* Time variation of the force on the riser's upper elbow during the accumulated liquid displacement, obtained by the numerical simulations and the mechanistic model for $Q^0_G/Q^0_{G|Crit}$ = 0.5 and 1.2 and for various outlet air pressures. *(2)* The comparison between the numerical simulation and the mechanistic model showing the contribution of the components for the case of $Q^0_G/Q^0_{G|Crit}$ = 1.2, $D$ = 50 mm, $\varepsilon_{i\text{-}hor}$ = 75%, $t_{ramp}$ = 2 s and for various outlet air pressures: $P_O$ = *(a)* 1, *(b)* 2 and *(c)* 5 atm.

The slug model predictions are based on the momentum balance on the fluid in the elbow (as elaborated in **Section 3.2**). The simplified momentum balance (**Eq. 3.5**) suggests that the force contains two components: the pressure $(P_6\text{-}P_O)A$, and the fluid inertia, $\rho U^2 A$. Yurishchev et al., 2024 demonstrated the validity of the model assumptions for atmospheric outlet pressure by comparing the force obtained in numerical simulations via two different approaches (see **Section 2.5**). The numerical simulation results presented in this section are based on pressure integration on the elbow's surfaces. As shown in **Figure 4.8.1**, these data are somewhat overpredicted by the mechanistic model.

The mechanistic model predictions and its components are shown in **Figure 4.8.2**, along with the numerical simulation results. The comparison suggests that none of the components can be ignored, and the inertia component is higher than the pressure component. Also, the time-variation nature of the components is different. Before the liquid plug arrives at the elbow, the pressure component is 0 since $P_6 \approx P_O$, when ignoring the gas frictional pressure drop. The inertia component is ~0 because of the low gas density and velocity. When the liquid plug front reaches the elbow, the pressure component increases mildly as the liquid part that passed the elbow accelerates and increases the wall friction in the downstream section. However, it is the fluid inertia component that rises sharply, as its density instantly increases from gas to liquid density. Finally, when the liquid plug tail leaves the elbow and then the jumper domain, both components decline to ~0.

**Figure 4.4** and **Figure 4.8** indicate that increasing the gas pressure level results in a relatively slight decrease in the maximum pressure drop and force values. The reason for this is that the tested flow rates are scaled with the critical gas velocity, which becomes smaller with the rise of the pressure level ($\propto Z^{0.5}P_G^{-0.5}$, **Eq. 3.14**). As a result, the acceleration and friction components of the pressure drop as well as the pressure and inertia components of the force acting on the elbow are lower (see **Figure 4.7** and **Figure 4.8**). Note that the effect of increasing the pressure is less than expected according to the scaling (**Eq. 3.14**), since at the time the liquid plug flows through the elbow, it is still at the acceleration stage, and its velocity is lower than the terminal velocity.

### *4.2.2 Methane-Liquid high-pressure system*

A typical liquid displacement process obtained by the numerical simulations for high-pressure methane is presented in **Figure 4.9**. The process is generally similar to that observed in the relatively low-pressure displacement processes presented **in Section 4.2.1**. At $t = 0$, the liquid rests in the horizontal section. Then, upon increasing the *inlet* gas mass flow, the liquid is pushed toward the riser's lower elbow. However, since the critical gas flow rate values decrease with gas pressure (see **Eq. 3.14**), the gas and consequently the liquid in-situ velocities are relatively low in high-pressure systems. During the longer residence time of the phases in the system, the gas flow penetrates the liquid plug, whereby the liquid is displaced in the riser as a highly aerated slug, in contrast to an almost whole liquid plug observed at low gas pressures (see **Figure 4.5**). Analysis of the simulation results for the average gas volume fraction in the riser suggests that an aeration level of $\alpha \approx 0.7$ represents the gas volume fraction in the slug. (for the case presented in **Figure 4.9**). However, as shown in **Figure 4.10** the aeration level may vary under different operational conditions. Apparently, the gas mass flow ramp-up rate and its final value does not affect the aeration level. The gas pressure level, pipe diameter, and liquid viscosity due to different MEG concentrations have a rather mild effect on the aeration level. Yet, reducing the initial liquid amount leads to a higher aeration level. The calculated aeration levels are used in the aerated slug model (see **Section 3.1**).

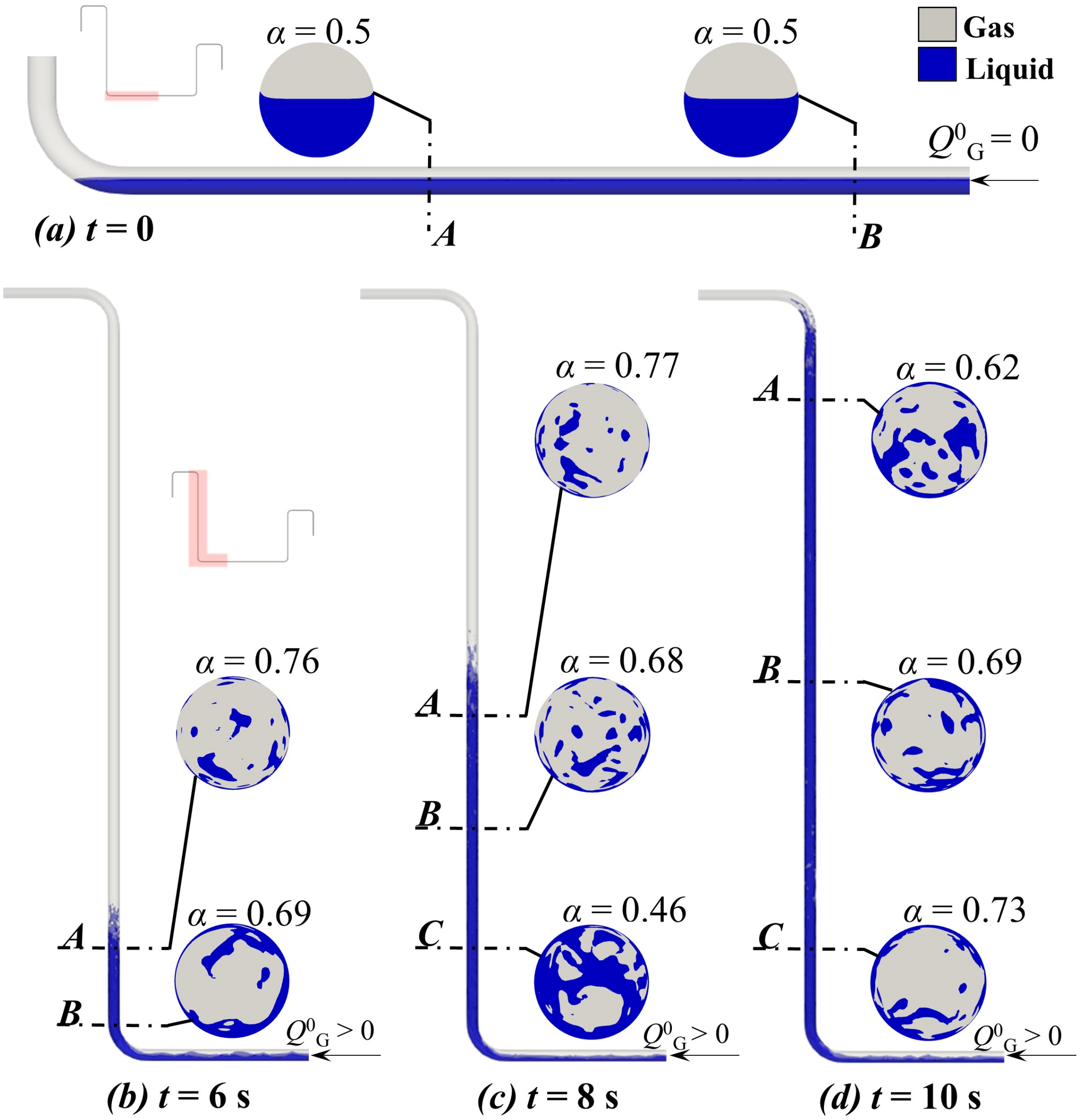

**Figure 4.9:** Demonstration of accumulated liquid displacement by methane flow obtained by the numerical simulation for the case of $P_O$ = 100 atm, $D$ = 100 mm, $Q^0{}_G/Q^0{}_{G|Crit}$ = 1.2, $\varepsilon_{i\text{-}hor}$ = 50% and $t_{ramp}$ = 15 s. The methane and water distributions at various cross-sections correspond to highly aerated slugs or wavy-annular flow patterns. Blue $\alpha$<0.5, gray $\alpha$>0.5.

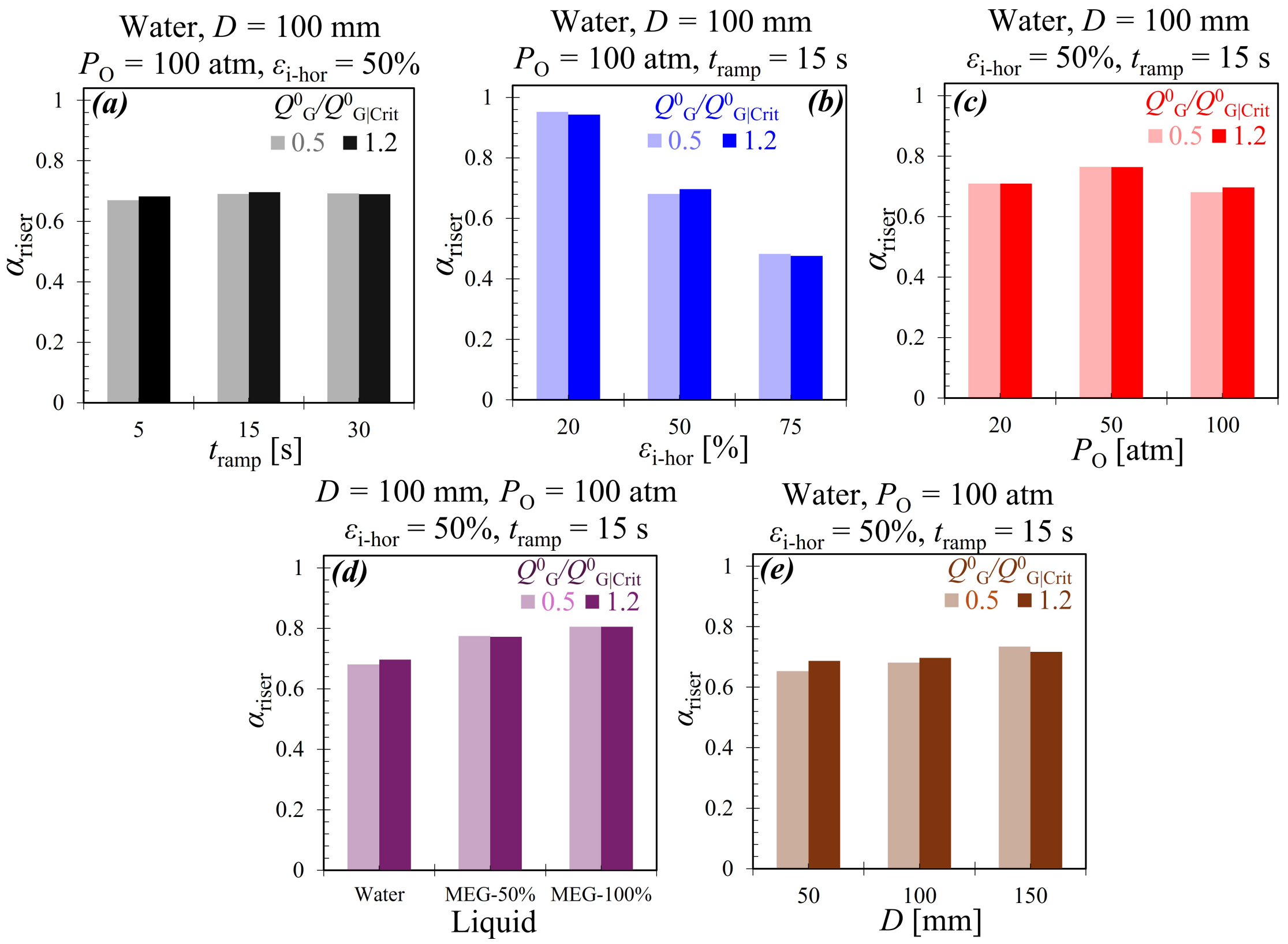

**Figure 4.10:** The average aeration level in the riser, $\alpha_{riser}$: Effects of *(a)* gas mass flow ramp up rate, $t_{ramp}$; *(b)* initial liquid amount, $\varepsilon_{i\text{-}hor}$; *(c)* gas pressure level, $P_O$; *(d)* liquid viscosity due to different MEG concentrations, and *(e)* pipe diameter, $D$.

As shown in **Figure 4.11.a**, the aerated slug model reasonably predicts the pressure drop time variation obtained in the simulations. The most important contribution to the pressure peak is associated with the hydrostatic component (see **Figure 4.11.b**). Note that the nonzero value of the initial pressure drop is due to the hydrostatic pressure of the gas in the riser. An additional and optional modification to the aerated slug model is suggested, and it concerns a time shift to improve the pressure rise timing (compare green and black lines in **Figure 4.11.a**). Usually, the model predicts an early pressure rise due to the different initial conditions considered in the model and the numerical simulations. In the numerical simulations as in reality, some time is needed to assemble the liquid in the elbow region, while in the model, the liquid is already posed there at $t = 0$ (see **Figure 3.1**, and compare **a.1** to **a.2**). The studied cases showed that a time shift of $\Delta t_0 = \sim 0 - 7$ s is needed to better match the model predictions with the numerical simulation results. The time shift decreases with the gas mass flow rate, gas mass flow ramp-up rate, and with increasing the amount of accumulated liquid.

The numerical simulations offer us a clear image of the two-phase flow phenomena in the system that cannot be obtained from the mechanistic model. The latter can, however, provide information on the different components responsible for the total pressure drop (as discussed in **Section 4.2.1**). Notably, the time variation of the pressure drop exhibits clear time-fluctuations over the general trend both in the numerical simulations and in the mechanistic model predictions (see **Figure 4.11**). The mechanistic model reveals that the periodic part of the pressure time variation originates from the liquid acceleration component (see **Figure 4.11.b**). The observed dominant pressure fluctuation frequencies are discussed below (**Section 4.3**).

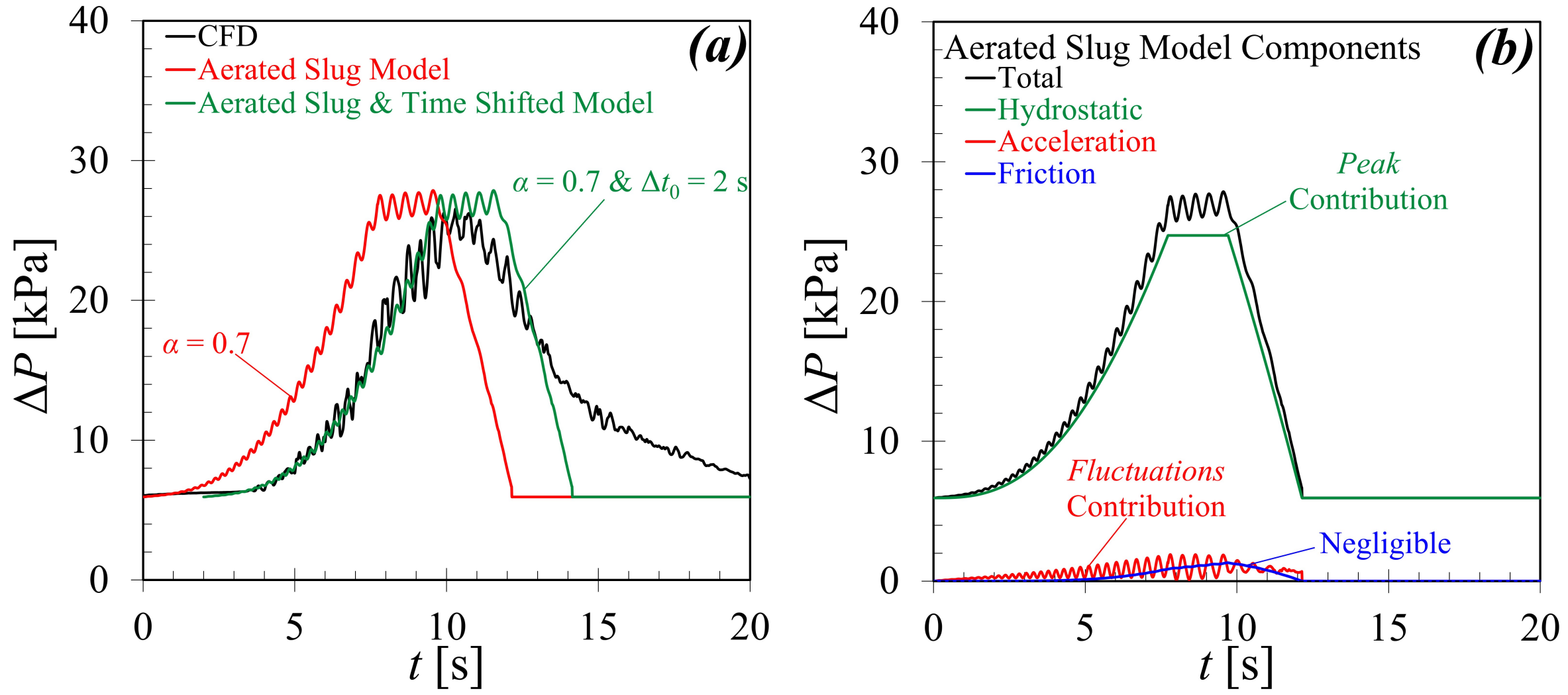

**Figure 4.11:** High-pressure methane-water system - time variation of the pressure drop during the accumulated liquid displacement *(a)* Comparison between the numerical simulation data (black), and the predictions of the aerated slug model (solid red), and the aerated slug model with a time shift that accounts for the delay of slug formation (green). *(b)* The pressure drop components: hydrostatic, friction, and acceleration are shown to demonstrate their contribution to the total pressure gradient. $P_O$ = 100 atm, $D$ = 100 mm, $\varepsilon_{i\text{-}hor}$ = 50%, $Q^0_G/Q^0_{G|Crit}$ = 1.2, and $t_{ramp}$ = 15 s.

When examining the force acting on the riser's upper elbow, one should note that the elbow is located above the jumper's outlet (20$D$, see **Figure 2.1**). The height difference between the elbow and the outlet results in pressure difference due to the hydrostatic pressure, which becomes significant at elevated pressure due to the relatively high gas density. As a result, a force is acting on the elbow also under no-flow conditions. For instance, in the case of $D$ = 100 mm, $P_O$ = 100 atm, the initial force at $t$ = 0 s is ~ 13 N. To eliminate this effect on the force calculations, the pressure in the elbow obtained in the simulations was corrected by adding the pressure drop over the down-comer section. Accordingly, the down-comer section at the outlet (downstream of point 7) was not included also in the mechanistic model (see **Figure 3.2**).

**Figure 4.12.a** shows the horizontal force acting on the elbow obtained in the numerical simulation (blue – **Eq. 2.1** (ignoring the friction component), and red – **Eq. 2.2**) in comparison to that predicted by the aerated slug model (black – **Eq. 3.5**). As shown, ignoring the friction in the force calculation is justified as the deviation between the two calculations approaches is insignificant (blue vs. red). The peak value of the force is practically reached at the same time

as the maximal pressure drop in the riser (see **Figure 4.11**). This is expected since as the liquid slug reaches the upper elbow, the hydrostatic pressure, which is the major component in the pressure drop, is maximal. The model predicts a sharp force increase upon the arrival of the aerated liquid slug to the elbow, then a mild increase of the force during its passage through the elbow, followed by a sharp decrease when the slug leaves the elbow. The force then keeps increasing until the gas velocity reaches its final constant value. However, in the numerical simulation, the elbow experiences a slightly different reaction. Initially, we observe a sharp force increase similar to the model prediction, but the subsequent force decline is more moderate. This behavior is explained by the differences between the aeration level distribution of the liquid slug in the model and the simulation. The aeration level in the model is based on the average level obtained from the simulations and is assumed to be constant as elaborated above. Therefore, a sharp force decline after the slug tail exits the elbow is expected. However, the actual aeration level in the simulation is lower in the slug's front and higher in its tail due to the gas penetration into the liquid slug during its displacement. Thus, a more moderate force decline is observed in the simulation results. Both the model and the numerical simulation show that liquid displacement is associated with force fluctuations.

**Figure 4.12.b** presents the force components obtained by the aerated slug model (**Eq. 3.5**). The results suggest that the inertia component, $\rho U^2 A$, is dominant compared to the much lower pressure component $(P_6\text{-}P_O)A$. As the fluid in the slug is still at its acceleration stage while flowing through the elbow, the force experienced by the elbow increases with time. **Figure 4.12.c** displays the simulation results obtained by the two methods that can be used for the total force calculation (red and blue) and their corresponding force components (green, orange, and purple). The figure reveals that when using the momentum equation on the liquid for the force calculation (**Eq. 2.2**) , the inertia term, $\iint \rho U_x^2 dA_c$, is indeed the dominant one, while the pressure term, $\iint (p_{out}\text{-}P_O) dA_c$ that represents the effect of the fluid flow through the elbow (see **Section 2.5**), and the time derivative of the fluid momentum in the elbow, $\partial/\partial t(\iiint \rho U_x dV_{el})$, are practically negligible. In fact, considering only the liquid inertia term practically yields the same force that is obtained by the pressure integration on the elbow surface (**Eq. 2.1**, see the blue curve **Figure 4.12.c**). This is also reflected in **Figure 4.12.d** that shows a comparison of the force components obtained by the model and by the numerical simulations.

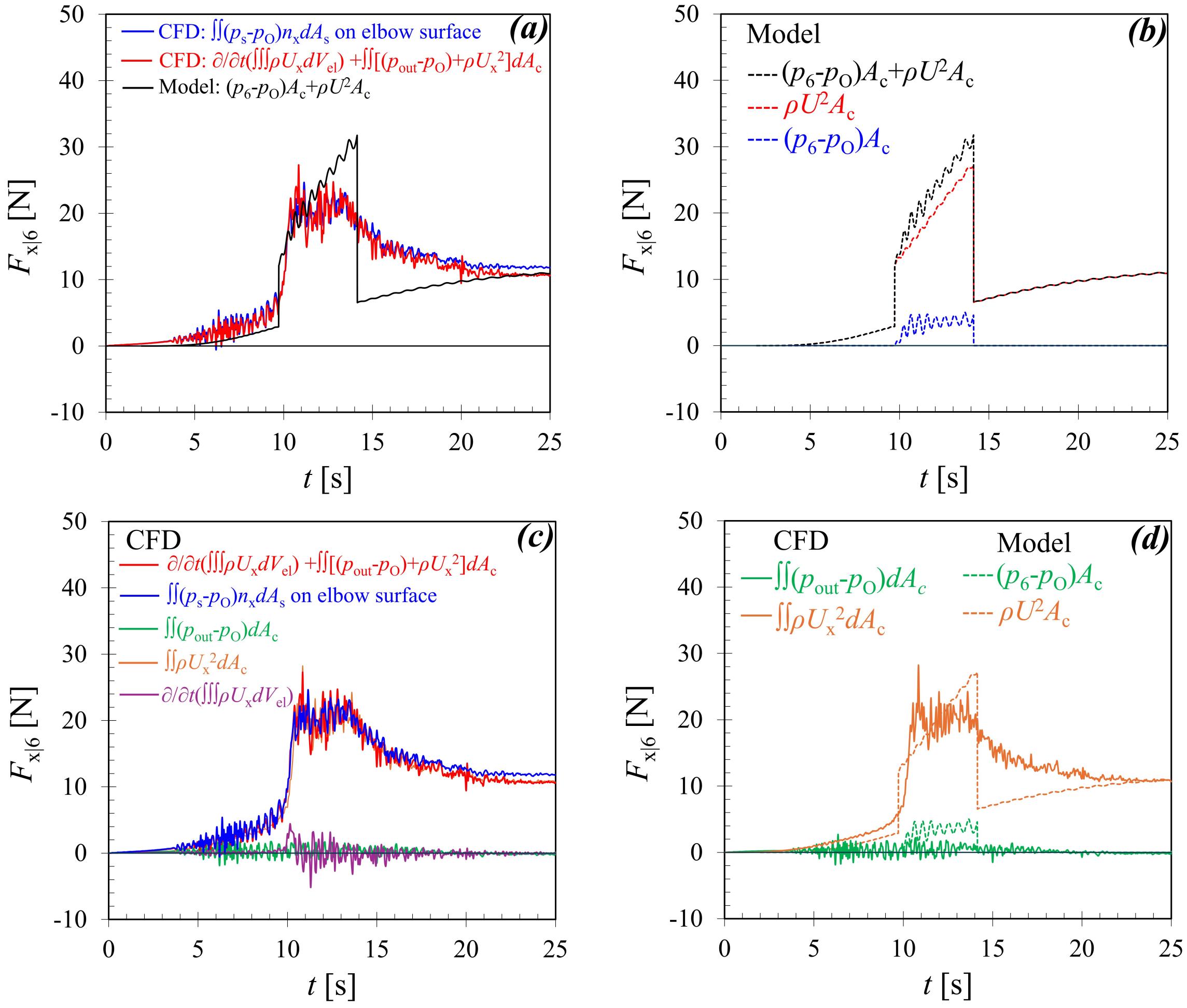

**Figure 4.12:** *(a)* Time variation of the force on the riser's upper elbow during the accumulated liquid displacement, obtained by the numerical simulations (**Eq. 2.1** or **Eq. 2.2**) and by the aerated slug model (**Eq. 3.5**); *(b,c)* Force components via the model and the simulation, respectively. *(d)* Comparison between the force components in the model and in the simulation. $D$ = 100 mm, $P_O$ = 100 atm, $\varepsilon_{i\text{-}hor}$ = 50%, $t_{ramp}$ = 2 s, and $Q^0{}_G/Q^0{}_{G|Crit}$ = 1.2.

### *Gas flow ramp-up rate effect*

The gas production start-up protocols seek to prevent harmful consequences as a result of sudden changes in the gas mass flow. Usually, the gas mass flow is gradually increased to protect the pipeline and other structures from the risks of sudden pressure rise. The effects of the gas flow ramp-up rate on the time variation of the pressure drop and force on the riser's upper elbow are examined for three ramp-up rates corresponding to $t_{ramp}$ = 5, 15 and 30 s.

A good agreement between the aerated slug model and the numerical simulations data is shown in **Figure 4.13**. The presented results are for the methane-water system, $D$ = 100 mm, $P_O$ = 100 atm, and $\varepsilon_{i\text{-}hor}$ = 50 %. Note that $t_{ramp}$ can be scaled by the gas residence time (based on the final gas in situ velocity), which is $t_{res}$ = 14.85 s and 6.19 s for $Q^0{}_G/Q^0{}_{G|Crit}$ = 0.5 and 1.2, respectively. The results show that the ramp-up rate does not significantly affect the maximal pressure rise values (see **Figure 4.13.upper row**). This is expected since the pressure peak values result mainly from the hydrostatic pressure contribution (see **Figure 4.11.b**). Thus, similar pressure peak values are anticipated for a given initial liquid amount. The pressure peak timing is in accordance with the aerated slug formation and its penetration into the riser, which occurs earlier at a higher ramp-up rate. In addition, with a higher gas mass flow rate (**Figure 4.13.b.2**), we documented sharper pressure rise and decline.

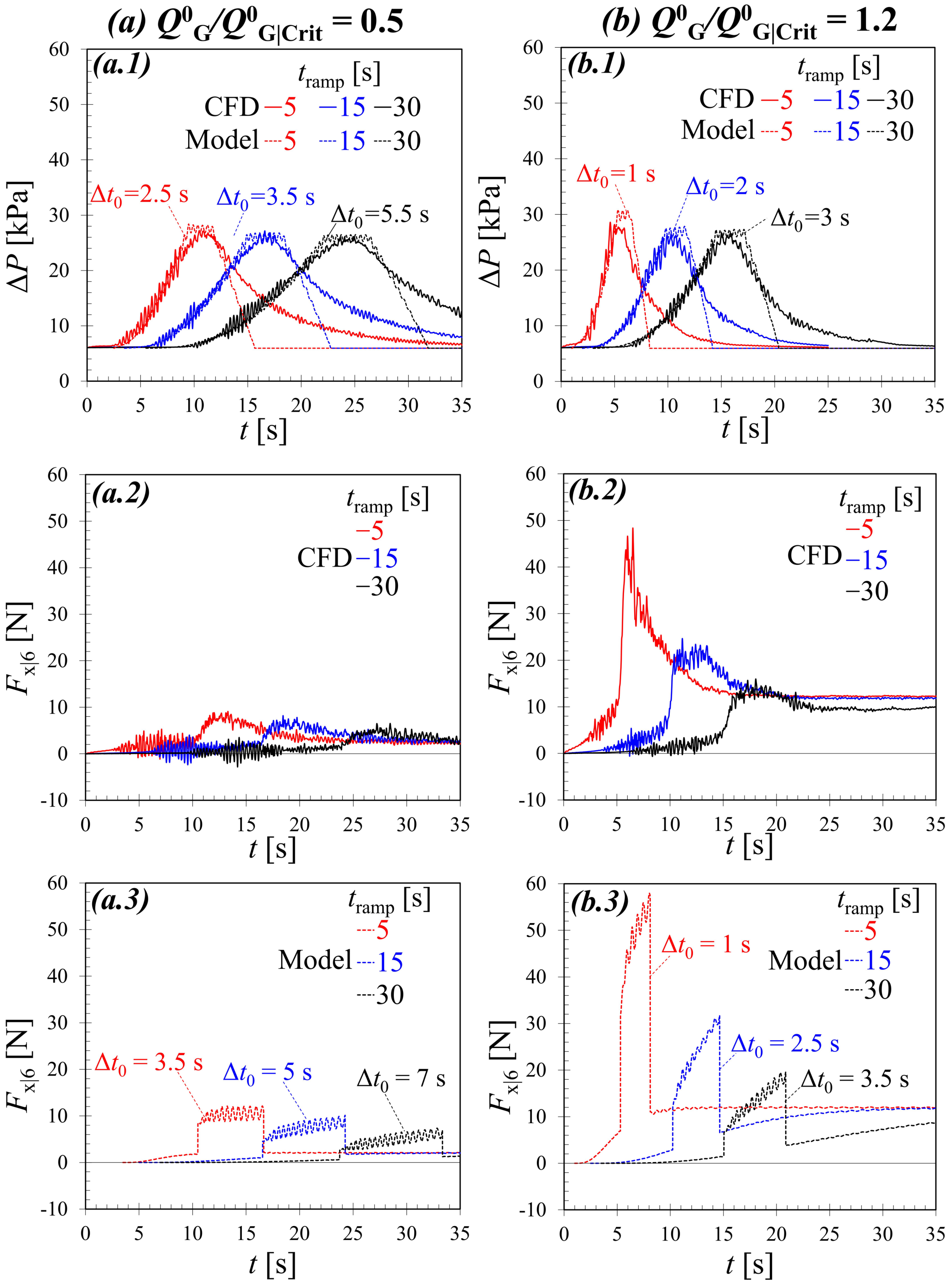

**Figure 4.13**: Effect of the gas mass flow ramp-up rate corresponding to $t_{ramp}$ of methane inlet flow. Comparison between the numerical simulations data and mechanistic model predictions (aerated slug & time shift correction): time variation of the pressure drop (upper figures) and force acting on the riser's upper elbow (middle and lower figures) during the accumulated water displacement for $Q^0_G/Q^0_{G|Crit}$ = 0.5, 1.2 (*a*,*b*); $P_O$ = 100 atm, $D$ = 100 mm, and $\varepsilon_{i\text{-hor}}$ = 50%.

The effect of the gas mass flow ramp-up rate on the $F_x$ force component acting on the riser upper elbow is presented in **Figure 4.13.(middle and lower rows)**, where the results of the simulation are compared with those predicted by the aerated slug model. As shown, an increase in the ramp-up rate (decrease of $t_{ramp}$) leads to a clear increase in the maximal force exerted on the elbow. The effect is more significant at supercritical gas flow rates. For instance, the numerical simulations show 54% and 96% augmentation when changing from $t_{ramp}$ = 30 s to 15 s and from $t_{ramp}$ = 15 s to 5 s, respectively. This is due to the higher liquid slug velocity upon its arrival to the riser's upper elbow with a faster gas mass flow ramp-up rate and the dominancy of the fluid inertia term, $\iint \rho U_x^2 dA_c$. As anticipated, the force peak occurs earlier for the lower $t_{ramp}$, since the liquid slug reaches the elbow faster. In general, similar trends are predicted by the aerated slug model, however, it tends to overpredict the maximal force obtained in the numerical simulation.

***Initial liquid amount effect***

The effect of the initial amounts of accumulated liquid on the pressure drop and force time-variation is demonstrated by considering three initial liquid amounts, corresponding to $\varepsilon_{i\text{-}hor}$ = 20, 50, and 75%. Comparisons between the predictions of the aerated slug model and the numerical simulations data are presented in **Figure 4.14** for the methane-water system, $D$ = 100 mm, $P_O$ = 100 atm, and $t_{ramp}$ = 15 s. The aeration levels $\alpha$ used in the model are 0.7 and 0.5 to $\varepsilon_{i\text{-}hor}$ = 50 and 75%, respectively, according to the simulation results (see **Figure 4.10.b**). The findings indicate that the maximum pressure values tend to rise in tandem with increasing the initial liquid amount. This is attributed to the augmented hydrostatic pressure component resulting from the larger amount of liquid present in the system and therefore also the riser, and the lower aeration level (see **Figure 4.10.b**). Also, a higher liquid amount leads to an earlier timing of the pressure peak, as the formation of the liquid slug is faster (see **Figure 4.14.upper row**). Note that the accuracy of the model decreases upon reducing the amount of liquid since, with low amounts of accumulated liquid, the assumption of the model regarding slug flow is less applicable. Therefore, the model calculations for $\varepsilon_{i\text{-}hor}$ = 20% are not shown.

As larger initial liquid amounts result in the faster formation of longer and less aerated slugs, the force impact is observed earlier and lasts for a longer time. Also, due to the lower aeration level, the slug inertia term is higher, and the force peak value is larger (see **Figure 4.14.lower row**). Obviously, upon removing most of the liquid from the jumper, the force on the elbow does not vanish, and it is a result of the gas flow. Similar trends are predicted by the slug flow model (not shown).

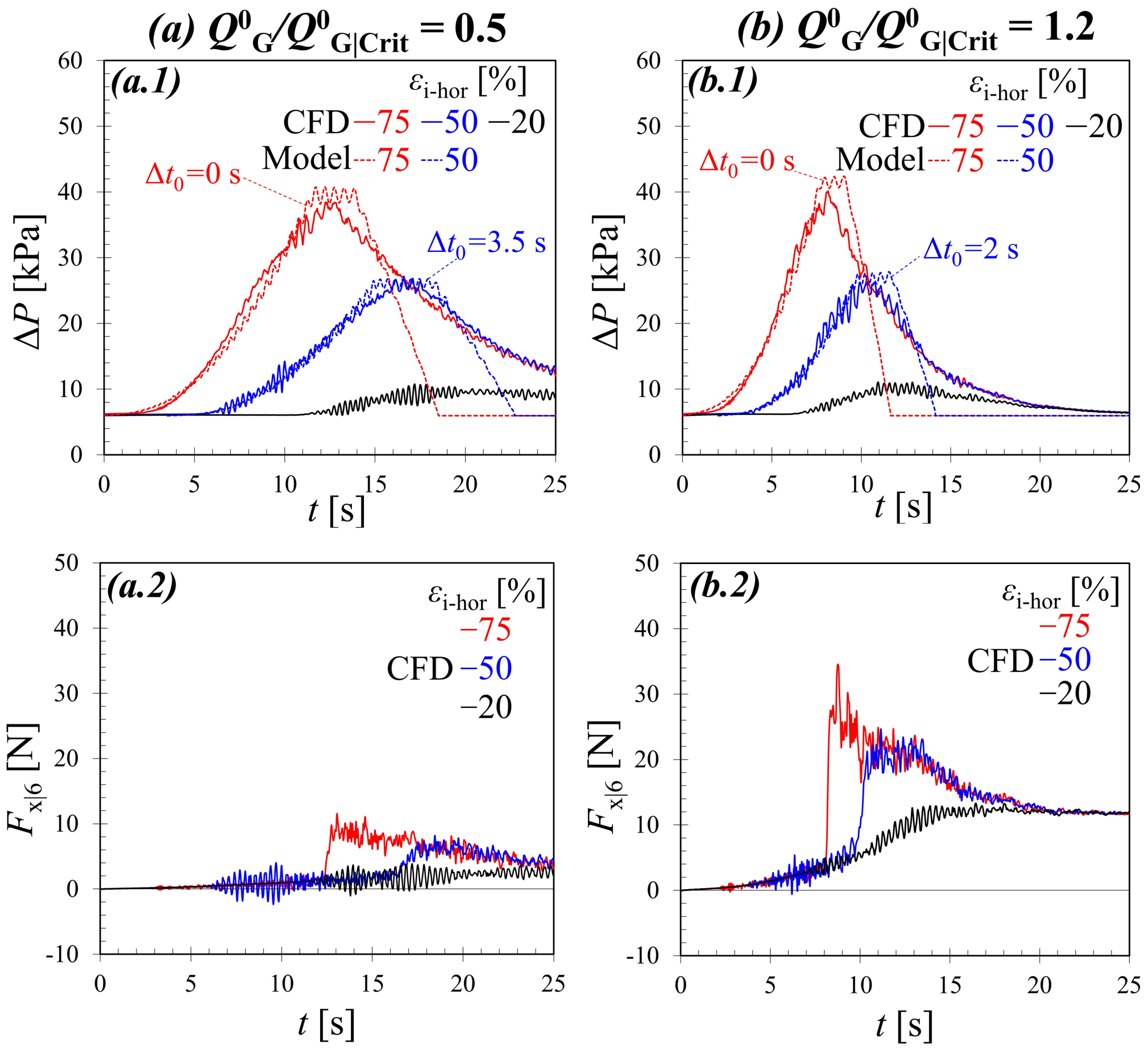

**Figure 4.14**: Effect of the initial amount of accumulated water. Comparison between the numerical simulations data and mechanistic model predictions (aerated slug & time shift correction): time variation of the pressure drop (upper figures) and force acting on the riser's upper elbow (lower figures) during the accumulated water displacement for $Q^0_G/Q^0_{G|Crit} = 0.5$, 1.2 (*a*,*b*). $P_O = 100$ atm, $D = 100$ mm, and $t_{ramp} = 15$ s.

***Gas pressure effect***

The gas pressure level in the jumper may vary depending on the downstream conditions and can alter during the lifespan of the production wells. The main impact of the pressure is on the gas density based on the gas equation of state (see **Table 2.3**). Note that the gas viscosity also increases with pressure, however, it is considered to have a weak effect in turbulent flow. The effect of the pressure level is demonstrated in **Figure 4.15** by examining the results obtained for methane-water flow at three outlet pressures $P_O = 20$, 50, and 100 atm for $D = 100$ mm, $\varepsilon_{i\text{-}hor} = 50\%$ and $t_{ramp} = 15$ s. The gas residence times based on the final gas in situ velocity are $t_{res,20\ atm} = 6.94$ s and 2.89 s; $t_{res,50\ atm} = 9.91$ s and 4.12 s; $t_{res,100\ atm} = 14.85$ s and 6.19 s, for $Q^0_G/Q^0_{G|Crit} = 0.5$ and 1.2, respectively. The figure suggests that the gas pressure level mildly affects the maximal pressure drop values (see **Figure 4.15.upper row**). Despite the large difference between the gas densities in the range of the tested gas pressures (density ratio of up to ~ 6), the aerated slug effective densities are similar, and so are the hydrostatic pressure components, which dictate the maximal pressure drop values (e.g., the ratio between the aerated slug densities for $P_G$ = 100 atm and 20 atm is ~ 1.2). In addition, the tested in-situ gas velocities decrease with the gas pressure since they are scaled with the critical gas flow rate corresponding to the gas pressure ($\propto Z^{0.5} P_G^{-0.5}$, **Eq. 3.14**).

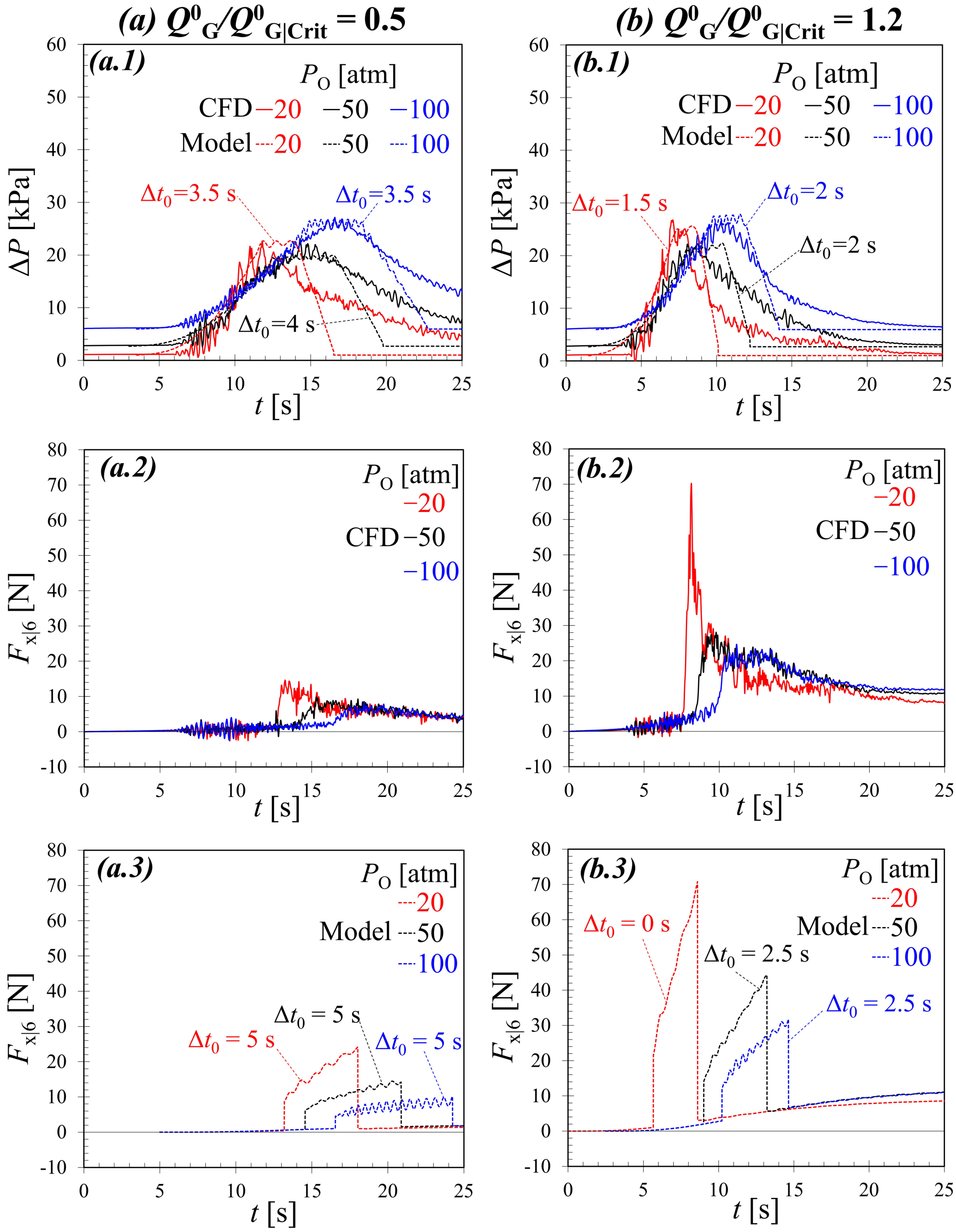

**Figure 4.15:** Effect of the methane pressure level. Comparison between the numerical simulations data and mechanistic model predictions (aerated slug & time shift correction): time variation of the pressure drop (upper figures) and force acting on the riser's upper elbow (middle and lower figures) during the accumulated water displacement for $Q^0_G/Q^0_{G|Crit}$ = 0.5, 1.2 (*a,b*). $D$ = 100 mm, $\varepsilon_{i\text{-}hor}$ = 50%, and $t_{ramp}$ = 15 s.

The effect of the pressure, $P_O$, on the force exerted by the flow on the upper elbow is presented in **Figure 4.15.middle and lower rows**. The mild variation of the maximal pressure (**Figure 4.15.a.1** and **b.1**) with the pressure level is consistent with the variation of the aeration level in the riser in the tested pressure range (see **Figure 4.10.c**). Both the numerical simulations and the model predictions indicate that higher gas pressure results in a decrease in the maximal force exerted by the flow. A change in the gas pressure impacts not only the in-situ critical gas velocity, but also the aerated slug density (as elaborated above in **Sections 4.1**). While the former decreases, the latter increases with the gas pressure. Therefore, a priori prediction of the total effect on the force in the elbow is not obvious. Examining the fluid inertia term reveals a stronger impact of the velocity and, consequently, a decrease of the force with the gas pressure both in the simulations and the model predictions. Also, as a result of the lower in situ velocity of the fluid at higher pressure, the increase of the pressure and force acting on the elbow are delayed.

***Liquid viscosity effect***

The liquid in the jumper may originate mainly from condensates, produced water, and injected hydrate inhibitors (e.g., Mono-Ethylene-Glycol, MEG). The physical properties of the liquid are according to their compositions. The effect of the liquid properties on the pressure drop in the jumper was studied by examining three types of liquids: pure water, pure MEG, and a 50% MEG water volumetric solution. **Table 2.1** implies that the key difference between the tested liquids is their dynamic viscosity, with some increase in the solution density as the MEG concentration increases.

The comparison between the pressure drop predictions of the aerated slug model and the numerical simulation data for methane-liquid flow is presented in **Figure 4.16.upper row**. $D$ = 100 mm, $P_O$ = 100 atm, $\varepsilon_{i\text{-}hor}$ = 50% and $t_{ramp}$ = 15 s, the gas residence times based on the final gas in situ velocity are $t_{res}$ = 14.85 s and 6.19 s for $Q^0_G/Q^0_{G|Crit}$ = 0.5 and 1.2, respectively. The figure suggests that the liquid viscosity has a weak effect on the maximal pressure drop values. The numerical simulations show a slight decrease in pressure peak values with a reduction in the water fraction in the MEG solution, albeit the higher liquid density ( MEG100% >MEG50% > water). Examining the flow in the simulations reveals that the higher liquid viscosity of the richer MEG solution prolongs the liquid displacement process into the riser. At the time of the pressure drop peak, MEG solution is still observed in the horizontal section, and thereby, smaller amounts of MEG solution are present in the riser. Indeed, the longer residence time of the liquid in the system results in higher aeration of the liquid in the riser (see **Figure 4.10.d**) and a lower hydrostatic pressure head. However, at longer times, when most of the water has already left the riser, the trend is reversed. The effect of the liquid properties on the force in the elbow is examined in **Figure 4.16.lower row** by comparing the results obtained with water, MEG50% and MEG100%. The figures suggest that practically, with the tested liquids, the maximal force and the overall time over which the force is affected by the displaced liquid are practically the same. Similar trends are predicted by the slug flow model (not shown).

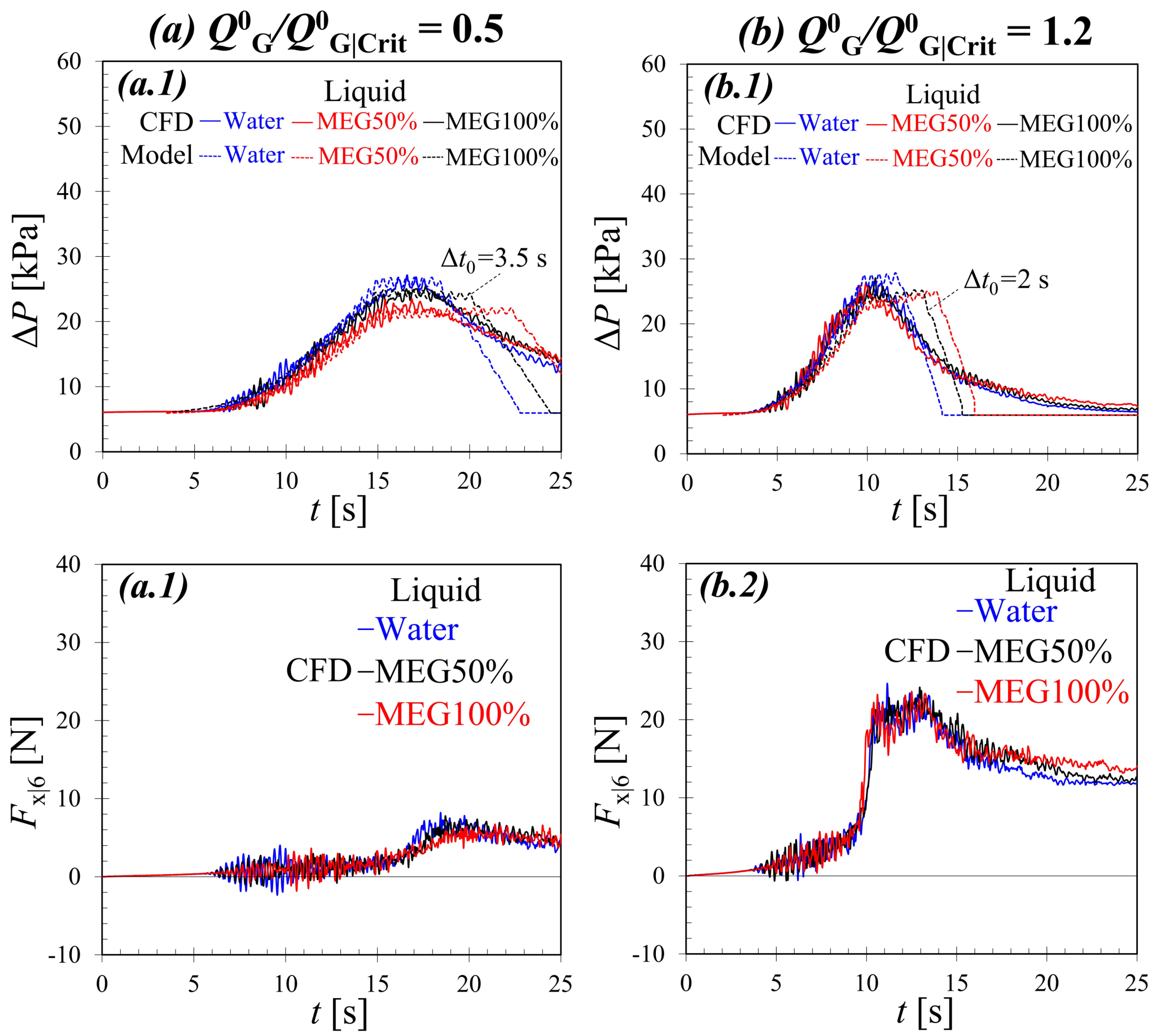

**Figure 4.16:** Effect of the liquid properties (via MEG solution concentration 0, 50% 100%). Comparison between the numerical simulations data and mechanistic model predictions (aerated slug & time shift correction): time variation of the pressure drop (upper figures) and force acting on the riser's upper elbow (middle and lower figures) during the accumulated water displacement for $Q^0_G/Q^0_{G|Crit}$ = 0.5, 1.2 ($a$,$b$); $P_O$ = 100 atm, $D$ = 100 mm, $\varepsilon_{i\text{-hor}}$ = 50%, and $t_{ramp}$ = 15 s.

### *Pipe diameter effect*

To demonstrate the effect of the internal pipe diameter on the time-variation of the pressure drop, the water displacement by methane in three pipe diameters, $D$ = 50, 100, and 150 mm, was tested. The jumper's geometry with $D$ = 50 and 100 mm is scaled according to **Figure 2.1**. On the other hand, in the jumper with $D$ = 150 mm, the scaling is not preserved, and the lengths of the different jumper sections are the same as those of the jumper with a diameter of 100 mm. In this way, we tested the impact of increasing the pipe diameter while retaining the same overall pipeline length.

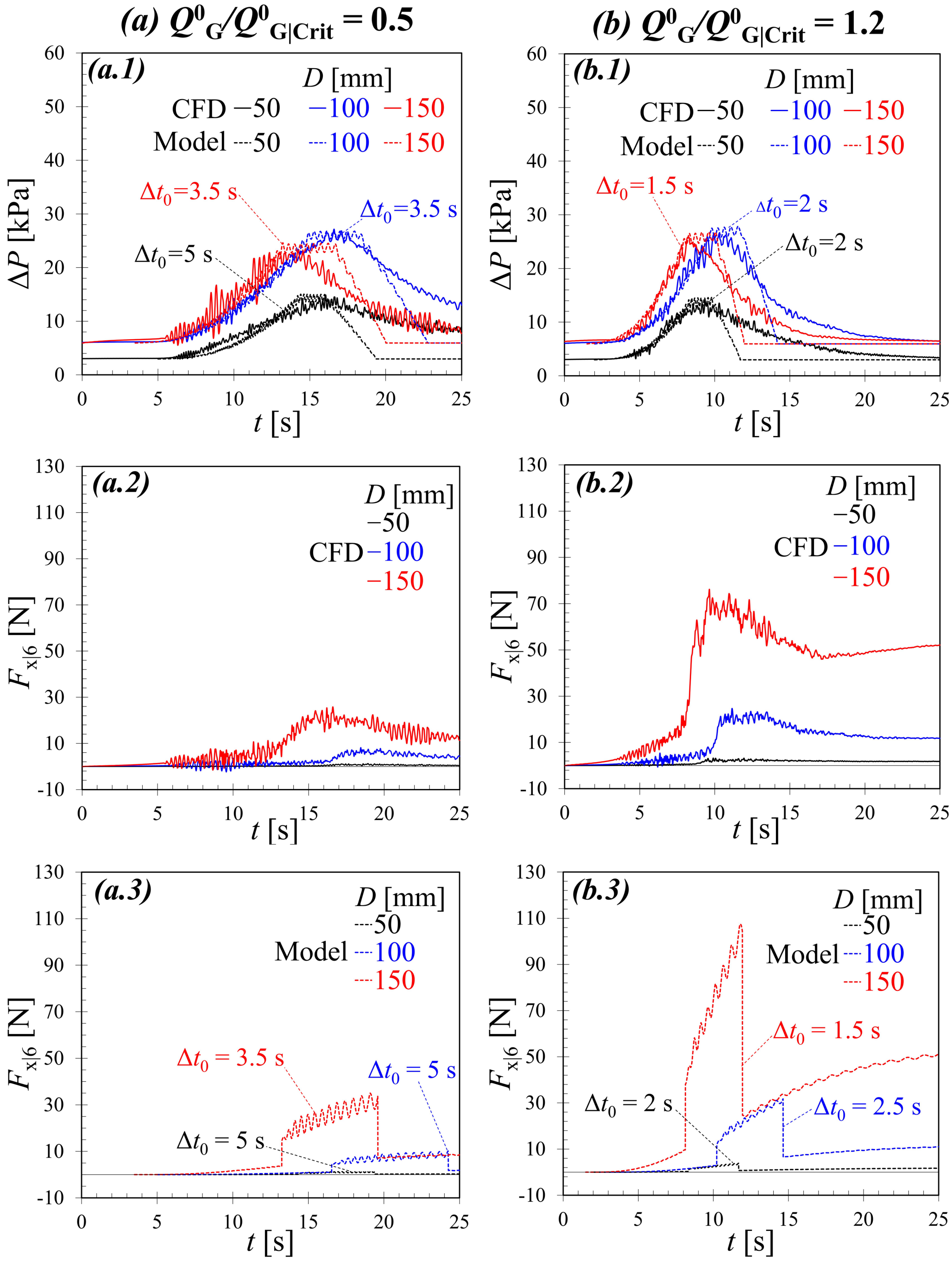

**Figure 4.17:** Effect of the pipe diameter. Comparison between the numerical simulations data and mechanistic model predictions (aerated slug & time shift correction): time variation of the pressure drop (upper figures) and force acting on the riser's upper elbow (middle and lower figures) during the accumulated water displacement for $Q^0_G/Q^0_{G|Crit} = 0.5, 1.2$ (*a,b*); $P_O = 100$ atm, $\varepsilon_{i\text{-}hor} = 50\%$, and $t_{ramp} = 15$ s

The comparison between the aerated slug model and the numerical simulation data is presented in **Figure 4.17.upper row**. $P_O$ = 100 atm, $\varepsilon_{i\text{-}hor}$ = 50% and $t_{ramp}$ = 15 s; the gas residence times based on the final gas in situ velocity are $t_{res,50\ mm}$ = 10.4 s and 4.33 s; $t_{res,100\ mm}$ = 14.85 s and 6.19 s; $t_{res,150\ mm}$ = 11.56 s and 4.81 s, for $Q^0{}_G/Q^0{}_{G|Crit}$ = 0.5 and 1.2, respectively. The figure shows that the diameter increase (while retaining the length scaling, from $D$ = 50 to 100 mm, namely, doubling the pipe lengths) leads to an increase in the maximal pressure rise. This is due to a larger liquid amount in a longer riser, which results in an increased hydrostatic pressure component. However, if the jumper length is kept the same, but the diameter is increased from $D$ = 100 to 150 mm, the maximal pressure drop slightly decreases both in the numerical simulations and in the mechanistic model predictions. Apparently, a slightly higher aeration level in the larger diameter pipes (see **Figure 4.10.e**) reduces the effective slug density and the hydrostatic pressure component.

**Figure 4.17.(middle and lower rows)** illustrates that the maximal force acting on the riser's upper elbow increases with the increase of the pipe diameter. It is expected since the inertia term depends on the fluid velocity and the cross-section area of the pipe, and both increase with the pipe diameter. The gas velocity, which is scaled to the critical gas superficial velocity, is proportional to $D^{0.5}$. Hence, the fluid inertial term is roughly proportional to $D^3$. Hence, the fluid inertial term is roughly proportional to $D^3$. This power law variation of the maximal force is in accordance with the simulation results and with model predictions (with only a slightly different power, $D^{3.2}$), in the range of the tested pipe diameters (see **Figure 4.18**).

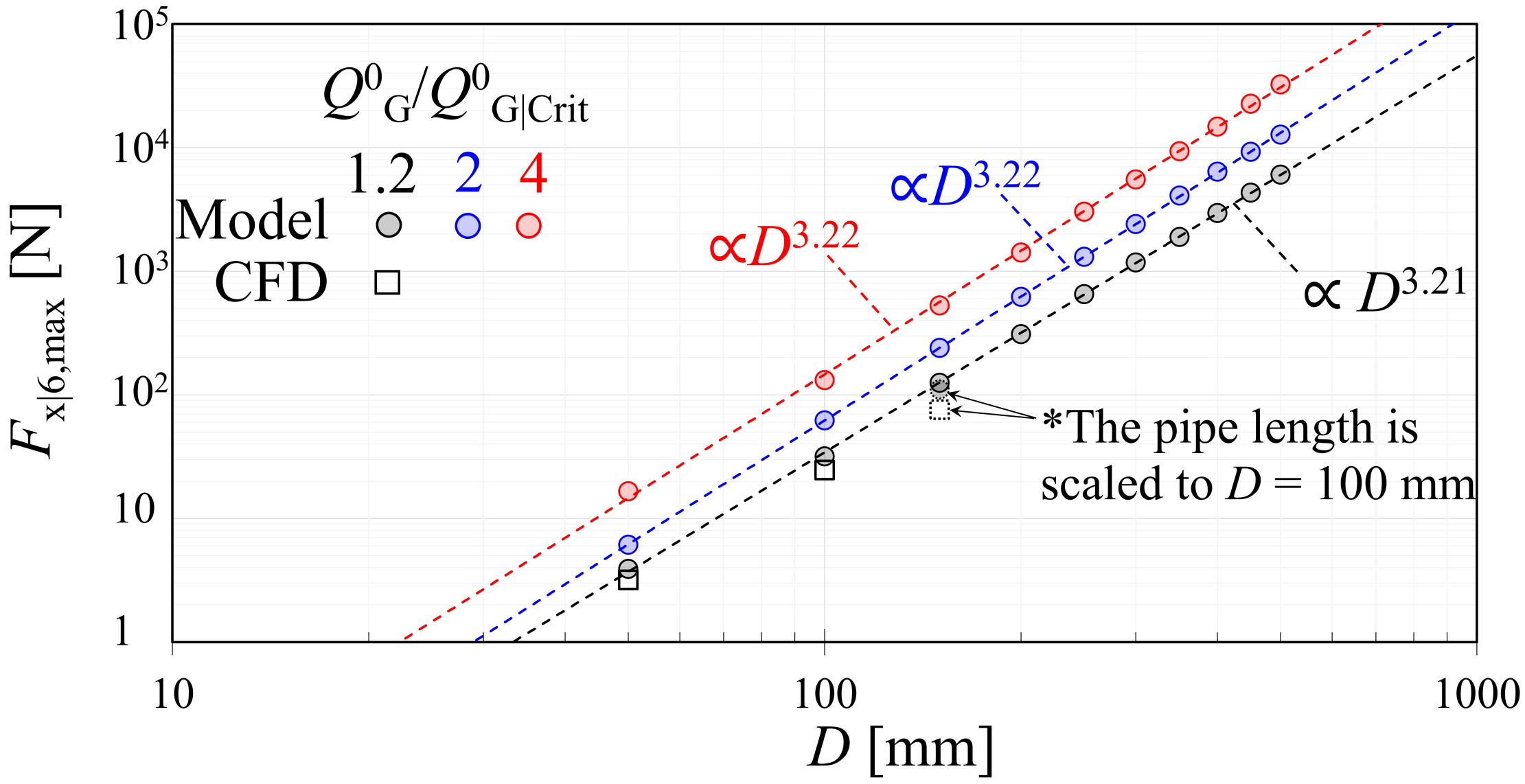

**Figure 4.18:** Effect of the pipe diameter on the maximal force acting on the riser's upper elbow during the accumulated liquid displacement. Comparison between the numerical simulations data and mechanistic model predictions for $Q^0{}_G/Q^0{}_{G|Crit}$ = 1.2, 2 and 4.

## 4.3 Pressure oscillation frequencies

The fluctuation frequencies of both pressure drop and force on the upper elbow of the riser were derived through the application of the FFT algorithm to the time-varying data obtained from numerical simulations. In **Figure 4.19**, the impact of various parameters, including gas mass flow ramp-up rate, initial liquid amount, gas pressure level, liquid viscosity, and pipe diameter, on the dominant oscillation frequencies of pressure drop is illustrated.

It is noteworthy that the force frequencies exhibit similar trends to the presented pressure drop frequency values. Consequently, the specific data on force oscillations are not included in the figure. The analysis indicates that the dominant frequencies tend to increase with a decrease in the liquid amount and an increase in the gas pressure level. However, they do not show significant sensitivity to gas flow ramp-up rate and liquid viscosity.

Interestingly, the dominant frequency experiences a notable decrease by approximately a factor of two when the pipe diameter increases from $D$ = 50 to 100 mm, which results in a doubling of the pipe length. Conversely, when the pipe length remains constant and the diameter increases from $D$ = 100 to 150 mm, the frequency remains practically unaffected. These findings suggest a strong dependence of the dominant frequency of pressure fluctuations on the overall pipe length while showing independence from the pipe diameter.

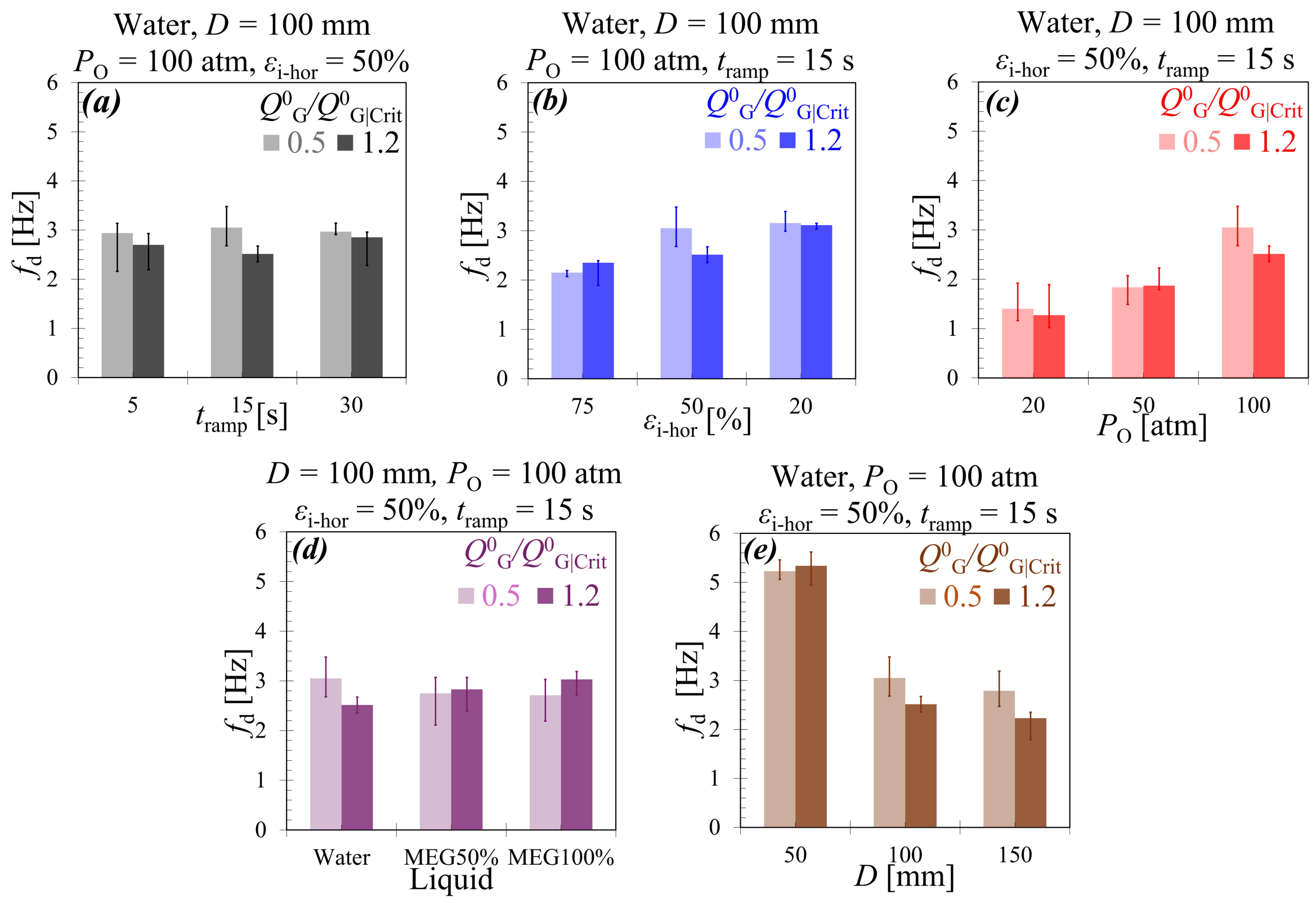

**Figure 4.19:** The dominant frequency of the pressure drop fluctuations, $f_d$ : Effects of *(a)* mass gas flow ramp-up rate, $t_{ramp}$; *(b)* initial liquid amount, $\varepsilon_{i\text{-hor}}$; *(c)* gas pressure level, $P_O$; *(d)* liquid viscosity, and *(e)* pipe diameter, $D$. Error bars correspond to ± 30% range of a peak frequency.

In gas-liquid flow, pressure fluctuations can stem from two primary sources: acoustic waves and the dynamic behavior of liquid displacement within the riser and its corresponding flow pattern. Notably, the impact of overall pipe length on the frequency of pressure drop observed in numerical simulations suggests that these oscillations can be ascribed to the acoustic wave behavior of a compressible fluid in response to a change in the inlet flow rate while maintaining a constant pressure at the outlet (i.e., an open-closed system). In such cases, the first harmonic of these oscillations is given by $f_1 = C/(4L)$, where $C$ represents the speed of sound, and $L$ denotes the overall pipe length.

In the absence of a liquid phase, pressure fluctuations resulting from alterations in mass flow rate are attributed to acoustic waves. Simulations conducted with a single-phase flow of gas, such as air or methane, in the jumper geometry, responding to changes in the inlet flow rate, exhibited pressure oscillations characterized by a dominant frequency close to the expected frequency of the first harmonic of an acoustic wave (see **Figure 4.20.b&d**). As shown, the oscillation frequency is indeed proportional to $1/L$. Altering the gas inlet temperature revealed that the frequency of pressure oscillations was proportional to $T^{0.5}$, a characteristic of acoustic waves. These oscillations demonstrated rapid decay, diminishing to less than 5% of the initial amplitude within 5 seconds (**Figure 4.20.a&c**). Interestingly, altering time-dependent boundary conditions, such as different ramp-up rates for the inlet mass flow, did not significantly impact the dominant frequency or decay time (not shown). Notably, an abrupt change (i.e., step change) in the inlet mass flow rate resulted in initial pressure change values (**Figure 4.20.a**), which aligned favorably with Joukowski, 1904 equation - $\Delta p = \rho_G C_G \Delta V$, where $\Delta p$ and $\Delta V$ are pressure and velocity change, respectively. As anticipated, for a constant input mass flow rate of the gas, the oscillation amplitude increases with reducing the pipe diameter **Figure 4.20.c**), but the frequency remains unchanged. Higher frequency harmonics as $f_2 = 3C_G/(4L)$ and $f_3 = 5C_G/(4L)$ were also detected in the gas flow simulations, albeit with considerably lower amplitudes than the first harmonic. It is worth noting that the amplitude of the initial pressure oscillations following the change in the input gas flow rate can be mitigated by extending the ramp-up time.

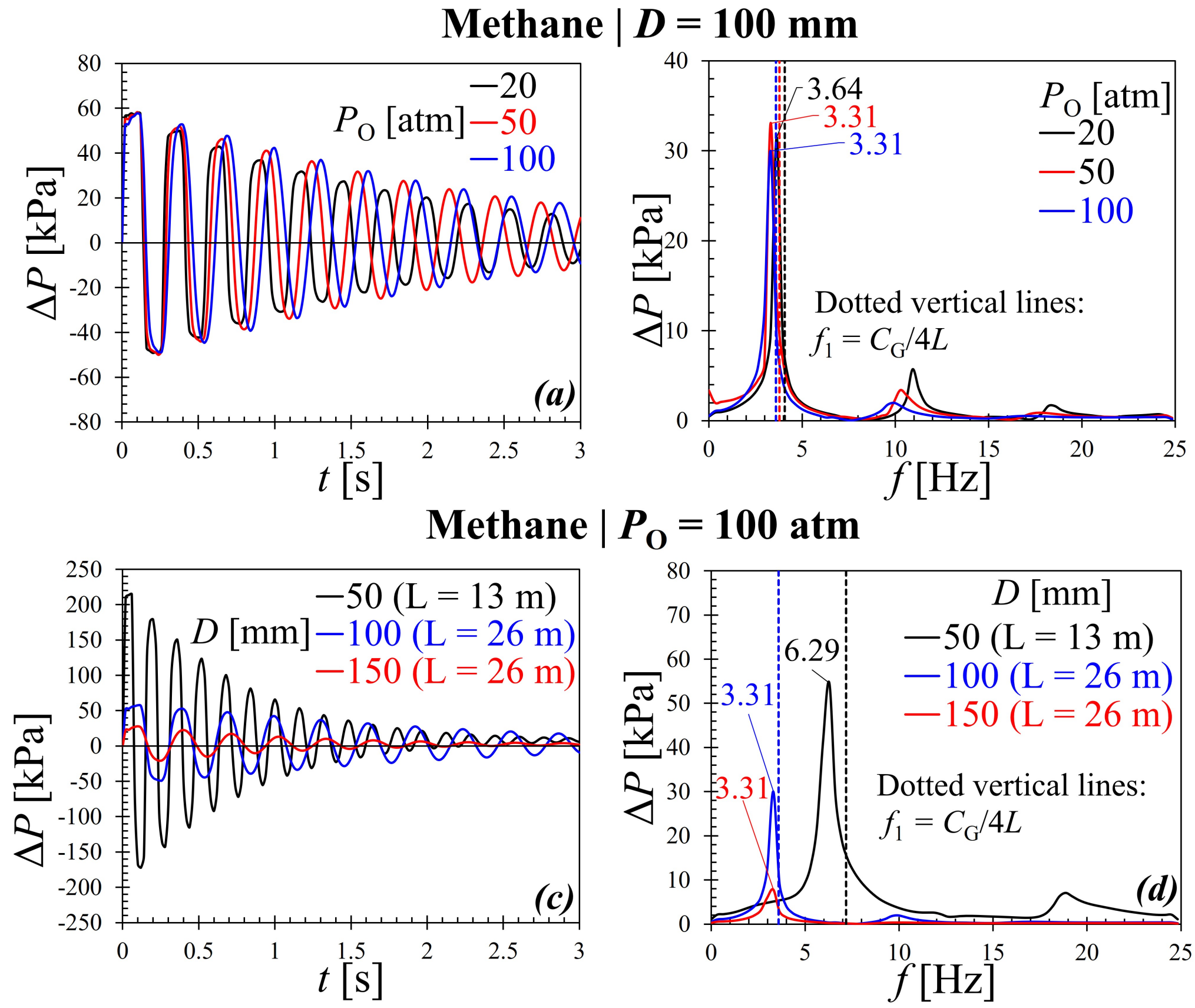

**Figure 4.20:** Pressure fluctuation in transient simulations of single-phase gas flow through the jumper in response to a step change in the input mass flow rate. Effect of pressure level on *(a)* the pressure oscillation amplitude and *(b)* frequency. Effect of diameter and length on *(c)* the pressure oscillation amplitude and *(d)* frequency. The results presented for tramp = 0 s (step change) and $\dot{m}_{GI}$ = 1.2 kg/s corresponds to $Q^0_{GS}/Q^0_{GS|crit}$ = 1.2 for $P_O$ = 100 atm, $D$ = 100 mm).

To investigate the potential impacts of bends, we conducted simulations of single-phase gas flow in a straight pipe with an overall length identical to that of the jumper. We observed nearly identical first harmonic frequencies ($f_1 = C_G/(4L)$) in pressure oscillations compared to those recorded in the jumper geometry. However, the simulations in the straight pipes exhibited higher harmonics with slightly greater amplitudes in comparison to the jumper geometry. This suggests that the bends in the jumper geometry may have a dissipating effect on the higher harmonic oscillations.

The prediction of acoustic wave frequency in transient gas-liquid flow through the jumper is considerably more complex due to the dependence of sound speed on both time and spatial variations of the holdup (e.g., Benjelloun and Ghidaglia, 2021). To address this challenge, we employ Wood's equation (Wood, 1941) to estimate the sound velocity in the jumper sections occupied by the gas-liquid mixture. Wood's equation predicts the sound velocity of a homogeneous gas-liquid mixture ($C$) based on the gas void fraction and the sound velocities of the gas ($C_G$) and the liquid ($C_L$) at the system's temperature and pressure. It is given by:

$$\frac{1}{(\alpha\rho_G + (1-\alpha)\rho_L)C^2} = \frac{\alpha}{\rho_G C_G^2} + \frac{1-\alpha}{\rho_L C_L^2} \qquad \textbf{4.1}$$

In fact, the simulations utilize the VOF method, which treats the mixture as homogeneous with a local value of $\alpha$ assigned to each cell. In the simulation, the liquid is modeled as incompressible. However, within the gas void fraction range encountered in the jumper, the contribution of a high sound velocity in the liquid in **Eq. 4.1** to the value of $C$ is practically negligible. Also, the sound velocity of the gas-liquid mixture is rather insensitive to the value of the void fraction in a wide range of $0.2 < \alpha < 0.85$, but is much more sensitive to the pressure than in single-phase gas flow (Kieffer, 1977; Benjelloun and Ghidaglia, 2021).

As the jumper sections between the inlet and the horizontal section are occupied by gas only, the predicted first harmonic acoustic frequency in a jumper of an overall length $L$ is given by:

$$f_{P-J} = \frac{\frac{L_G}{L}C_G + \left(1 - \frac{L_G}{L}\right)C}{4L} \qquad \textbf{4.2}$$

Here $L_G = L_{12} + L_{23} + L_{34}$ (see **Figure 2.1**). The values of $C_G$ are given in **Table 2.3**.

A comparison of the dominant pressure oscillation frequency obtained in the simulations with the predicted first harmonic acoustic wave frequency, $f_{P-J}$, is presented in **Figure 4.21**. The predicted values are derived from representative gas void fraction values during the liquid displacement through the riser (**Figure 4.10**), which were also utilized in the aerated slug model. Considering the complicated transient two-phase flow in the jumper during the liquid displacement, the acoustic wave frequencies predicted by **Eq. 4.2** exhibit reasonable agreement with the dominant pressure fluctuation frequencies obtained in the simulations. Consequently, the latter can be reasonably attributed to acoustic waves. It is worth noting that in the simulation, higher harmonics of the acoustic waves were not detected, suggesting their dissipation, or their presence at levels comparable to the "noisy" fluctuations arising from the time variations of the flow phenomena.

Compared to single-phase gas flow, the presence of liquid in the jumper results in prolonged disturbances in gas pressure, especially at sub-critical gas velocities. The temporal changes in local liquid holdup and the consequent alterations in gas velocity contribute to variations in the pressure field. Consequently, acoustic waves persist as long as the liquid remains in the jumper. We also conducted an analysis of pressure oscillation frequencies in the simulations at different stages of liquid purging. The results (not shown) indicate that during the late stage of liquid purging, the gas void fraction in the riser increases with time. Consequently, the sound speed increases and the pressure oscillation frequency exhibits a moderate increase toward the single-phase gas flow value. Similarly, at very short simulation times, when the liquid layer is still in the horizontal section, the sound speed through the horizontal section is very close to that of the gas (refer to **Eq. 8** in Benjelloun and Ghidaglia, 2021, for the speed of sound in stratified flow). Thus, the initial pressure oscillation frequency in the simulation is close to that obtained in single-phase gas flow.

Joukowski's equation elucidates why the acoustic wave was not detected in the low-pressure air-water simulations. The input gas flow rate, when scaled by the critical value for liquid purging, is roughly proportional to $P^{0.5}$. Consequently, with this scaling of the flow rate change, the amplitude of the acoustic wave is considerably smaller in low-pressure systems. Conversely, water purging in low-pressure systems is characterized by the flow of water plugs and chunks in the riser (refer to **Figure 4.4** and **Figure 4.6** in **Section 4.2.1**). These water-related phenomena are accountable for the observed dominant fluctuation frequencies. In contrast, in high-pressure systems, the amplitudes of fluctuations due to the time variation of the flow pattern are much smaller. In such cases, the dominant frequency observed in the simulations is that of the acoustic wave.

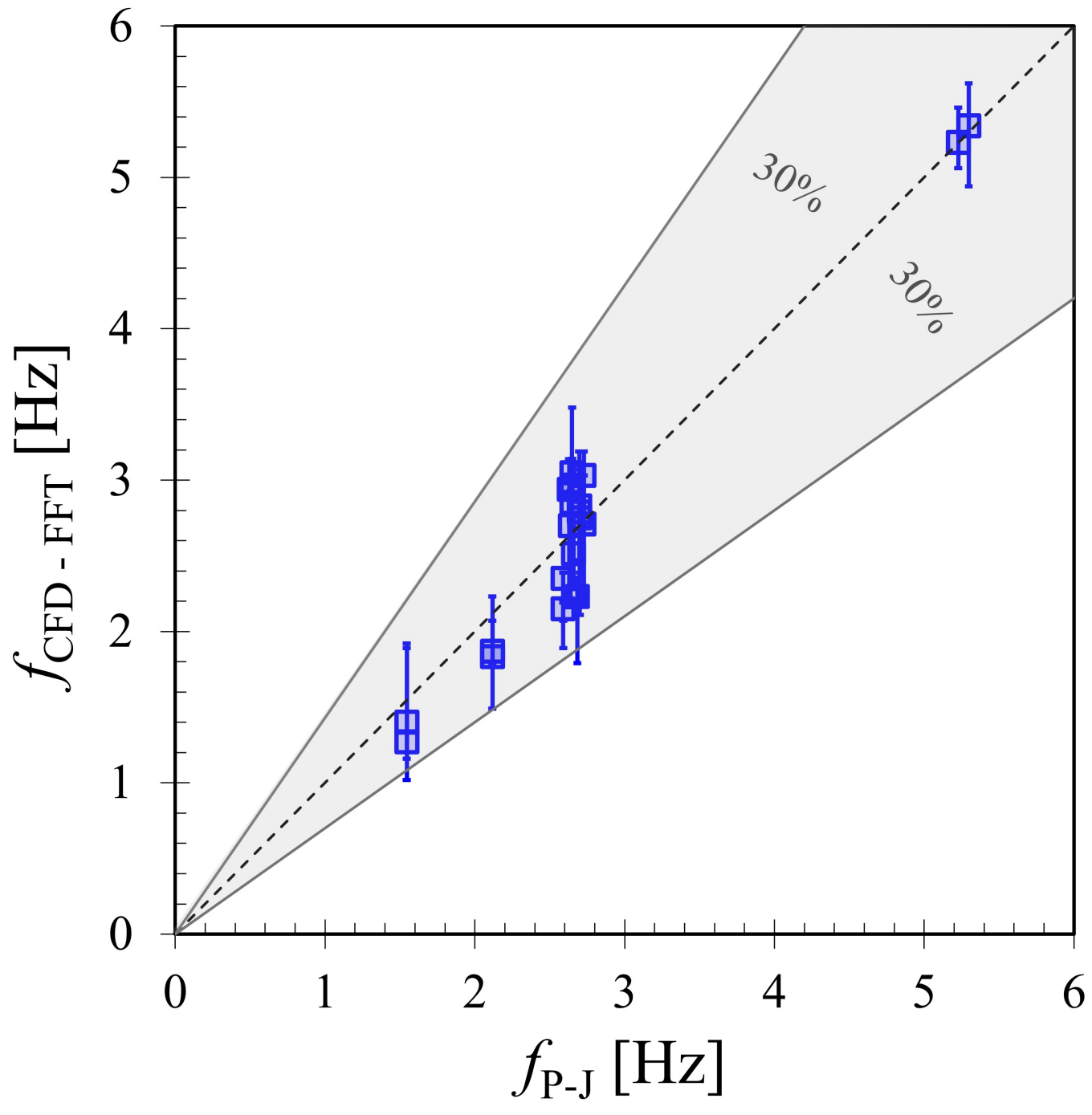

**Figure 4.21:** Comparison of the pressure fluctuation frequencies obtained in the numerical simulations with those predicted by **Eq. 4.2** denoted as $f_{P-J}$. Error bars in the CFD data correspond to ± 30% range of the dominant frequencies.

In high-pressure systems, the pressure drop (and force) calculated by the aerated slug mechanistic model also exhibit oscillations during liquid displacement. These oscillations are linked to the dynamics of liquid displacement in the jumper and are connected to the acceleration of the liquid by the compressible gas, but they are not acoustic waves. The dominant frequencies obtained by the aerated slug model in various tested cases were determined by applying the FFT algorithm to the time variation of the predicted pressure drop. The influence of system parameters on fluctuation frequency can be inferred from the natural frequency calculated based on the analysis of the linearized aerated slug model (see **Section 3.3**, **Eq. 3.8**).

**Figure 4.22** presents the fluctuation frequency predicted by the aerated slug model alongside the dominant frequencies obtained in the CFD simulations. The figure shows that the frequencies derived from the model equations are slightly lower than the dominant frequency observed in the CFD simulations, albeit exhibiting a similar trend. The results are presented in terms of the St vs. Eu dimensionless parameters, deduced from the linearized slug flow model (**Eq. 3.10**). While the linearized model captures the general trend implied by the simulation results, it tends to overpredict the dominant pressure fluctuation frequencies. As depicted, adjusting the constant in **Eq. 3.9** by multiplying RHS by a factor of $C_M$=0.55 allows the linearized model prediction to be fine-tuned, resulting in a more reasonable agreement with the frequency data obtained from the CFD simulations.

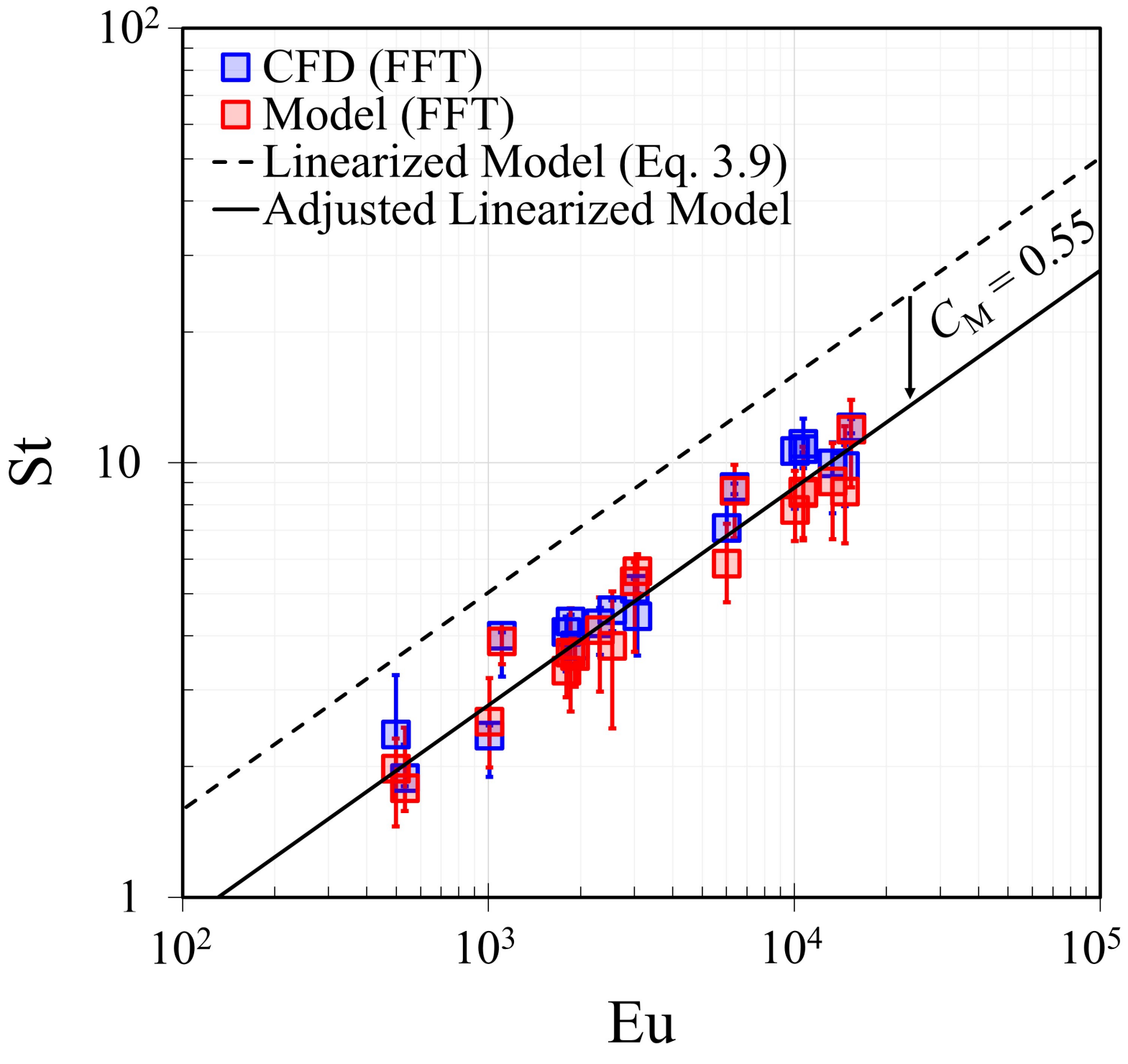

**Figure 4.22:** St vs. Eu obtained from the numerical simulations results (blue □) and the slug flow model (red □). Error bars correspond to ± 30% range of a St number. The solid (black) line is obtained by adjusting the constant of the linearized model (**Eq. 3.10**, dashed black).

## 5 Conclusions

The focus of this study is to gain insights into the transient gas-liquid flow within an M-shaped jumper, a crucial component of subsea high-pressure gas production pipelines. Employing 3D numerical simulations and mechanistic models, we aim to illustrate the impacts of various parameters, such as pipe diameter, gas pressure level, initial amount of accumulated liquid, liquid properties, and gas mass flow rate on the temporal variations of pressure drop across the jumper and the forces acting on the riser's upper elbow during liquid displacement by gas flow. Furthermore, the study endeavors to determine the critical gas mass flow rate necessary for the complete removal of initially accumulated liquid in the jumper.

The numerical simulations reveal that the critical gas flow rate corresponds to the minimum value preventing liquid flow reversal at the jumper's riser, and this value is found to be independent of the initial liquid amount in the jumper, and practically unaffected by the liquid viscosity and ramp-up rate.. The corresponding critical gas production rate, $Q^0_{G|Crit}$, demonstrates an increase with increasing both the gas pressure and pipe diameter. A scaling rule, $Q^0_{G|Crit} \propto (ZM)^{-0.5} P_G^{0.5} D^{2.5}$, based on the flow reversal criteria, proves effective in predicting critical gas flow rate values, as it aligns with the data from numerical simulations.

The pressure drop on the jumper rises during the purging of accumulated liquid to overcome the hydrostatic pressure in the jumper’s riser, the frictional pressure drop, and to accelerate the initial stationary fluid (gas and liquid). The effect of the system parameters and operational condition on the system pressure drop and forces on the riser’s upper elbow were examined for various gas mass flow rates.  This analysis is conducted for gas mass flow rates that are scaled by the corresponding critical value necessary for liquid purging, such as 50% (subcritical) and 120% (supercritical) of the critical flow rates.

In general, the simulations indicate that the maximum pressure drop across the jumper increases with a rise in the initially accumulated liquid amount and gas pressure level. However, it appears to be practically independent of the liquid properties (e.g., MEG-water solutions) and of the pipe diameter when the lengths of the jumper's sections are scaled up or down proportionally to the pipe diameter. On the other hand, the maximum force acting on the riser's upper elbow due to the gas-liquid flow increases with the gas mass flow and ramp-up rate, initially accumulated liquid amount, and with the pipe diameter. It was found to decrease with an increase in the gas pressure level and remains practically unaffected by the tested liquid properties.

Pressure and force fluctuations are observed during the liquid displacement process in both low and high-pressure systems. In low-pressure systems, the dominant fluctuation frequencies are approximately 1 Hz and seem to be associated with significant temporal variations in liquid holdup within the riser. Conversely, 13 through an open-closed-ended pipe of length *L*, where $f_1 = C/(4L)$, where *C*, the speed of sound in gas-liquid systems is contingent on the gas void fraction and gas pressure. The prevalence of the acoustic wave frequency at high pressure is attributed to the considerably higher change in the gas mass flow rate considered, associated with the higher critical gas mass flow rate. The acoustic wave frequency is detectable in the simulation as long as liquid is present in the jumper.

While the acoustic wave fluctuations are of low amplitude, they could pose a potential risk to the structural integrity of the jumper over an extended exposure time, especially when their frequency aligns closely with the structure's natural frequency. It is worth noting that in a field system, the relevant length scale *L* for determining the acoustic wave frequency is the distance from the valve location to the constant pressure separator, and the acoustic wave frequency is lower than that predicted based on the jumper length.

## Acknowledgments

This study was supported through generous funds provided by the Israel Ministry of Energy under contract number 220-17-005. We thank the Agar family for their donation in support of the research.

## Appendix A: Convergence tests and discretization error estimation

***Spatial and temporal convergence tests***

Grid and time convergence tests were conducted to evaluate the numerical accuracy and stability of the computational models utilized. In the grid convergence tests, the computational mesh was refined while keeping all other parameters invariable, and the changes in the numerical solution were observed. Various grid resolutions were explored, ranging from coarse to fine, as detailed in **Table A.1** and **Table A.2**, for methane and air respectively. First, these tests were performed with a maximum Courant number of 0.5. To further assess the impact of temporal discretization, a sensitivity analysis was conducted by reducing the maximum Courant number to 0.25 while using the selected grid. This methodology was systematically repeated for three pipe diameters: $D$ = 50, 100, and 150 mm, considering the methane as the gaseous phase. Note that spatial and temporal convergence tests for air as the gaseous phase were reported in our previous study, Yurishchev et al., 2024.

The convergence tests (grid and time) focused on the main parameters that were investigated in the present study (1) the critical gas flow rate needed to purge the accumulated liquid (see **Figure A.1**); (2) the pressure drop on the jumper during the liquid displacement (see **Figure A.2**); and (3) the horizontal force applied on the riser's upper elbow during the liquid passage (see **Figure A.3**). The tests revealed that grid refinement (relatively to the selected grid) and reducing the maximal Courant number to 0.25 (from 0.5) usually led to lower fluctuations in time (in terms of amplitude). However, in general, it did not significantly affect the results. The discretization error estimation is presented below.

**Table A.1:** Axial, lateral, and total number of elements of grids used to obtain spatial convergence with methane (gas pressure 20÷300 atm) in $D$ = 50, 100, and 150 mm* domains. *In the case of $D$ = 150 mm, the sections' lengths are scaled to $D$ = 100 mm to reduce the grid size.

Methane, $P_O$ = 20 ÷ 300 atm

| ***Grids*** | ***Axial* x *Lateral Elements (Total Elements)*** | | |
|---|---|---|---|
| | $D$ = 50 mm | $D$ = 100 mm | $D$ = 150 mm* |
| Fine | 3245 x 2400 (7,788,000) | 4842 x 2628 (12,724,776) | 4938 x 3564 (17,599,032) |
| Coarse | 1315 x 1105 (1,453,075) | 2668 x 900 (2,401,200) | 2635 x 2057 (5,420,195) |
| Selected | 1625 x 2400 (3,900,000) | 3425 x 1984 (6,795,200) | 3533 x 2960 (10,457,680) |

**Table A.2:** Axial, lateral, and total number of elements of grids used to obtain spatial convergence in $D$ = 50 mm domain with air (gas pressure 1 ÷ 5 atm).

Air, $D$ = 50 mm, $P_O$ = 1 ÷ 5 atm

| ***Grids*** | ***Axial* x *Lateral Elements (Total Elements)*** |
|---|---|
| Fine 1 | 5295 x 1125 (5,956,875) |
| Fine 2 | 3735 x 2000 (7,470,000) |
| Coarse | 1795 x 720 (1,292,400) |
| Selected | 2730 x 1125 (3,071,250) |

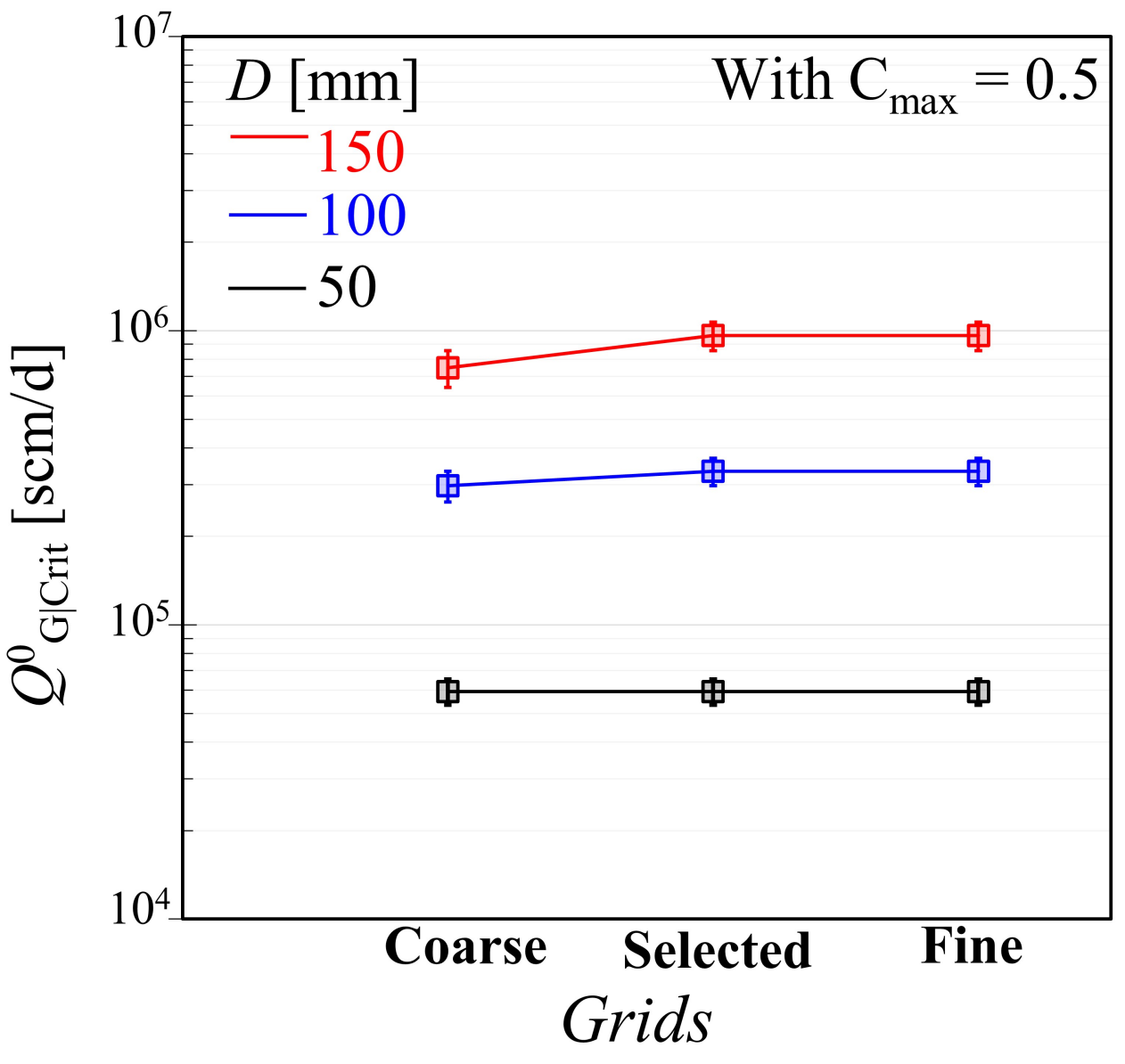

**Figure A.1:** Grid convergence tests regarding the critical gas flowrate, $Q^0_{G|Crit}$, needed to purge the accumulated liquid from the jumper for different pipe diameters, $D$ = 50, 100 and 150 mm, considering methane as the gaseous phase at gas pressure level, $P_O$ = 100 atm.

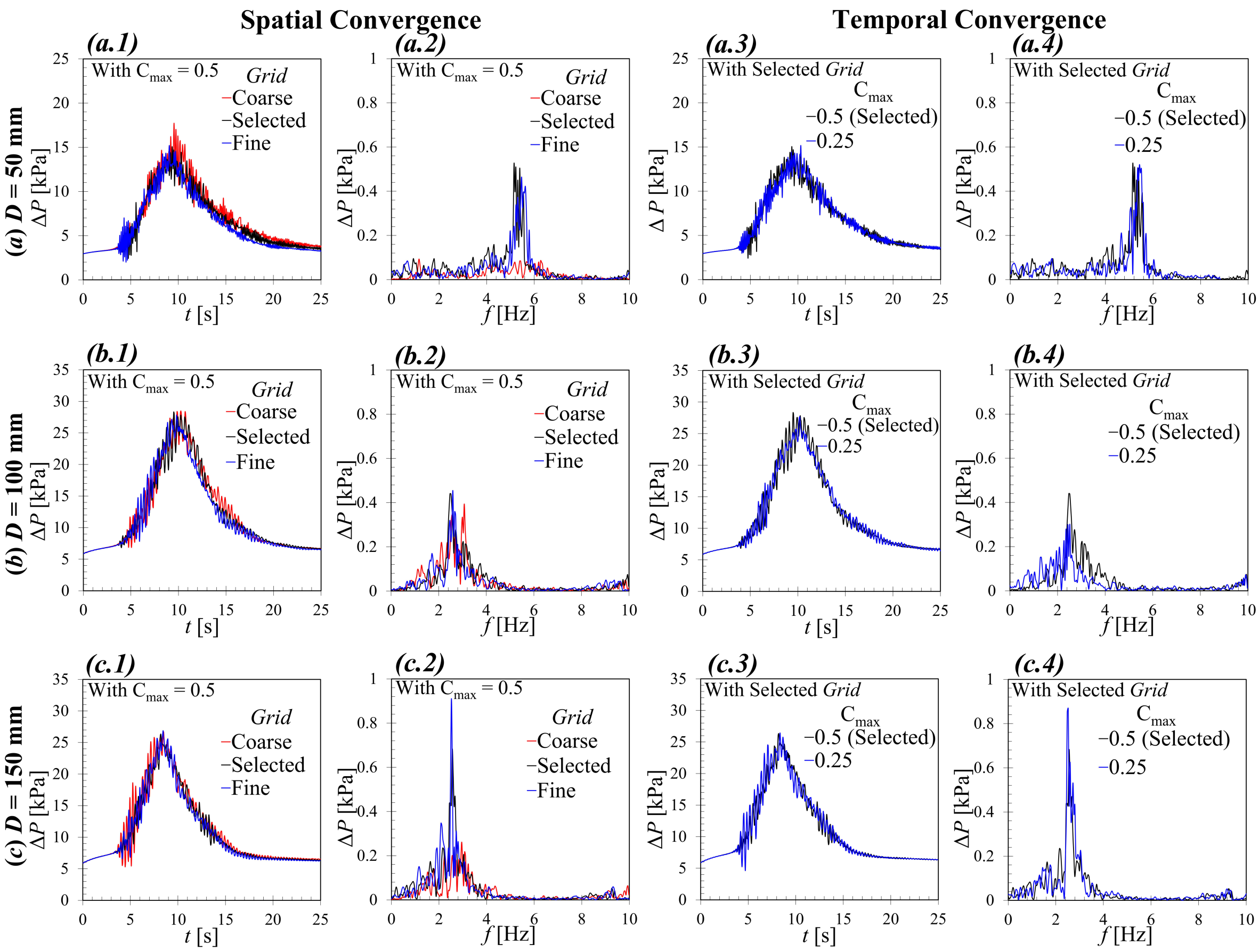

**Figure A.2**: Grid and time convergence tests regarding time and frequency response of pressure drop for a specific case of methane as gaseous phase and pipe diameter, $D$ = *(a)* 50 mm, *(b)* 100 mm, and *(c)* 150 mm. $P_O$ = 100 atm, $Q^0_G/Q^0_{G|Crit}$ = 1.2, $\varepsilon_{i\text{-}hor}$ = 50%, $t_{ramp}$ = 15 s.

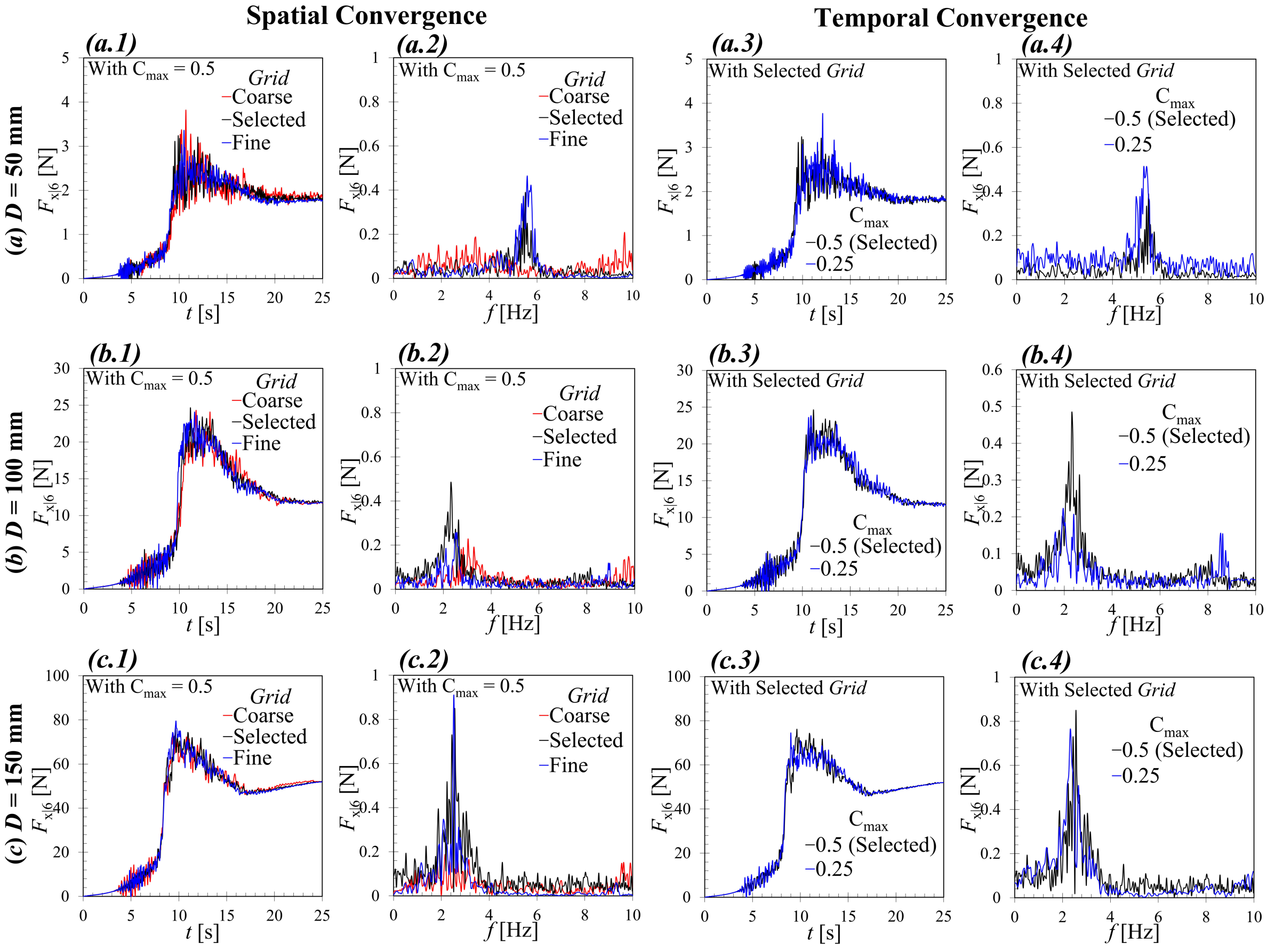

**Figure A.3**: Grid and time convergence tests regarding time and frequency response of horizontal force acting on the riser's upper elbow for a specific case of methane as gaseous phase and pipe diameter, $D$ = *(a)* 50 mm, *(b)* 100 mm, and *(c)* 150 mm. $P_O$ = 100 atm, $Q^0_G/Q^0_{G|Crit}$ = 1.2, $\varepsilon_{i\text{-}hor}$ = 50%, $t_{ramp}$ = 15 s.

### *Discretization error estimation*

The Grid Convergence Index (GCI) method was applied to estimate the discretization error in the present study. The method serves as a measure of the percentage by which the computed value deviates from the value of the asymptotic numerical solution and thereby offers insights into how the solution might change with further grid refinement. The GCI method is based on Richardson, 1911; Richardson and Gaunt, 1927. In the following, the recommended procedure for estimating the discretization error presented by Celik et al., 2008 is applied.

A representative cell size, $h$ for the grids used in the grid convergence tests is defined:

$$h = \left[\frac{1}{N}\sum_{\mathrm{i}=1}^{N}\Delta V_{\mathrm{i}}\right]^{\frac{1}{3}} \qquad \textbf{A.1}$$

where $\Delta V_i$ is the volume of the i$^{th}$ cell, and $N$ is the total number of cells used for the computations.

The apparent order of the method accuracy, $p$ is given by:

$$p = \frac{1}{\ln r_{21}}\left|\ln\left|\frac{\varepsilon_{32}}{\varepsilon_{21}}\right| + q(p)\right|$$

$$q(p) = \ln\left(\frac{r_{21}^{p} - s}{r_{32}^{p} - s}\right) \qquad \textbf{A.2}$$

$$s = 1 \cdot \text{sign}\left(\frac{\varepsilon_{32}}{\varepsilon_{21}}\right)$$

where $r_{21}=h_2/h_1$ and $r_{32}=h_3/h_2$, with $h_1<h_2<h_3$, and $\varepsilon_{21} = \varphi_2 - \varphi_1$, $\varepsilon_{32} = \varphi_3 - \varphi_2$, with $\varphi_k$ denoting the solution on the k$^{th}$ grid. The extrapolated value resulting from further grid refinement, $\varphi_{\text{ext}}^{21}$ is then:

$$\varphi_{\text{ext}}^{21} = \frac{\left(r_{21}^{p}\varphi_1 - \varphi_2\right)}{\left(r_{21}^{p} - 1\right)} \qquad \textbf{A.3}$$

The corresponding relative errors, $e_{\text{a}}^{21}$ and $e_{\text{ext}}^{21}$ are:

$$e_{\text{a}}^{21} = \left|\frac{\varphi_1 - \varphi_2}{\varphi_1}\right|$$

$$e_{\text{ext}}^{21} = \left|\frac{\varphi_{\text{ext}}^{21} - \varphi_1}{\varphi_{\text{ext}}^{12}}\right| \qquad \textbf{A.4}$$

Finally, the fine-grid convergence index is given by:

$$GCI_{\text{fine}}^{21} = \frac{1.25e_{a}^{21}}{r_{21}^{p} - 1} \qquad \textbf{A.5}$$

**Table A.3** presents three examples of calculations for the pressure drop on the jumper. The cases considered are $D$ = 50, 100, and 150 mm with methane-water, $P_{\text{O}}$ = 100 atm, $Q^0_{\text{G}}/Q^0_{\text{G|Crit}}$ = 1.2, $\varepsilon_{\text{i-hor}}$ = 50% and $t_{\text{ramp}}$ = 15 s. The shown times represent the peak pressures for each case. As shown in the table, the numerical uncertainties in the fine-grid solution ($CGI^{21}_{\text{fine}}$) for the pressure drop are 4.91%, 1.16%, and 2.03%, for $D$ = 50 mm, 100 mm, and 150 mm, respectively. Furthermore, the medians of the local order of accuracy $p_{1/2}$, discretization uncertainty $GCI_{\text{fine}}{}^{21}{}_{1/2}$ and oscillatory convergence percentage for the same cases are presented in **Table A.4**.

**Table A.3:** Sample calculations of discretization error for the pressure drop on the jumper. The simulated cases of Methane-water, $P_O$ = 100 atm, $Q^0_G/Q^0_{G|Crit}$ = 1.2, $\varepsilon_{i\text{-}hor}$ = 50% and $t_{ramp}$ = 15 s, with $D$ = 50 mm, 100 mm, and 150 mm.

**$\varphi$ is the pressure drop on the jumper**

| $D$ [mm] | 50 | 100 | 150 |
|---|---|---|---|
| $N_1$; $N_2$; | 7,788,000; 3,900,000; | 12,724,776; 6,795,200; | 17,599,032; 10,457,680; |
| $N_3$ | 1,453,075 | 2,401,200 | 5,420,195 |
| $r_{21}$ | 1.26 | 1.23 | 1.19 |
| $r_{32}$ | 1.39 | 1.41 | 1.25 |
| $t$ [s] | 9.5 | 10.3 | 8.7 |
| $\varphi_1$ [kPa] | 13.19 | 24.94 | 22.75 |
| $\varphi_2$ [kPa] | 14.68 | 26.99 | 24.09 |
| $\varphi_3$ [kPa] | 15.47 | 26.12 | 23.68 |
| $\varepsilon_{32}$ | 0.79 | -0.873 | -0.41 |
| $\varepsilon_{21}$ | 1.49 | 2.05 | 1.34 |
| $\varepsilon_{32}/\varepsilon_{21}$ | 0.53 | -0.42 | -0.31 |
| $p$ | 5.87 | 10.91 | 8.81 |
| $\varphi_{ext}^{21}$ [kPa] | 12.68 | 24.71 | 22.51 |
| $e_a^{21}$ [%] | 11.28 | 8.23 | 5.86 |
| $e_{ext}^{21}$ [%] | 4.09 | 0.94 | 1.16 |
| $GCI_{fine}^{21}$ [%] | 4.91 | 1.16 | 2.03 |

**Table A.4:** The medians of the local order of accuracy, discretization uncertainty, and oscillatory convergence percentage. The simulated cases of Methane-water, $P_O$ = 100 atm, $Q^0_G/Q^0_{G|Crit}$ = 1.2, $\varepsilon_{i\text{-}hor}$ = 50% and $t_{ramp}$ = 15 s, with $D$ = 50 mm, 100 mm, and 150 mm.

| $D$ [mm] | 50 | 100 | 150 |
|---|---|---|---|
| $p_{1/2}$ | 3.98 | 3.38 | 4.63 |
| $GCI_{fine}{}^{21}{}_{1/2}$ [%] | 5.69 | 1.14 | 3.02 |
| Oscillatory Convergence Percentage | 11.2 | 19.4 | 14.4 |

**Table A.5** presents three examples of calculations for the horizontal force acting on the riser's elbow. The cases considered are $D$ = 50, 100, and 150 mm with methane-water, $P_O$ = 100 atm, $Q^0_G/Q^0_{G|Crit}$ = 1.2, $\varepsilon_{i\text{-}hor}$ = 50% and $t_{ramp}$ = 15 s. The shown times represent the force values corresponding to liquid passage through the elbow for each case. As shown in the table, the numerical uncertainties in the fine-grid solution ($CGI^{21}{}_{fine}$) for the pressure drop are 4.61%,

2.92%, and 2.36%, for $D$ = 50 mm, 100 mm, and 150 mm, respectively. Furthermore, the medians of the local order of accuracy $p_{1/2}$, discretization uncertainty $GCI_{fine}{}^{21}{}_{1/2}$ and oscillatory convergence percentage for the same cases are presented in **Table A.6**.

**Table A.5:** Sample calculations of discretization error for the horizontal force acting on the riser's upper elbow. The simulated cases of methane-water, $P_O$ = 100 atm, $Q^0_G/Q^0_{G|Crit}$ = 1.2, $\varepsilon_{i\text{-}hor}$ = 50% and $t_{ramp}$ = 15 s, with $D$ = 50 mm, 100 mm, and 150 mm.

**$\varphi$ is the horizontal force acting on the riser's elbow**

| $D$ [mm] | 50 | 100 | 150 |
|---|---|---|---|
| $N_1$; $N_2$; | 7,788,000; 3,900,000; | 12,724,776; 6,795,200; | 17,599,032; 10,457,680; |
| $N_3$ | 1,453,075 | 2,401,200 | 5,420,195 |
| $r_{21}$ | 1.26 | 1.23 | 1.19 |
| $r_{32}$ | 1.39 | 1.41 | 1.25 |
| $t$ [s] | 12.2 | 12.25 | 12.05 |
| $\varphi_1$ [N] | 2.15 | 21.07 | 62.78 |
| $\varphi_2$ [N] | 2.31 | 18.13 | 63.32 |
| $\varphi_3$ [N] | 3.06 | 19.51 | 64.39 |
| $\varepsilon_{32}$ | 0.767 | 1.37 | 1.07 |
| $\varepsilon_{21}$ | 0.145 | -2.93 | 0.55 |
| $\varepsilon_{32}/\varepsilon_{21}$ | 5.27 | -0.47 | 1.95 |
| $p$ | 4.51 | 9.27 | 2.18 |
| $\varphi_{ext}{}^{21}$ [N] | 2.08 | 21.57 | 61.83 |
| $e_a{}^{21}$ [%] | 6.74 | 13.93 | 0.87 |
| $e_{ext}{}^{21}$ [%] | 3.83 | 2.28 | 1.52 |
| $GCI_{fine}{}^{21}$ [%] | 4.61 | 2.92 | 2.36 |

**Table A.6:** The medians of the local order of accuracy, discretization uncertainty, and oscillatory convergence percentage regarding the horizontal force acting on the riser's upper elbow. The simulated cases of methane-water, $P_O$ = 100 atm, $Q^0_G/Q^0_{G|Crit}$ = 1.2, $\varepsilon_{i\text{-}hor}$ = 50% and $t_{ramp}$ = 15 s, with $D$ = 50 mm, 100 mm, and 150 mm.

| $D$ [mm] | 50 | 100 | 150 |
|---|---|---|---|
| $p_{1/2}$ | 4.35 | 3.27 | 5.42 |
| $GCI_{fine}{}^{21}{}_{1/2}$ [%] | 4.56 | 3.21 | 2.07 |
| Oscillatory Convergence Percentage | 9.4 | 14.2 | 12.6 |

## Appendix B: Linearization of the aerated slug model equations

To obtain some basic understanding of the system dynamics (e.g., time constant, resonant frequency and stability), the set of non-linear differential equations (**Eq. 3.1, 3.2 and 3.4)** are linearized around the steady state solution at *t*=0, where :

$$H^0 = U_\mathrm{S}^0 = \dot{m}_\mathrm{GI} = L_\mathrm{S}^0 = {H_\mathrm{r}^\mathrm{S}}^0 = 0$$

$$P_\mathrm{G}^0 = P_\mathrm{O} + \rho_\mathrm{G} g L_\mathrm{r},\ \ L_\mathrm{G}^0 = L_{12} + L_{23} + L_\mathrm{d} + L_\mathrm{h}(1-\varepsilon_\mathrm{i-hor}),\ \ {L_\mathrm{h}^\mathrm{L}}^0 = L_\mathrm{h}\varepsilon_\mathrm{i-hor},\ \ {H_\mathrm{r}^\mathrm{G}}^0 = L_\mathrm{r}$$

The linearized set of equations in terms of deviations variables from the steady state solution (denoted by upper bar) are:

Velocity continuity at the boundary of the aerated slug front reads:

$$\frac{d\overline{H}}{dt} = \overline{U}_\mathrm{S} \qquad \mathbf{B.1}$$

Momentum balance on the aerated slug and liquid plug reads:

$$\frac{d\overline{U}_\mathrm{S}}{dt} = \frac{\overline{P}_\mathrm{G} - (1-\alpha)\Delta\rho\overline{H}g - C_1\overline{U}_\mathrm{S}}{(1-\alpha)\rho_\mathrm{L}{L_\mathrm{h}^\mathrm{L}}^0} \qquad \mathbf{B.2}$$

where $\Delta\rho=\rho_\mathrm{L}-\rho_\mathrm{G}$ is the liquid-gas densities difference, and $C_1$ is a constant depends on the liquid flow regime:

$$C_1 = \frac{d\Delta P_\mathrm{f}}{dU_\mathrm{S}}|_{U_\mathrm{S}=0} \rightarrow C_{1\mathrm{Laminar}} = \frac{32{L_\mathrm{h}^\mathrm{L}}^0\mu_\mathrm{L}(1-\alpha)}{D^2};\ C_{1\mathrm{Turbulent}} = 0 \qquad \mathbf{B.3}$$

Mass balance on the gas upstream the liquid (slug/plug) tail reads:

$$\frac{d\overline{P}_\mathrm{G}}{dt} = \frac{\mathrm{Z}RT\overline{\dot{m}}_\mathrm{GI} - AP_\mathrm{G}^0\overline{U}_\mathrm{S}}{AL_\mathrm{G}^0} \qquad \mathbf{B.4}$$

The Laplace transform is applied to obtain the transfer functions representing the system response to a change in the gas flow, whereby:

$$H(s) = \frac{k_\mathrm{n}}{\tau^2 s^2 + 2\tau\xi s + 1}\frac{\dot{m}_\mathrm{GI}(s)}{s} \qquad \mathbf{B.5}$$

$$U_\mathrm{S}(s) = \frac{k_\mathrm{n}}{\tau^2 s^2 + 2\tau\xi s + 1}\dot{m}_\mathrm{GI}(s) \qquad \mathbf{B.6}$$

$$P_\mathrm{G}(s) = \left(k_\mathrm{p1} - \frac{k_\mathrm{p2}}{\tau^2 s^2 + 2\tau\xi s + 1}\right)\frac{\dot{m}_\mathrm{GI}(s)}{s} \qquad \mathbf{B.7}$$

where the gains, $k_n$, $k_{p1}$ and $k_{p2}$ are:

$$k_{\mathrm{n}} = \frac{\dfrac{ZRT}{\Delta\rho(1-\alpha)g}}{AL_{\mathrm{G}}^{0}\left(1+\dfrac{P_{\mathrm{G}}^{0}}{\Delta\rho(1-\alpha)gL_{\mathrm{G}}^{0}}\right)};\ k_{\mathrm{p1}} = \frac{ZRT}{AL_{\mathrm{G}}^{0}};\ k_{\mathrm{p2}} = \frac{P_{\mathrm{G}}^{0}}{L_{\mathrm{G}}^{0}}k_{\mathrm{n}} \qquad \textbf{B.8}$$

The time constant $\tau$, and the damping parameter, $\xi$ are given by:

$$\tau = \left(\frac{\dfrac{\rho_{\mathrm{L}}}{\Delta\rho}\dfrac{L_{\mathrm{h}}^{\mathrm{L}\,0}}{g}}{1+\dfrac{P_{\mathrm{G}}^{0}}{(1-\alpha)\Delta\rho g L_{\mathrm{G}}^{0}}}\right)^{0.5} \qquad \textbf{B.9}$$

$$\xi_{\mathrm{Laminar}} = \frac{16\mu_{\mathrm{L}}}{D^{2}}\left(\frac{\dfrac{L_{\mathrm{h}}^{\mathrm{L}\,0}}{g\rho_{\mathrm{L}}\Delta\rho}}{1+\dfrac{P_{\mathrm{G}}^{0}}{(1-\alpha)\Delta\rho g L_{\mathrm{G}}^{0}}}\right)^{0.5};\ \xi_{\mathrm{Turbulent}} = 0 \qquad \textbf{B.10}$$